\documentclass[aps,prb,preprint,onecolumn,amsmath,amssymb,superscriptaddress,eqsecnum,floatfix,scrartcl]{revtex4-2}
\pdfoutput=1      
\usepackage{amssymb}  
\usepackage{color,soul} 
\usepackage{graphicx}  
\usepackage[normalem]{ulem}
\usepackage[tight]{subfigure}
\usepackage{mathrsfs}
\usepackage[pdftex,pdfusetitle,bookmarks=true,colorlinks=true,citecolor=blue,urlcolor=blue,linkcolor=magenta]{hyperref}
\usepackage{hypcap}
\usepackage{multirow}
\usepackage{float}
\usepackage{bm}

\definecolor{OliveGreen}{cmyk}{0.64, 0, 0.95, 0.40}
\definecolor{forestgreen}{rgb}{0.13, 0.55, 0.13}

\DeclareMathOperator{\Tr}{Tr}

\newcommand{\ket}[1]{\lvert#1\rangle}
\newcommand{\bra}[1]{\langle#1\rvert}

\begin{document}
\title{Entanglement spectra of non-chiral topological (2+1)-dimensional phases with strong time-reversal breaking, Li-Haldane state counting, and PEPS}
\date{\today}

\author{Mark J.~Arildsen}
\email{marildse@sissa.it}
\affiliation{Department of Physics, University of California, Santa Barbara, California 93106, USA}
\affiliation{SISSA, via Bonomea 265, 34136 Trieste, Italy}

\author{Norbert Schuch}
\affiliation{University of Vienna, Faculty of Mathematics,
Oskar-Morgenstern-Platz 1, 1090 Wien, Austria}
\affiliation{University of Vienna, Faculty of Physics,
Boltzmanngasse 5, 1090 Wien, Austria}

\author{Andreas W.~W.~Ludwig}
\affiliation{Department of Physics, University of California, Santa Barbara, California 93106, USA}

\begin{abstract}
    The Li-Haldane correspondence [PRL 101, 010504 (2008)] is often used to help identify wave functions of (2+1)-D chiral topological phases (i.e., with non-zero chiral central charge), by studying low-lying entanglement spectra (ES) on long cylinders of finite circumference. Here we consider such ES of states [in fact, of wave functions of certain Projected Entangled Pair States (PEPS)] that are not chiral (i.e., having zero chiral central charge), but which strongly break time-reversal as well as reflection symmetry, while preserving their product, thus  exhibiting the same symmetry properties as a chiral topological state.
    This leads to ES which have branches of both right- and left-moving chiralities, but with vastly different velocities. For circumferences much smaller than the inverse entanglement gap scale, the low-lying ES appear chiral in some topological sectors, and precisely follow the Li-Haldane state counting of a truly chiral phase. On its face, this could lead one to mistakenly identify the phase as chiral. However, by considering the ES in all possible sectors, one can observe distinct differences from a chiral phase. We explore this phenomenon in the setting of an SU(3) spin liquid PEPS studied by Kure\v{c}i\'c, et al.~[PRB 99, 045116 (2019)], where the topologically trivial sector appeared to have the characteristic Li-Haldane counting of a chiral $\mathrm{SU}(3)$-level-one [$\mathrm{SU}(3)_1$] Conformal Field Theory (CFT). In fact, the PEPS has $D(\mathbb{Z}_3)$ topological order, with 9 sectors. We compute the ES in minimally entangled states corresponding to all these sectors, which map to the 9 anyon types of doubled $\mathrm{SU}(3)$-level-one Chern-Simons Topological Field Theory. The state countings of the ES in all sectors are exactly consistent with our expectation: specifically, the ES contain representations of global SU(3) symmetry from the tensor products of the (lowest-lying) multiplet of primary states of a ``high-velocity" chiral $\mathrm{SU}(3)_1$ CFT with the full content of a ``low-velocity" chiral $\mathrm{SU}(3)_1$ CFT sector, a non-chiral structure beyond that observable in the topologically trivial sector of the ES.
\end{abstract}

\maketitle

\newpage
\tableofcontents
\newpage

\section{Introduction}
Identifying chiral topological order in quantum states in 2+1 dimensions that break time-reversal symmetry is a problem of substantial import. By ``chiral" here, and in the rest of this work, we mean specifically that the chiral central charge~\footnote{See, e.g., Ref.~\cite{KitaevAnnalsPhysics2006}} associated to the bulk topological state is non-zero. Thus, even a state that strongly breaks time-reversal symmetry may indeed be non-chiral by virtue of possessing a zero chiral central charge, a point which is crucial to keep in mind for what follows.
One powerful tool for understanding the topological order of such chiral states is the entanglement spectrum, which for chiral topological states will be a ``chiral" entanglement spectrum, with distinct chiral branches.
The Li-Haldane entanglement-edge correspondence can then be used to relate the low-lying entanglement spectrum at finite size and properties of a corresponding theory of a physical edge of the topological state~\cite{Li2008}, and so the chiral branches in such a finite size entanglement spectrum will additionally exhibit characteristic ``Li-Haldane''
multiplicities of particular types of low-lying states at fixed momentum.
Yet more generally, however, calculation of the low-lying entanglement spectrum in only a subset of the available topological sectors can fail to uniquely identify the correct nature, i.e. chiral or not chiral, of the topological state. 
Diagnostic features of the entanglement spectrum at finite size such as the characteristic ``Li-Haldane'' multiplicities can appear {\it the same} in some sectors for distinct types of topological state. 
This can lead to a low-lying entanglement spectrum in those sectors which appears chiral, even for a non-chiral topological state. Looking at all of the sectors of the entanglement spectrum, however, can reveal the full picture. 
This approach provides a way to correctly diagnose topological states that are ``close-to-chiral", in the sense that they possess some features consistent or nearly consistent with chirality (such as, e.g., possessing identical symmetry properties, namely broken time-reversal as well as reflection symmetries, while preserving their product), as non-chiral.

Topological states satisfy an entanglement entropy area law~\cite{Kitaev2006}, and one way to build quantum states that obey, by construction, the entanglement entropy area law is by the use of  Projected Entangled Pair States (PEPS) tensor network methods~\cite{Verstraete2006,schuch:peps-sym}.
PEPS tensors are able to encode the symmetries of quantum states directly at the microscopic level and have been constructed to represent a wide variety of topological quantum states. 
A famous result~\cite{Dubail2015,Wahl2013}
states that in the case of non-interacting fermions, PEPS wavefunctions {\it can} describe corresponding chiral non-interacting (2+1)-dimensional topological phases. (A Hamiltonian possessing
such a wave function of non-interacting fermions as a ground state is required to be either
long-ranged, or gapless if local, a statement referred to as the ``no-go theorem".)
The generalization to interacting PEPS, however, is an open question.
Thus, the question of whether ``close-to-chiral" PEPS can, perhaps in certain limits, represent truly chiral states is one of significant particular interest.

The type of chiral topological state whose representation by PEPS is addressed in this work is the $\mathrm{SU}(3)$ chiral spin liquid. $\mathrm{SU}(N)$ chiral spin liquids with $N > 2$ have experimental relevance: there is a long-standing proposal for implementation of $\mathrm{SU}(N)$ magnetism in ultracold alkaline-earth atoms in optical lattices through decoupling of nuclear spin and electronic angular momentum \cite{Gorshkov2010}. Theoretical work has identified an Abelian $\mathrm{SU}(N)$ chiral spin liquid state feasible on such a platform \cite{Nataf2016}. The ultracold alkaline-earth atom approach to $\mathrm{SU}(N)$ symmetric systems has seen extensive experimental activity \cite{Cazalilla2014}, including to implement an $\mathrm{SU}(N)$ fermionic liquid in 1D \cite{Pagano2014}, to access $\mathrm{SU}(N)$ Mott insulators \cite{Taie2012,Hofrichter2016,Taie2022}, and to investigate $\mathrm{SU}(N)$ quantum magnetism \cite{Zhang2014,Scazza2014,Cappellini2014,Ozawa2018}. Beyond ultracold atoms, $\mathrm{SU}(N)$ chiral spin liquids may also occur in condensed matter systems such as a proposed $\mathrm{SU}(4)$ chiral spin liquid in double-layer moir\'{e} superlattices \cite{Zhang2021}. As more such modalities come online, it will be important to be able to guarantee the presence of chiral topological order. Appropriate probes will be required to accomplish this goal in these experimental settings. However, in the context of numerical models, as discussed above, the entanglement spectrum has proven to be an accessible diagnostic.

Our goal in this work is thus to understand a PEPS through the content of its entanglement spectrum. The $\mathrm{SU}(3)$ spin liquid PEPS we analyze was first studied in Ref.~\cite{kurecic:su3_sl}, which found that the PEPS is non-chiral, but possesses left- and right-moving branches in the entanglement spectrum that differ significantly in velocity.
On the other hand, the same analysis also observed that the low-lying entanglement spectrum of this PEPS on the cylinder, in a particular charge and flux sector, has $\mathrm{SU}(3)$ representation content, at given momentum, that is precisely consistent with the presence of the (1+1)D \emph{chiral} $\mathrm{SU}(3)$-level-1 [$\mathrm{SU}(3)_1$] Wess-Zumino-Witten (WZW) Conformal Field Theory (CFT).
Ordinarily, the presence of this counting of representation content
 is regarded as an indicator of chiral topological behavior due to the Li-Haldane correspondence, as discussed above. 
Through a full accounting of the representation content of the low-lying levels of the entanglement spectrum in \emph{all} charge and flux sectors of the PEPS, however, we are able to understand that the counting results in this case are actually consistent with the non-chiral topological order of the PEPS, in a sense extending the result of Li and Haldane, and we are able to use the entanglement spectrum to quantitatively validate the non-chiral picture above.

The first task in this effort is to understand the mapping between the topological order reflected in the PEPS and a corresponding (2+1)D topological quantum field theory. 
Specifically, we wish to establish the correspondence between the charge and flux topological sectors of the PEPS (accessible by charge projection and flux threading, as discussed below), and the corresponding anyonic sectors of the topological quantum field theory. 
The PEPS we consider has $D(\mathbb{Z}_3)$ ($\mathbb{Z}_3$ Drinfeld double) topological order, which is the same topological order as that of doubled $\mathrm{SU}(3)$-level-one Chern-Simons theory (in short: ``$\mathrm{SU}(3)_1 \otimes \overline{\mathrm{SU}(3)_1}$ doubled Chern-Simons theory").
Therefore, in Section \ref{sec:anyons}, we write down a mapping between the $D(\mathbb{Z}_3)$ charge and flux sectors and the topological (anyonic) sectors of $(2+1)$D $\mathrm{SU}(3)_1 \otimes \overline{\mathrm{SU}(3)_1}$ doubled Chern-Simons theory. 
We will then use this mapping to understand the appropriate correspondence of the content of this doubled Chern-Simons theory to each of the nine charge and flux sectors of $D(\mathbb{Z}_3)$.

Next, we must consider how the entanglement spectrum, computed on a bipartitioned (surface of an) infinite cylinder, is able to detect these topological sectors.
In Section \ref{sec:esmes}, we describe the entanglement Hamiltonian (the spectrum of which is the entanglement spectrum) and first outline why we might expect that this spectrum must exhibit the content of a (1+1)D Conformal Field Theory (CFT) reflecting the theory underlying the description of a physical boundary (``edge state'') of the topological state
\footnote{In the case under consideration, the entanglement spectrum, just as a physical boundary of our topological state, will eventually be gapped in the limit of large system size (cylinder circumference). However, we will investigate the entanglement spectrum at system sizes much smaller than the inverse entanglement gap (proportional to the correlation length). In this limit the entanglement spectrum (as well as the spectrum of a physical boundary) reflects the spectrum of the underlying gapless CFT, and thus the entanglement spectrum serves as a direct diagnostic of the underlying Topological Field Theory.}.
The minimally entangled states (MES) are also explained. 
These allow us direct access to states in the space of ground states that can be assigned particular topological quantum numbers. 
The construction of the MES is briefly outlined as well.
This discussion explains how the entanglement spectrum is able to obtain the topological data necessary to realize the mapping of Section \ref{sec:anyons}. 

The construction of the PEPS wavefunction itself is set out in Section \ref{sec:peps}, where it is shown to be describable as a tensor network on the square lattice. 
Section \ref{sec:peps} additionally discusses the method of actually constructing the minimally entangled sectors in the PEPS through charge projection and flux threading, taking advantage of the symmetries of the PEPS. 
Finally, we describe how to compute the entanglement spectrum from the PEPS, including the numerical methods used to do this.

We detail our expectations for that entanglement spectrum in Section \ref{sec:cftes}.
Each of the nine topological sectors of the entanglement spectrum as understood in the charge and flux basis of the $D(\mathbb{Z}_3)$ Drinfeld double  description of the PEPS can be mapped by our mapping of Section \ref{sec:anyons} to the anyonic basis of $\mathrm{SU}(3)_1 \otimes \overline{\mathrm{SU}(3)_1}$ doubled $(2+1)$D Chern-Simons theory.
In this formulation we can, building on the results of Refs.~\cite{Qi2012}, and \cite{ArildsenLudwig2022}, understand the CFT content of the low-lying entanglement spectrum, which will be seen to consist of two chiral, left- and right-moving branches of $\mathrm{SU}(3)_1$ WZW CFT.
We calculate, based on this CFT description, the countings of irreducible representations under the global $\mathrm{SU}(3)$ symmetry that we expect to be present in the spectrum due to the two chiral, left- and right-moving $\mathrm{SU}(3)_1$ branches of WZW CFT.
In our PEPS, since one of these two chiral branches turns out to have a substantially lower velocity than the other, we will see that the content of the low-lying ES should consist of the tensor products of $\mathrm{SU}(3)$ representations of the full content (primary and descendant states) of the entire low-velocity chiral $\mathrm{SU}(3)_1$ branch, with only the lowest-level primary $\mathrm{SU}(3)$ representation from the high-velocity chiral $\mathrm{SU}(3)_1$ branch. 
We also clarify why this CFT content shows up in the entanglement spectrum, making use of the conformal boundary state approach to the Li-Haldane correspondence outlined in Refs.~\cite{Qi2012} and \cite{ArildsenLudwig2022}, generalized to the non-chiral, but ``close-to-chiral" case that we have here.

Finally, in Section \ref{sec:results}, we exhibit the low-lying states of the numerical entanglement spectra of the PEPS, computed as discussed in Section \ref{sec:peps}.
These confirm the expected countings of Section \ref{sec:countings} in all nine sectors.
We are also able to extract the velocity ratio of the velocities of the high- and low-velocity branches, which allows us to confirm that none of the descendant representations from the high-velocity chiral $\mathrm{SU}(3)_1$ branch appear in numerical ES data that are available to us for analysis.
These results show that we have correctly diagnosed the nature of the non-chiral topological order in the PEPS we analyze.

\section{Anyons}
\label{sec:anyons}
We consider $(2+1)$-dimensional $\mathrm{SU}(3)_1 \otimes \overline{\mathrm{SU}(3)_1}$ doubled Chern-Simons theory, which possesses, as already mentioned, the same $D(\mathbb{Z}_3)$ ($\mathbb{Z}_3$ quantum double) topological order as the PEPS model.
We can then write down an explicit mapping between the anyons present in the PEPS [and $D(\mathbb{Z}_3)$] and the corresponding anyons of $\mathrm{SU}(3)_1 \otimes \overline{\mathrm{SU}(3)_1}$ Chern-Simons theory that will (as discussed below) determine the characteristic state countings of the entanglement spectrum.

To write down the mapping, we must understand the expression of the topological order on both sides. 
In $D(\mathbb{Z}_3)$ (as realized, for example, in the $\mathbb{Z}_3$ generalization of the toric code \cite{Kitaev2003}, though what matters for our purposes here will be the anyonic statistics) we have anyons of type $e^q m^\phi$, where $e$ and $m$ are unit $\mathbb{Z}_3$ gauge charge and flux, respectively, and $q$ and $\phi$ are integers modulo 3 (i.e., $e^2$ is equivalent to $\bar{e} = e^{-1}$, and similarly for $m$).
There are thus nine such anyons, which we can denote by $(q,\phi)$.
The associated modular $S$-matrix is given by (see, e.g., Ref.~\cite{Bakalov2001})
\begin{equation}
    \label{eq:sdz3}
	S^{D(\mathbb{Z}_3)}_{(q_1,\phi_1)(q_2,\phi_2)} = \omega^{-q_1 \phi_2 - q_2 \phi_1},
\end{equation}
where $\omega = e^{\frac{2\pi i}{3}}$. 
In turn, we consider $(2+1)$-dimensional doubled Chern-Simons theory
\footnote{Its topological order can be related, e.g., to the topological order of the stacking of a fractional quantum Hall state at filling fraction $\nu=1/3$ and its time-reversed partner.}.
The Chern-Simons theory will again have nine types of anyon, which can be written as $(\kappa,\bar{\kappa})$ for $\kappa, \bar{\kappa}$ integers modulo 3, corresponding to the anyons in chiral $\mathrm{SU}(3)_1$ and anti-chiral $\overline{\mathrm{SU}(3)_1}$ Chern-Simons theories, respectively. 
We can then write the modular $S$-matrix for the doubled Chern-Simons theory as (see, e.g., Ref.~\cite{Nayak2008})
\begin{equation}
    \label{eq:sdsu3}
    S^{\mathrm{SU}(3)_1 \otimes \overline{\mathrm{SU}(3)_1}}_{(\kappa_1,\bar{\kappa}_1)(\kappa_2,\bar{\kappa}_2)} = \omega^{\kappa_1\kappa_2-\bar{\kappa}_1\bar{\kappa}_2}.
\end{equation} 
In the same notation, we can write the modular $T$-matrices as 
\begin{align}
    \label{eq:tdz3}
    T^{D(\mathbb{Z}_3)}_{(q_1,\phi_1)(q_2,\phi_2)} &= \delta_{q_1 q_2} \delta_{\phi_1\phi_2} \omega^{q_1 \phi_1} \\
    \label{eq:tdsu3}
    T^{\mathrm{SU}(3)_1 \otimes \overline{\mathrm{SU}(3)_1}}_{(\kappa_1,\bar{\kappa}_1)(\kappa_2,\bar{\kappa}_2)} &= \delta_{\kappa_1 \kappa_2} \delta_{\bar{\kappa}_1\bar{\kappa}_2} \omega^{\kappa_1^2-\bar{\kappa}_1^2}.
    \quad
\end{align} 
From equating the entries of the $S$ and $T$ matrices of Eqs.~\eqref{eq:sdz3}-\eqref{eq:tdsu3} we can obtain equations relating the $(q,\phi)$ anyon type in the $D(\mathbb{Z}_3)$ theory to the corresponding $(\kappa,\bar{\kappa})$ type in $\mathrm{SU}(3)_1 \otimes \overline{\mathrm{SU}(3)_1}$ doubled Chern-Simons theory.
This is worked out in Appendix \ref{app:matrixmap}.
Up to a choice of the sign of $\kappa$ and $\bar{\kappa}$, which amounts to the choice of the assignment of the sign of $\kappa$ and $\bar{\kappa}$ to the non-trivial topological sectors of chiral $\mathrm{SU}(3)_1$ and $\overline{\mathrm{SU}(3)_1}$ Chern-Simons theories, respectively, we have
\begin{align}
    \label{eq:charge}
	q &= \kappa+\bar{\kappa} \  ({\rm mod} \ 3) \\
    \label{eq:flux}
	\phi &= \kappa-\bar{\kappa}  \ ({\rm mod} \ 3).
\end{align}
This correspondence between the nine $D(\mathbb{Z}_3)$ anyons of type $e^q m^\phi$ and those of $\mathrm{SU}(3)_1 \otimes \overline{\mathrm{SU}(3)_1}$ doubled Chern-Simons theory of type $(\kappa,\bar{\kappa})$ is made explicit in Table \ref{table:anyoncomparison}.
\begin{table}
	\begin{tabular}{cc|c|c|c}
		\multicolumn{2}{c|}{\multirow{2}{*}{$(\kappa, \bar{\kappa})$}} & \multicolumn{3}{c}{Charge $q$} \\
		 \multicolumn{2}{c|}{} & 0 & 1 & 2 \\
		 \hline
		 \multirow{3}{*}{Flux $\phi$} & 0 & $(0,0)$ & $(-1,-1)$ & $(+1,+1)$ \\
		 \cline{2-5}
		 & 1 & $(-1,+1)$ & $(+1,0)$ & $(0,-1)$ \\
		 \cline{2-5}
		 & 2 & $(+1,-1)$ & $(0,+1)$ & $(-1,0)$ 
	\end{tabular}
    \caption{This table illustrates the relation between the anyons of the quantum double $\mathbb{Z}_3$ of type $e^q m^\phi$ (which identify the rows and columns) and the anyons of the $\mathrm{SU}(3)_1 \otimes \overline{\mathrm{SU}(3)_1}$ doubled Chern-Simons theory of type $(\kappa,\bar{\kappa})$ that possess the same statistics (which occupy the table entries).}
    \label{table:anyoncomparison}
\end{table}

\section{Entanglement spectrum and minimally entangled states}
\label{sec:esmes}

In this section, we introduce entanglement spectra, and explain how they decompose into topological sectors, how these sectors can be constructed, and how they are labeled by the anyons of the theory.

\subsection{Entanglement spectrum}

\begin{figure}[t]
    \includegraphics[width=246pt]{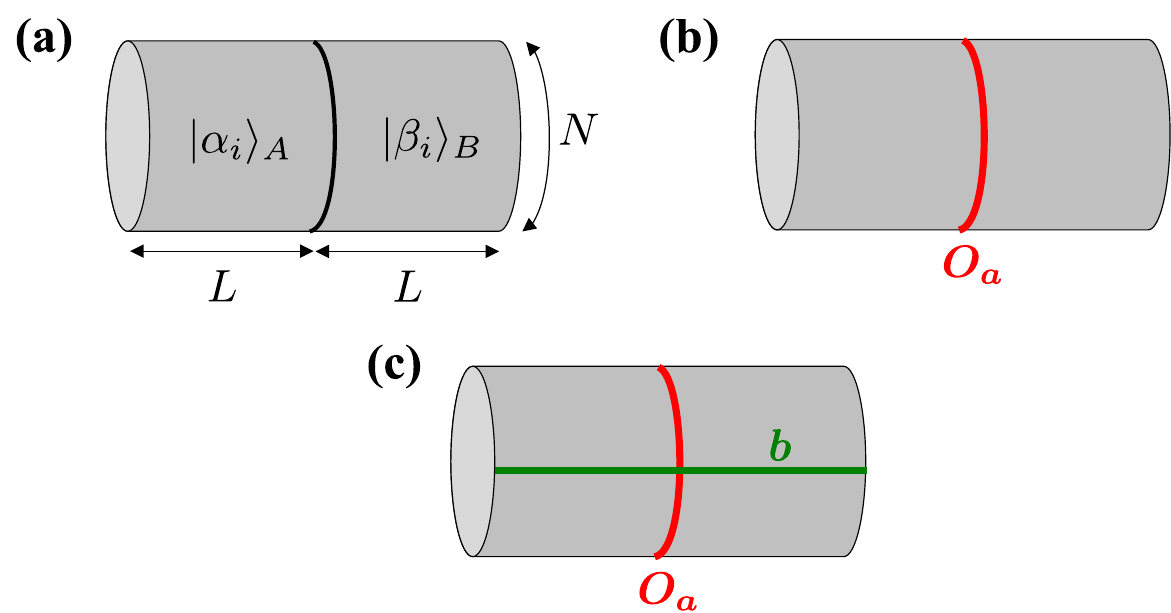}
    \caption{Construction of minimally entangled states, cf.\ text. 
    \textbf{(a)} Bipartition on a cylinder.
    \textbf{(b)}~The operator $O_a$ measures the effect of transporting an anyon $a$ along a loop around the cylinder; its eigenspaces (for a complete set of $O_a$) yield the minimally entangled states.
    \textbf{(c)}~Minimally entangled states can be transformed into each other by moving an anyon $b$ from one end of the cylinder to the other (``threading it through the cylinder''), which changes the eigenvalue of $O_a$ as determined by the anyonic braiding statistics. }
    \label{fig:entspec}
\end{figure}

Let us consider a system on a cylinder of length $2L$ and circumference $N$, where $L\gg N$, and -- for now -- a unique ground state of a gapped Hamiltonian on that system, or an otherwise uniformly generated state (e.g.,\ a uniform PEPS) with exponentially decaying correlations, which we denote by $\ket\Psi$. The gap (or decay of correlations) ensures that the properties in the bulk become independent of the choice of boundary conditions as $L$ becomes sufficiently large. Now consider a bipartition of the cylinder in the middle, Fig.~\ref{fig:entspec}(a), and consider its Schmidt decomposition
\begin{equation}
    \label{eq:schmidt}
    \ket\Psi = 
    \sum_i e^{-E_i} \ket{\alpha_i}_A\ket{\beta_i}_B\ ,
    \quad E_i\in\mathbb R\ ,
\end{equation}
where we have expressed the Schmidt coefficients in terms of ``entanglement energies'' $E_i$.
The system (as the ground state of a gapped Hamiltonian, or a PEPS)  
obeys an area law for the entanglement entropy: 
the entanglement between any region and its complement 
scales as the length of the boundary, and
thus the amount of entanglement in the considered
bipartition scales linearly with $N$, but is independent of $L$. 
Thus, only on the order of $c^{N}$ Schmidt coefficients $e^{-E_i}$ dominate the Schmidt decomposition, i.e., have a low entanglement energy $E_i$. At the same time, the local distribution of the entanglement implies that those dominant $E_i$ capture the entangled degrees of freedom close by to the cut, and will thus not depend on the boundary conditions as $L$ increases.

The fact that the number of low-lying $E_i$ scales as the number of degrees of freedom along the 
cut suggests that those $E_i$ can be interpreted as the spectrum of a 1D Hamiltonian $H_{1D}$ associated to the boundary at the cut,  the \emph{entanglement Hamiltonian}. To further fix $H_{1D}$, we can make use of quantum numbers in the Schmidt decomposition. For any symmetry of $\ket\Psi$, we can choose a Schmidt decomposition where multiplets $\{\ket{\alpha_i}\}$ and $\{\ket{\beta_i}\}$ transform as conjugate irreducible representations (irreps) of the symmetry group, and thus, the $E_i$ can correspondingly be labeled by those irreps. In particular, this applies to translational symmetry around the cylinder (i.e., momentum), as well as to on-site symmetries such as the $\mathrm{SU}(3)$ symmetry relevant in this work. 
The resulting spectrum $E_i$, labeled by the appropriate quantum numbers, is accordingly termed \emph{entanglement spectrum}.
Indeed, using momentum labels it has first been observed by Li and Haldane~\cite{Li2008} in a slightly different setup, considering fractional quantum Hall wavefunctions on a sphere, that the resulting entanglement spectrum given by the low-lying $E_i$ reproduces the spectrum of a chiral (1+1)D CFT. 
In PEPS, this connection of the entanglement spectrum to a one-dimensional theory can be made explicit, and a corresponding $H_{1D}$ can be directly computed (see later). 
Further, in Section \ref{sec:cftes}, we are able to clearly describe the relation of that $H_{1D}$ to the Hamiltonian of a (1+1)D non-chiral, ``doubled'' CFT, where ``doubled'' refers to the presence of both chiral sectors, ``right-moving" and ``left-moving": 
upon employing a conformal boundary state description \cite{Qi2012,ArildsenLudwig2022}, we will then see why the structure of $H_{1D}$ ought to have this form, an extension of the observation of Li and Haldane~\cite{Li2008}.

\subsection{Entanglement spectrum of topologically ordered systems and minimally entangled states}
\label{sec:mes}

Let us now turn towards the entanglement spectrum of systems with topological order, as before on a cylinder. 
Topologically ordered systems on the cylinder exhibit a degenerate space of ground states, spanned by topological sectors. As before, the properties in the bulk, including the entanglement spectrum, are independent of the details of the boundary conditions, \emph{except} for the selection of the topological sectors. Clearly, an entanglement spectrum can be obtained from any state in the space of ground states.
However, it turns out that there is a discrete set of \emph{extremal} states which spans the ground space manifold, the so-called \emph{minimally entangled states} (MES) $\ket{\Psi^\theta}$. They are characterized by the fact that in the Schmidt decomposition, the states in either half of the system are entirely supported in a subspace characterized by a topological quantum number $\theta$,
\begin{equation}
    \label{eq:schmidtdec-mes}
    \ket{\Psi^\theta} = 
    \sum_i e^{-E_{i,\theta}} \ket{\alpha^\theta_i}\ket{\beta^\theta_i}\ ,
\end{equation}
where $\bra{\alpha^{\theta}_i}\alpha^{\theta'}_{i'}\rangle=\delta_{\theta,\theta'}\delta_{i,i'}$ (and correspondingly for the $\ket{\beta_i^\theta}$), and states with a different label $\theta$ are distinguished by a \emph{topological} property. It thus follows that a general ground state is of the form
\begin{equation}
    \label{LabelEqShiftedEntanglementEnergies}
    \ket\Psi = \sum c_\theta \ket{\Psi^\theta} = 
    \sum_{\theta,i} e^{-(E_{i,\theta}-\log |c_\theta|)} 
       \, \big(
        \tfrac{c_\theta}{|c_\theta|}
        \ket{\alpha^\theta_i}\big)\,\ket{\beta^\theta_i}\ ,
\end{equation}
and thus, the full entanglement spectrum of $\ket{\Psi}$ is obtained by mixing the (shifted) entanglement spectra of the individual topological sectors $\ket{\Psi^\theta}$. (This also confirms that the $\ket{\Psi^\theta}$ are the states with the minimal entanglement across the bipartition \cite{Zhang2012}.)

How can the MES $\ket{\Psi^\theta}$ be constructed? To this end, let us first define loop operators $O_a$: These are operators which transport an anyon of type $a$ once around the cylinder [Fig.~\ref{fig:entspec}(b)].
Due to the topological nature of the system, it does not matter along which specific path $O_a$ is applied.
For the given scenario, where the anyons transform as representations of an Abelian group, we have that some $k$ copies of $a$ fuse to the vacuum, and thus, the eigenvalues of $O_a$ are $k$'th roots of unity. 
We can thus decompose any state $\ket\Psi$ into eigenspaces of $O_a$. In order to obtain an exhaustive decomposition, we have to consider a maximal set $\mathcal A$ of anyons where no $a\in \mathcal A$ can be obtained by fusing any of the other anyons in $\mathcal A$.
Specifically, for $D(\mathbb Z_3)$, it is sufficient to consider e.g.,\ $\mathcal A=\{(q=1,\phi=0),(q=0,\phi=1)\}$, while for chiral $\mathrm{SU}(3)_1$ Chern-Simons theory, considering a single anyon $\kappa = 1$ suffices. 
[For doubled $\mathrm{SU}(3)_1 \otimes \overline{\mathrm{SU}(3)_1}$ Chern-Simons theory, one then accordingly chooses $\mathcal A=\{(\kappa=1,\bar \kappa=0),(\kappa=0,\bar \kappa=1)\}$.] 
Any such sector is then labeled by the eigenvalues of the $O_a$ for all $a\in\mathcal A$, which we denote by $\theta$. We can then write
\begin{equation}
    \ket\Psi = \sum_\theta c_\theta \ket{\Psi^\theta}\ .
\end{equation}
Since the $O_a$ act the same way anywhere and the eigenspace projections can be applied more than once -- specifically, once in each half of the system -- the label $\theta$ equally applies to the basis vectors in the Schmidt decomposition, cf.~\eqref{eq:schmidtdec-mes}. Moreover, the different $\ket{\Psi^\theta}$ (and the corresponding Schmidt vectors) live in different eigenspaces of the $O_a$, and thus in different topological sectors. 
Finally, since the number of different $\theta$ equals the number of anyons and thus the dimension of the space of degenerate ground states on the cylinder, they cannot be decomposed further. They thus form a basis of minimally entangled states spanning the whole space, as introduced above. 

It remains to discuss how the different MES can be related; this will be in particular relevant to the construction of the MES states in the PEPS representation in Section~\ref{sec:peps}. To this end, take $\ket{\Psi^{\theta_0}}$, where $\theta_0$ denotes the sector with trivial irrep for all $a\in\mathcal A$. Now create an anyon $b$ at the left boundary and move it to the right boundary, Fig.~\ref{fig:entspec}(c). As the trajectory of anyon $b$ crosses the loop operator $O_a$, this will give rise to a phase $S_{b,a}$ in the eigenvalue of $O_a$, and thus transform $\ket{\Psi^{\theta_0}}$ to another $\ket{\Psi^{\theta'}}$, where $\theta'$ is obtained from updating the eigenvalues of all $a\in\mathcal A$ which do not braid trivially with $b$ according to the modular $S$-matrix. It is immediate to see that this way, all MES $\ket{\Psi^\theta}$ can be obtained by threading an arbitrary anyon $b$ through the cylinder, starting from $\ket{\Psi^{\theta_0}}$; this establishes a one-to-one correspondence between MES (and thus the sectors of the entanglement spectrum) and the anyons of the theory.

Overall, this discussion provides two different ways to construct the MES: Either, one starts from a state $\ket\Psi$ which has overlap with all MES sectors, and applies projections onto the eigenspaces of the $O_a$, $a\in\mathcal A$, or one starts from a specific MES, such as $\ket{\Psi^{\theta_0}}$, and creates the other MES by threading anyons along the cylinder. Moreover, both schemes can be combined, that is, one starts from a superposition of a subset of MES $\ket\Psi$ and constructs the full set of MES by a combination of applying irrep projections and threading anyons for suitable subsets of anyons; this approach will be used for the PEPS in the next section.

Finally, the labelling of MES, and thus entanglement spectra, by the anyons of the theory, combined with the mapping from Section~\ref{sec:anyons} between the anyons of the Drinfeld double $D(\mathbb Z_3)$ and the $\mathrm{SU}(3)_1\otimes \overline{\mathrm{SU}(3)_1}$ doubled Chern-Simons theory, provides us with a mapping between the different sectors of the entanglement spectra of the respective theories.

\section{The PEPS model and its entanglement spectrum}
\label{sec:peps}

In this section, we introduce the topological $\mathrm{SU}(3)$ spin liquid
which we will study. We give its construction in terms of Projected
Entangled Pair States (PEPS), discuss how to construct the minimally
entangled sectors, and explain how to extract the entanglement spectrum.

\subsection{The model}

The model wavefunction which we focus on has been constructed and studied
in Ref.~\cite{kurecic:su3_sl} as an $\mathrm{SU}(3)$ topological spin
liquid on the kagome lattice. It was found numerically that it describes a
gapped system with $D(\mathbb Z_3)$ topological order. It possesses a
gapped entanglement spectrum and is thus non-chiral; yet, the left-
and right-moving branches of the entanglement spectrum possess very
different velocities, giving rise to a clearly recognizable right-moving
branch which was found to exhibit a Li-Haldane state-counting 
compatible with the trivial sector of a chiral $\mathrm{SU}(3)_1$ CFT 
in the sectors studied.

\begin{figure}
    \includegraphics[width=246pt]{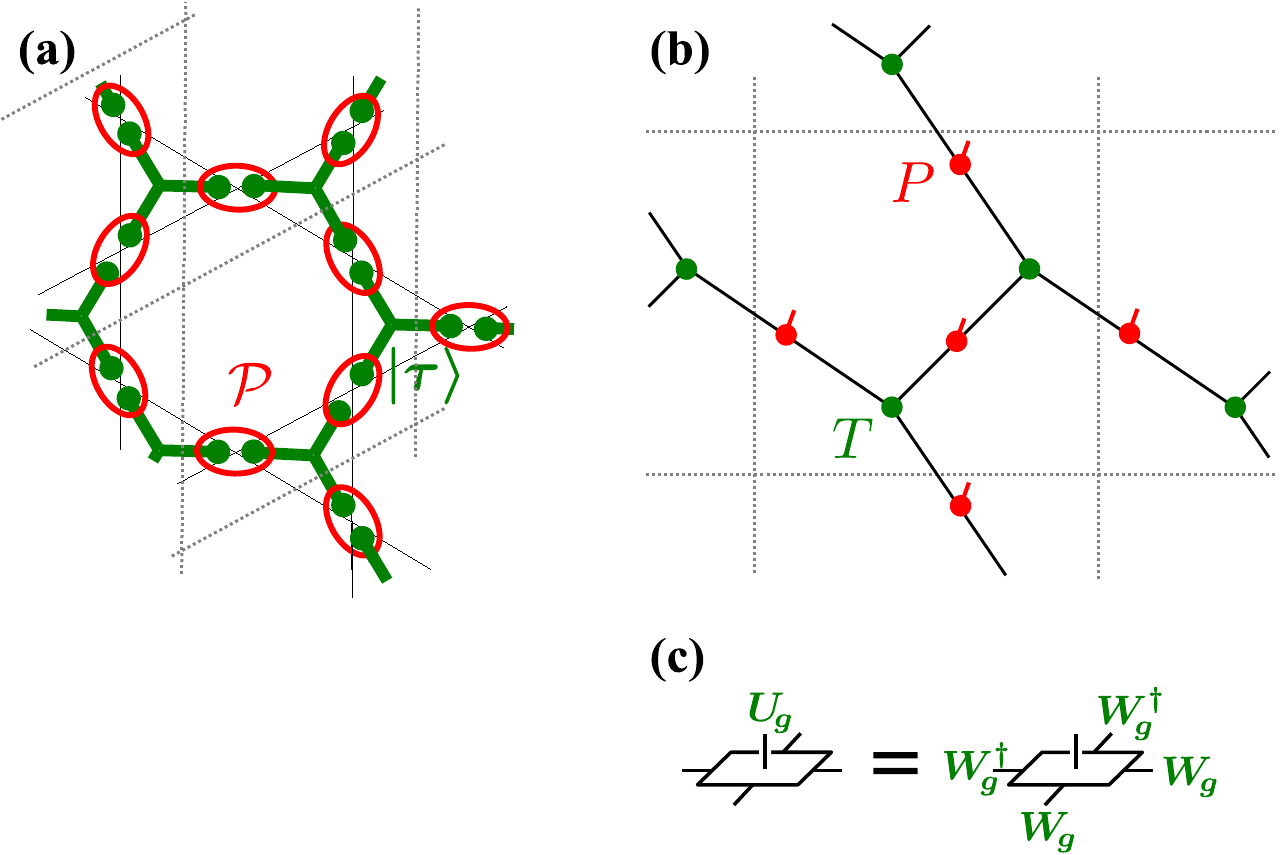}
    \caption{\textbf{(a)} The PEPS model considered is constructed from
    tripartite $\ket\tau$ tensors, Eq.~\eqref{eq:def-tau}, aligned clockwise,
    to which the projection $\mathcal P$, Eq.~\eqref{eq:def-P}, is applied as
    indicated, resulting in a wavefunction on the kagome lattice. The dashed
    lines indicate the unit cell. 
    \textbf{(b)} Tensor network representation
    of model, after shearing around the cylinder axis to obtain a square
    lattice from the indicated unit cell. The cylinder has periodic boundaries
    in vertical direction, with the cylinder axis aligned horizontally.
    \textbf{(c)} The resulting square lattice tensor has an $\mathrm{SU}(3)$
    symmetry, with physical representation $U_g\equiv \bm 3\otimes\bm 3\otimes
    \bm 3$, and virtual representation $W_g=\bm 1\oplus \bm 3 \oplus \bar{\bm
    3}$.
    \label{fig:peps}}
\end{figure}

We start by briefly reviewing the construction of the wavefunction. We
start from a tripartite state $\ket\tau\in\mathbb V^{\otimes 3}$, with
the seven-dimensional \emph{virtual space} $\mathbb V$ transforming as
the $\bm 1 \oplus \bm 3 \oplus \overline{\bm{3}}$ irrep of $\mathrm{SU}(3)$,
with orthonormal basis vectors $\ket{0}$,
$\ket{s}$, and $\ket{\bar s}$, $s=1,2,3$. It is defined as
\begin{align}
    \ket\tau  &= 
    \ket{0,0,0} 
    -\frac{i}{\sqrt{6}} \sum \varepsilon_{pqr}\big(\ket{p,q,r} +\ket{\bar p,\bar q,\bar r}\big)
    \nonumber
    \\
    \label{eq:def-tau}
    &
    +\frac{1}{\sqrt{3}} \sum \big(\ket{s,\bar s,0}+\ket{\bar s,s,0}
    \\
    &\qquad\quad
    +\ket{s,0,\bar s}+\ket{\bar s,0,s}+\ket{0,s,\bar s}+\ket{0,\bar s,s}\big)
\nonumber
\end{align}
(with $\varepsilon$ the fully antisymmetric tensor),
that is, it is an equal weight superposition of all nine $\mathrm{SU}(3)$
singlets in the tensor product $\mathbb V^{\otimes 3}$. 
Here, the phases
are chosen such that $\ket\tau$ is rotationally invariant and acquires a
complex conjugation under reflection (a ``chiral'' symmetry also used in
the construction of chiral PEPS~\cite{poilblanc:kl-peps-1,Poilblanc2016}).
We then arrange states $\ket\tau$ on the triangles of the kagome lattice as
shown in Fig.~\ref{fig:peps}(a), with the sites oriented clockwise,
and subsequently apply a map 
$\mathcal P:\mathbb V\otimes \mathbb V\to \mathbb C^3$
at each vertex, which maps the two virtual systems adjacent to the vertex
to a physical system equipped with the $\bm 3$ representation. Here,
$\mathcal P$ is given by
\begin{equation}
    \label{eq:def-P}
    \mathcal P = \frac13 \sum_s \:\ket{s} \Big[ \bra{0,s} - \bra{s,0}
    -\sqrt{\frac32} \sum_{pq} \varepsilon_{spq}\bra{\bar p,\bar q}\Big]\ .
\end{equation}
As $\mathcal P$ transforms odd under reflection, its alignment does not
matter (up to a global phase). Since $\mathcal P$ commutes with the
action of $\mathrm{SU}(3)$, and $\ket\tau$ transforms trivially under
$\mathrm{SU}(3)$, the construction thus yields an $\mathrm{SU}(3)$
invariant wavefunction $\ket\Psi$ on the torus.

The wavefunction can alternatively also be written as a tensor network, by
defining tensors $T_{pqr} := \langle p,q,r\ket\tau$ and $P^s_{pq} = \bra s
\mathcal P\ket{p,q}$. Then, the overall wavefunction (that is, the
expansion coefficient $\ket\Psi = \sum
c_{s_1,\dots,s_N}\ket{s_1,\dots,s_N}$ in the canonical local
basis) is obtained by contracting the tensors as shown in
Fig.~\ref{fig:peps}(b).
In particular, we can choose the unit cell indicated in the figure,
containing two $T$ and three $P$ tensors, and thus three physical spins.
By denoting the overall tensor of one unit cell by $A$, we can now
describe the model as a tensor network on a \emph{square} lattice
\footnote{Note that the mapping does not preserve lattice angles.
However, this is only relevant if we want to consider a torus, or if we
want to consider momenta with a component along the cylinder axis.
For our scenario, where we consider the system on
a cylinder and only care about momenta around the torus, the shearing of
the unit cell has no consequences.}.

\subsection{Symmetry and minimally entangled sectors}

\begin{figure}
    \includegraphics[width=246pt]{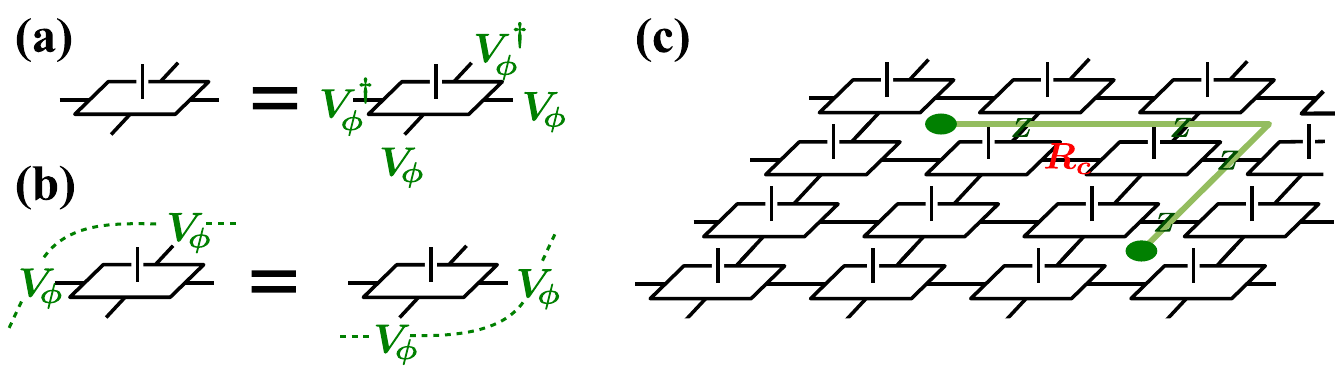}
    \caption{\textbf{(a)} The $\mathrm{SU}(3)$ PEPS defined in
    Fig.~\ref{fig:peps} possesses a virtual $\mathbb Z_3$ symmetry with
    representation $V_\phi=Z^\phi$, with $Z$ from
    Eq.~\eqref{eq:peps-virtual-Z}. 
    \textbf{(b)} This virtual symmetry can be
    interpreted as a ``pulling-through'' condition for strings of $V_\phi$'s.
    \textbf{(c)} Anyons can be described by (\emph{i}) strings of $V_\phi$ with open ends
    (illustrated here for $V_\phi=Z$), corresponding to fluxes $\phi$ and
    $-\phi$ at their endpoints, and
    (\emph{ii}) by localized defects $R_c$, transforming as an irreducible
    representation $c$ of $\mathbb Z_3$ when commuted with $Z$, corresponding
    to a single charge $c$.
    \label{fig:peps-sym}}
\end{figure}

Each unit cell is invariant under a virtual $\mathbb Z_3$ symmetry action
acting jointly on all indices Fig.~\ref{fig:peps-sym}(a)], with representation
$V_\phi=Z^\phi$, $\phi=0,1,2$,
\begin{equation}
    \label{eq:peps-virtual-Z}
    Z = \ket0\bra0 + \omega \sum_s \ket s\bra s + \bar \omega \sum_s \ket{\bar
    s}\bra{\bar s}\,,\ \omega = e^{2\pi i /3}\:,
\end{equation}
which assigns phases $\omega$ and $\bar\omega$ to the $\bm 3$ and
$\bar{\bm 3}$ subspace, respectively.
This can be seen immediately from the transformation properties
of $T$ and $P$ under a virtual $Z$ symmetry action: While $T$ transforms
trivially, each $P$ acquires a phase $\omega$, corresponding to its
physical $\mathrm{SU}(3)$ charge $\bm 3$;
as there are three $P$'s in each unit cell, the overall phase is trivial.
The trivial $\mathrm{SU}(3)$ charge per unit cell also indicates that a
possible no-go result for certain topological spin liquids in generalization of Zaletel
and Vishwanath's no-go result for
$\mathrm{SU}(2)$~\cite{zaletel:su2-semion} would not apply.

The virtual $\mathbb Z_3$ symmetry is reflected in a ``pulling through''
condition for strings of $V_\phi=Z^\phi$ operators on the virtual level -- that is,
such strings can be pulled through individual tensors
[Fig.~\ref{fig:peps-sym}(b)] and
thus through the tensor network, making them
\emph{topological} objects. This allows them to be used to build
anyonic excitations as well as to parametrize the ground space manifold,
and in particular to construct the minimally entangled
states~\cite{schuch:peps-sym,sahinoglu:mpo-injectivity,bultinck:mpo-anyons}.

Specifically, flux-like excitations are constructed through a string of
$Z$ operators (or $Z^2$ operators) with open ends, with fluxes $\pm1$
($\mp1$) associated to the endpoints, while charge-like excitations are
constructed by locally modifying the tensor network to transform as a
\emph{non-trivial} irreducible representation $R_c$ under the virtual $Z$
symmetry~\cite{schuch:peps-sym}, see Fig.~\ref{fig:peps-sym}(c).

\begin{figure}
    \includegraphics[width=246pt]{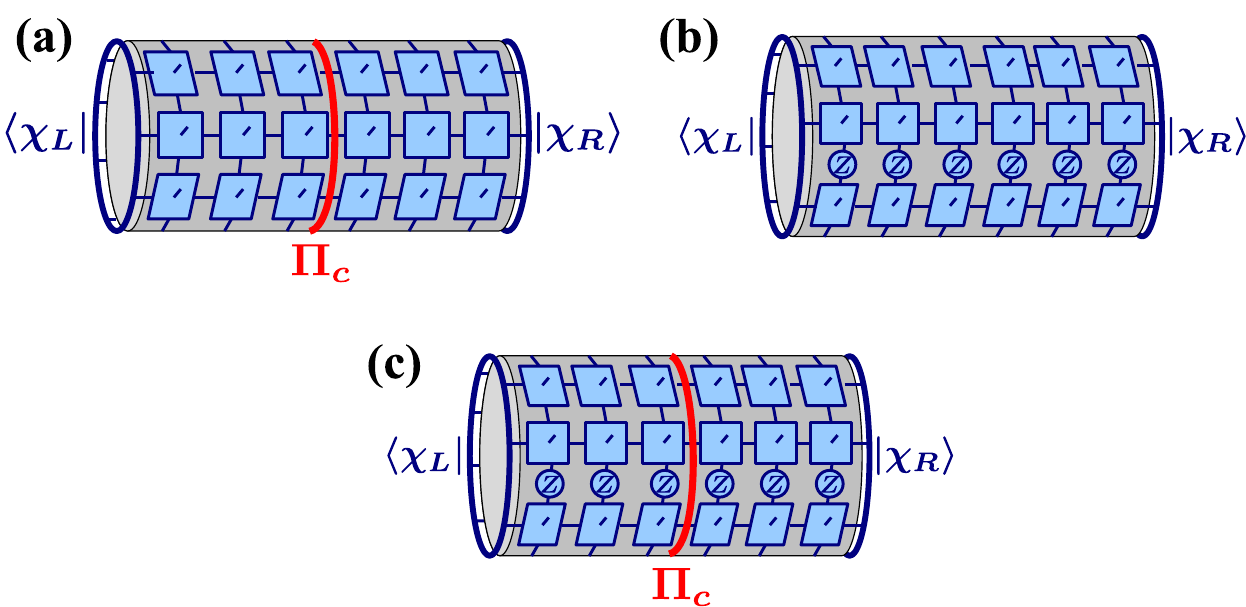}
    \caption{Construction of minimally entangled sectors on a cylinder with
    boundary conditions $\bra{\chi_L}$ and $\ket{\chi_R}$.
    \textbf{(a)}~A
    state with well-defined charge $c$ (and trivial flux) along the torus is
    obtained by applying a charge projector $\Pi_c=\sum (Z^{\otimes N_v})^\phi
    \omega^{c\phi}/3$ along a cut around the cylinder. 
    \textbf{(b)} A state with
    well-defined flux $\phi$ (and trivial charge) is obtained by threading a
    string $(Z^\phi)^{\otimes N_h}$ along the torus, illustrated here for
    $\phi=1$. 
    \textbf{(c)} A general minimally entangled state is obtained by
    combination of the two. Note that the charge projector $\Pi_c$ can
    be moved [using Fig.~\ref{fig:peps-sym}(b)] and thus alternatively be
    absorbed into the choice of at least one of the boundary conditions.
    \label{fig:peps-mes}}
\end{figure}

From this, we can directly see how to construct the minimally entangled sectors on
the cylinder and link them to the different anyons, see also
Ref.~\cite{schuch:topo-top}: First, recall that the minimally entangled
sectors are given by a cylinder with boundary conditions corresponding to
a well-defined anyon flux through the cylinder. This flux can be measured
through the phase observed when looping a dual anyon $a$ around the cylinder, denoted $O_a$ in Section~\ref{sec:esmes}.
Looping a magnetic particle around the cylinder corresponds to placing a
closed loop of $Z$'s around the cylinder~\cite{schuch:peps-sym}, and the
eigenspaces of the loop operator $Z^{\otimes N_v}$
with eigenvalue $1$, $\omega$, and $\omega^2$ 
thus correspond to the three
different charges threaded through the cylinder. The different sectors are thus
obtained by applying a charge projector 
\begin{equation}
    \label{eq:def-Pi-c}
    \Pi_c = \frac13 \sum_{\phi=0,1,2} (Z^{\otimes N_v})^\phi \omega^{c\phi}
\end{equation}
around the cylinder [Fig.~\ref{fig:peps-mes}(a)]. Importantly, $\Pi_c$ can be moved to
any position along the cylinder due to the pulling through condition, and
thus, a single $\Pi_c$ is sufficient to ensure a constant charge threaded through
the cylinder everywhere. Note that since $R_{c'}\Pi_c = \Pi_{c+c'}R_{c'}$
for an $R_{c'}$ which transforms as an irrep $c'$,
threading an additional charge $c'$ through the cylinder indeed changes
the label in the right way.

On the other hand, looping a charge around the cylinder does not
accumulate any phase, since in the chosen tensor network representation,
the charge is a point-like object transforming as an irrep with no string
attached~\cite{schuch:peps-sym}: We thus infer that the state we just constructed
is threaded by a trivial flux. In order to change this sector, we have to
thread a flux through the cylinder, that is, create a flux-antiflux pair
$\pm\phi$ at one end of the cylinder and move the flux to the other end.
This results in a flux-antiflux pair at the two ends of the cylinder,
connected by a string $(Z^\phi)^{\otimes N_h}$ along the cylinder, see
Fig.~\ref{fig:peps-mes}(b).
Indeed, looping a charge $c$ around the cylinder will now acquire a phase
$\omega^{\phi c}$ due to the commutation relation with $Z$, as required.

Overall, the minimally entangled states on the cylinder are thus
constructed as shown in Fig.~\ref{fig:peps-mes}(c), namely by 
(\emph{i})~projecting onto an
charge sector $c$ through the projector $\Pi_c$, Eq.~\eqref{eq:def-Pi-c},
and (\emph{ii}) placing a string $(Z^\phi)^{\otimes N_h}$ along the
cylinder. For a system on a sufficiently long cylinder, and in a gapped topological phase,
the resulting state deep in the bulk, as well as its entanglement
spectrum, will be independent of the specific boundary conditions chosen,
as long as they are compatible (i.e., have non-zero overlap) with the
chosen symmetry sectors.

\subsection{Entanglement spectrum}
\label{sec:pepses}
\begin{figure}
    \includegraphics[width=246pt]{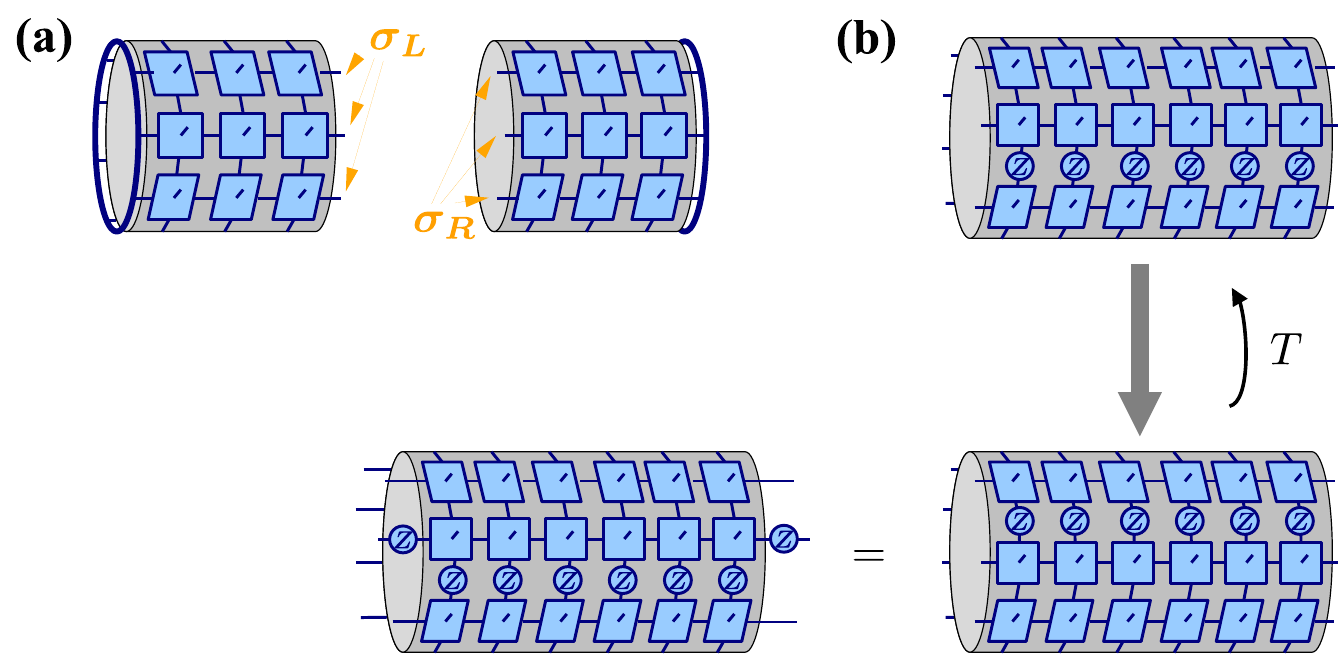}
    \caption{\textbf{(a)} In PEPS, the entanglement spectrum can be computed
    from the left and right virtual boundary states $\sigma_{L}$ and $\sigma_R$ at the
    respective cut (obtained by tracing the physical indices) as
    $\mathrm{spec}(\sigma_L\sigma_R^T)$. 
    \textbf{(b)}~In the case of a PEPS
    threaded by a flux string (top, here illustrated for $\phi=1$), the action
    of the translation operator (acting jointly on the physical and the virtual
    system, or equivalently -- after tracing the physical system -- at the virtual ket+bra
    layer) displaces the flux string (bottom right). The string can be moved
    back to its original position in the bulk using the virtual symmetry of
    Fig.~\ref{fig:peps-sym}(a), at the cost of one $Z^\phi$ operator appearing
    at the boundary (bottom left). This gives rise to a dressed translation operator
    $T_\phi=TZ_1^\phi$, and subsequently to a momentum shift
    $(c/3)(2\pi/N_v)$ in the sector with charge $c$.
    \label{fig:peps-entspec}}
\end{figure}

As explained in Section~\ref{sec:esmes}, the entanglement spectrum is obtained by
considering bipartitions of the minimally entangled states on the
cylinder, labeled by the quantum numbers for momentum (translation) and
$\mathrm{SU}(3)$. 
In PEPS, the entanglement spectrum can be
extracted by cutting the PEPS on the cylinder across the \emph{bonds}, and
computing the density operator obtained at the bond degrees of freedom
when tracing out the physical degrees of freedom, cf.\
Fig.~\ref{fig:peps-entspec}(a).
Let us denote these states by $\sigma_L$ and $\sigma_R$. The entanglement
spectrum is then obtained as
$\mathrm{spec}\big(\sigma_L(\sigma_R)^T\big)$, or equivalently 
$\mathrm{spec}\big(\sqrt{\sigma_L}(\sigma_R)^T\sqrt{\sigma_L}\big)$ (the
latter being Hermitian). From the $\mathrm{SU}(3)$ symmetry of the
tensors [Fig.~\ref{fig:peps}(c)], it follows that $\sigma_{L}$ and $\sigma_{R}^T$ commute with a
global $\mathrm{SU}(3)$ action
\footnote{as long as we either choose
boundary conditions with a fixed $\mathrm{SU}(3)$ irrep, or consider the
limit of an infinitely long
cylinder and use that the transfer operator has a unique fixed point in
each topological sector, which is given in the topological
phase~\cite{schuch:topo-top}},
and thus, the spectrum of
$\sigma_L(\sigma_R)^T$ can be labeled by $\mathrm{SU}(3)$ quantum
numbers.
Moreover, we can assume that both
$\sigma_L$ and $\sigma_R$ are supported on the sector onto which $\Pi_c$
projects (e.g.,\ by placing a $\Pi_c$ on both half-cylinders, or by
choosing suitable boundary conditions). 

It remains to discuss how to assign momentum quantum numbers to
$\sigma_{\bullet}$ ($\bullet=L,R$). The non-trivial point here lies in the
fact that the cylinder can be threaded by a flux, rendering the underlying
tensor network not translationally invariant -- while the physical
state is independent of where the flux string is located, the boundary
states $\sigma_\bullet$ are not. Rather, the different string locations are
related through a gauge transformation, just as for any 1D chain threaded
by a flux. Specifically, as illustrated in Fig.~\ref{fig:peps-entspec}(b), translation
corresponds to rotating the cylinder by one site, and thus displaces the
$(Z^\phi)$-string accordingly. By virtue of the pulling through condition,
this is again equal to the original string location up to a gauge
transformation $Z^\phi$ applied to the virtual index at the boundary just
adjacent to the location of the string (without loss of generality, we assume this is
position $i=1$, denoted $Z_1^\phi$). Thus, we have that 
in the limit of an infinitely long cylinder, both
$\sigma=\sigma_L,\sigma_R^T$ transform under the translation operator $T$
as $T\sigma T^\dagger = Z_1^\phi \sigma
(Z_1^\phi)^\dagger$, or $[TZ_1^\phi,\sigma]=0$. This implies that the
eigenvalues of $\sigma$ are labeled by quantum numbers of the dressed
translation operator $T_\phi=TZ_1^\phi$.
Since $T_\phi^{N_v}=(TZ_1^\phi)^{N_v}=(Z^\phi)^{\otimes N_v}$, we have that
$T^{N_v}_\phi\equiv \omega^c$ for the minimally entangled state with charge $c$,
for which $\sigma$ is supported on $\Pi_c$. Thus, we find that for the
sector with charge $c$, the allowed momenta are $k=(2\pi/N_v)(n+c/3)$,
$n=0,\dots,N_v-1$.

The described procedure allows us to obtain the boundary states $\sigma_\bullet=\sigma_L,\sigma_R$, and thus the entanglement spectrum $\mathrm{spec}(\sigma_L\sigma_R^T)$, for each of the minimally entangled topological sectors. However, it does not provide us with the normalization of $\sigma_\bullet$ and thus of the entanglement spectrum, as the latter depends on the chosen boundary condition. On the other hand, the relative normalization of the entanglement spectrum in different sectors contains important information about the underlying CFT, as it allows us to relate the entanglement energies in the different sectors. 
In order to fix the relative normalization of the different sectors in a canonical way, we thus follow the procedure set forth in Ref.~\cite{schuch:topo-top}, and normalize 
the sectors such that the trace of the entanglement spectrum is identical for all topological sectors. This is motivated by the fact that this is the only choice consistent with the entanglement spectrum, taken jointly for all sectors with the corresponding weight, being the Gibbs state of a local entanglement Hamiltonian (see Supplemental Material of Ref.~\cite{schuch:topo-top}), and thus an underlying CFT.

\subsection{Numerical methods}

In order to numerically compute the entanglement spectrum of the PEPS for the different sectors, we proceed as follows~\cite{Poilblanc2016,kurecic:su3_sl}. First, we determine the left and right fixed point of the PEPS transfer matrix on an \emph{infinite} system, by approximating each by a matrix product state (MPS) of some bond dimension $\chi$, described by a tensor $A$. To this end, we repeatedly apply the transfer matrix (with tensor $T$) to the MPS, that is, we replace $A$ by $A$ contracted with $T$. Subsequently, we truncate the bond dimension of the MPS, by discarding all Schmidt coefficients below some threshold $\delta$. This is iterated until convergence is reached. (In practice, we repeatedly ramp down the threshold $\delta\to\delta/2$.) For the data reported, we used $\delta = 10^{-6}\cdot2^{-8}\approx 3.9\cdot10^{-9}$, resulting in an MPS with $\chi=111$, and an accuracy (overlap squared per site of old and new MPS) of $\approx 10^{-12}$. 

From the MPS for left and right fixed point, we can then construct $\sigma_L$ and $\sigma_R$ for a finite cylinder, by separating ket and bra indices -- resulting in a Matrix Product Operator (MPO) -- and wrapping the MPO on a cylinder. (Note that this is only an approximation to the true fixed points on the cylinder, since the MPO has been optimized for an infinite system.) When closing the boundary, we can insert a flux, that is, a symmetry action ${\hat Z}^\phi$ on the virtual ket and bra index of the MPO, to select a topological sector. $\hat Z$ amounts to the ``compressed'' version of the symmetry action $Z\otimes \bar Z$ along the cylinder. While we do not encode the $\mathrm{SU}(3)$ or $\mathbb Z_3$ symmetry, ${\hat Z}$ can be numerically determined from the symmetry action $Z\otimes \bar Z$
on the ``physical'' indices of the MPO, using the fundamental theorem of MPS~\cite{cirac:tn-review-2021}. 
In order to determine the entanglement spectrum -- which, as we explained, equals the spectrum of $\sigma_L(\sigma_R)^T$ -- we however do not explicitly construct $\sigma_L$ and $\sigma_R$, as this would be too memory-intensive. Rather, we use a Krylov method to determine the leading eigenvalues of $\sigma_L(\sigma_R)^T$, where we only need to be able to \emph{apply} $(\sigma_R)^T$ and $\sigma_L$ to a given vector $\vec v$; this can be carried out by contracting one MPO tensor after the other with $\vec v$, and additionally manually summing over one of the open virtual MPO indices, and thus only requires roughly a vector of size $\chi\cdot 7^N$ ($D=7$ being the dimension of $\vec v$ per site) to be stored (which limits the $\chi$ we can use). In addition, we apply projections onto momentum sectors and onto charge sectors after each application of $\sigma_L(\sigma_R)^T$, as detailed above, to retrieve the entanglement spectrum in a given topological and momentum sector. For each sector, we determine this way the 1500 leading eigenvalues. Since we do not encode the $\mathrm{SU}(3)$ symmetry, the irrep labels of the $\mathrm{SU}(3)$ multiplets are determined from their multiplicities.
Note that the normalization (recall that all sectors have the identical normalization) can be computed exactly, without having to restrict to the Krylov data, as $\mathrm{tr}(\sigma_L\sigma_R^T)$ equals the trace of the $N$'s power of the transfer matrix for $\sigma_L\sigma_R^T$, which is a $\chi^2\times \chi^2$ matrix.

\section{Finding CFT in the entanglement spectrum}
\label{sec:cftes}

\subsection{State-countings in (1+1)-dimensional CFT}
\label{sec:countings}

From the construction and computation of the PEPS of Section \ref{sec:peps}, we are able to obtain the entanglement Hamiltonian $H_{1D}$.
We can define the excitation energies $\Delta E_i$ to be the values of the (entanglement) energy eigenvalues $E_i$ of $H_{1D}$ [as defined in Eq.~\eqref{LabelEqShiftedEntanglementEnergies}] less the ground state energy $E_0$: 
\begin{equation}
\label{eq:def-delta-ei}
    \Delta E_i = E_i - E_0.
\end{equation}
As will be elaborated on below, the $\Delta E_i$ of the low-lying entanglement Hamiltonian will scale inversely with $N$ in the system under consideration
\footnote{when, as was mentioned before, $N$ is much smaller than the inverse entanglement gap}.

The entanglement excitation energies $\Delta E_i$ will be those depicted in the spectra of Figs.~ \ref{fig:doubledes0}-\ref{fig:doubledes2} in Section \ref{sec:results}.
Using an extension of the arguments given in  Refs.~\cite{Qi2012} and \cite{ArildsenLudwig2022}, we will see that the $\mathrm{SU}(3)$ representation content of those spectra at a given momentum is in one-to-one correspondence with that of the spectrum of a physical boundary (``edge''), a correspondence which we will understand from the conformal boundary state point of view in Section \ref{sec:cftbdy}.
First, we will give the necessary preliminaries for understanding the theory of a physical boundary and ultimately its implications for the CFT content of the entanglement spectrum. 
The theory will be determined by the topological order of the system which, as discussed in Section \ref{sec:anyons}, can be described by $\mathrm{SU}(3)_1 \otimes \overline{\mathrm{SU}(3)_1}$ doubled Chern-Simons theory.
To describe a physical boundary (and also the entanglement spectrum) of the doubled Chern-Simons theory, we must first understand the chiral $\mathrm{SU}(3)_1$ WZW CFT taken on its own. 
The theory, being at level $k=1$, has three sectors, each with an associated set of primary states, which correspond to one of the three lowest-dimensional representations of $\mathrm{SU}(3)$: $\bm{1}$, $\bm{3}$, and $\overline{\bm{3}}$. 
There is also a Noether current $J^a(x)$ corresponding to the eight generators of $\mathrm{SU}(3)$ symmetry, where $a = 1, \ldots, 8$. 
Considering this chiral CFT as present on the circular edge of a cylinder of circumference $N$, we can write the mode expansion of this current as
\begin{equation}
    \label{eq:jmodeexpansion}
    J^a(x) = \frac{2\pi}{N}\sum_{n=-\infty}^\infty J^a_{n}e^{2\pi inx/N}.
\end{equation}
Then the Hilbert space of the chiral $\mathrm{SU}(3)_1$ WZW CFT is built by acting with the modes $J^a_{-n}$ as raising operators to build states of the form
\begin{equation}
    \label{eq:jbasis}
    J^{a_1}_{-n_1}\cdots J^{a_m}_{-n_m}\ket{\rho,s_\rho},
\end{equation}
where $\ket{\rho,s_\rho}$ is a primary state multiplet with representation $\rho \in \{\bm{1},\bm{3},\overline{\bm{3}}\}$, $s_\rho$ is the specific state in the representation, and the $n_i$ are positive integers. 
In the sector rooted in a given primary state multiplet, the Hilbert space consists of a tower of states at different (1D momentum) levels $K = \sum n_i$ above the primary state. 
These states are organized into multiplets (irreducible representations) of $\mathrm{SU}(3)$ which we label by their dimension. 
There will be a characteristic number of multiplets of each dimension at each level $K$ of the tower. 
Table \ref{table:chiralmultiplets} contains the expected countings for the multiplets in the chiral $\mathrm{SU}(3)_1$ WZW CFT.
\begin{table}[t]
	\centering
	\begin{tabular}{c|c|c}
	    Level $K$ & $\bm{1}$ primary sector multiplets &  $\bm{3}$ primary sector multiplets \\
		\hline 
	0 & $\bm{1}$ & $\bm{3}$ \\
	1 & $\bm{8}$ & $\bm{3}+\overline{\bm{6}}$ \\
	2 & $\bm{1}+2(\bm{8})$ & $2(\bm{3})+\overline{\bm{6}}+\bm{15}$ \\
	3 & $2(\bm{1})+3(\bm{8})+\bm{10}+\overline{\bm{10}}$ &
	$3(\bm{3})+3(\overline{\bm{6}})+2(\bm{15})$ \\
	4 & $3(\bm{1})+6(\bm{8})+\bm{10}+\overline{\bm{10}}+\bm{27}$ &
	$6(\bm{3})+4(\overline{\bm{6}})+4(\bm{15})+\bm{24}$ \\
	5 & $4(\bm{1})+10(\bm{8})+3(\bm{10})+3(\overline{\bm{10}})+2(\bm{27})$ &
	$9(\bm{3})+8(\overline{\bm{6}})+7(\bm{15})+\bm{15}'+2(\bm{24})$ 
	\end{tabular}
	\caption{The multiplet content of the $\bm{1}$ and $\bm{3}$ primary sectors of the chiral $\mathrm{SU}(3)_1$ WZW theory~\cite{Kass1990}. Note that the content of the $\overline{\bm{3}}$ primary sector can be obtained by taking the conjugates of the representations given for the $\bm{3}$ primary sector.}
	\label{table:chiralmultiplets}
\end{table}
The Sugawara expression of the energy-momentum tensor $T(x)$ of the chiral CFT can be written as \cite{Knizhnik1984} 
\begin{equation}
    \label{eq:sugawara}
    T(x) = \frac{1}{k+3}\sum_{a=1}^{8} (J^a J^a)(x),
\end{equation}
where we use round brackets $()$ around $J^a J^a$ to indicate normal ordering, and the level $k = 1$ for $\mathrm{SU}(3)_1$. 
We can then expand $T(x)$ in terms of modes $L_{-n}$:
\begin{equation}
    \label{eq:lmodeexpansion}
    T(x) =\left(\frac{2\pi}{N}\right)^2 \left(-\frac{c}{24}+\sum_{n=-\infty}^\infty L_{n}e^{2\pi inx/N}\right),
\end{equation}
where the central charge $c = 2$ for the $\mathrm{SU}(3)_1$ WZW CFT.
Thus we see that the Hamiltonian of this theory, which describes the chiral edge states on a physical boundary of a cylinder, will be
\begin{equation}
    \label{eq:cfth}
    H_L = \frac{v_L}{2\pi}\int_0^{N} T(x)dx = 
    \frac{2\pi v_L}{N} \left( L_0 - \frac{c}{24}\right),
\end{equation}
where the subscript $L$ on the Hamiltonian $H_L$ and the velocity $v_L$ indicates that these are left-moving edge states, and $L_0$ is the zero-mode of $T(x)$. 
The momentum $P_L$ is given by 
\begin{equation}
    P_L = \frac{2\pi}{N} L_0.
\end{equation}
The eigenvalues of $H_L$ and $P_L$ are determined by those of $L_0$, which are the states of Eq.~\eqref{eq:jbasis}, and we have
\begin{equation}
    \label{eq:l0eigen}
    L_0 \left(J^{a_1}_{-n_1}\cdots J^{a_m}_{-n_m}\ket{\rho,s_\rho}\right) 
     =\left(h_\rho + K_L\right)\left(J^{a_1}_{-n_1}\cdots J^{a_m}_{-n_m}\ket{\rho,s_\rho}\right),  
\end{equation}
where $h_\rho = \frac{C_2(\rho)}{k+3}$ is the conformal weight of the primary state at representation $\rho$, $C_2(\rho)$ is the quadratic Casimir of $\rho$, and $K_L = \sum_{i=1}^m n_i$ is the level of the state. [For the primary states of $\mathrm{SU}(3)_1$ we have $C_2(\bm{1}) = 0$ and $C_2(\bm{3}) = C_2(\overline{\bm{3}}) = \frac{4}{3}$, leading to $L_0$ eigenvalues for the primary states equal to their conformal weights $h_{\bm{1}} = 0$ and $h_{\bm{3}} = h_{\overline{\bm{3}}} = \frac{1}{3}$.]

We now discuss the non-chiral WZW CFT.
This theory consists of two time-reversed copies of the chiral WZW CFT, 
one left-moving (``holomorphic'') CFT denoted by $\mathrm{SU}(3)_1$, 
and one right-moving (``anti-holomorphic'') CFT denoted by $\overline{\mathrm{SU}(3)_1}$.
We can take the definitions so far presented for the chiral theory [Eqs.~\eqref{eq:jmodeexpansion}-\eqref{eq:l0eigen}] as describing the left-moving theory. 
The right-moving theory, then, is described in terms of conjugate currents and their modes:
\begin{align}
	\bar{J}^a(x) &= \frac{2\pi}{N}\sum_{n=-\infty}^\infty \bar{J}^a_{-n}e^{2\pi inx/N} \\
	\bar{T}(x)& = \frac{1}{k+3}\sum_{a=1}^{8} (\bar{J}^a \bar{J}^a)(x) \\
	\bar{T}(x) &=\left(\frac{2\pi}{N}\right)^2 \left(-\frac{c}{24}+\sum_{n=-\infty}^\infty \bar{L}_{-n}e^{2\pi inx/N}\right),
\end{align}
and 
\begin{align}
	\label{eq:cfthr}
	\bar{H}_R &= \frac{\overline{v}_R}{2\pi}\int_0^{N} \bar{T}(x)dx = \frac{2\pi\overline{v}_R}{N} \left( \bar{L}_0 - \frac{c}{24}\right) \\
	\bar{P}_R &= \frac{2\pi}{N} \bar{L}_0,
\end{align}
in contrast to the $H_L$ and $P_L$ of Eq.~\eqref{eq:cfth}.
The eigenvalues of $\bar{L}_0$ take the same form as those of $L_0$,
\begin{equation}
    \label{eq:RightDescendants}
	\bar{L}_0\bar{J}^{\bar{a}_1}_{-\bar{n}_1}\cdots \bar{J}^{\bar{a}_m}_{-\bar{n}_m}\ket{\bar{\rho},\bar{s}_{\bar{\rho}}}_R = \left( \bar{h}_{\bar{\rho}} + \bar{K}_R\right)\bar{J}^{\bar{a}_1}_{-\bar{n}_1}\cdots \bar{J}^{\bar{a}_m}_{-\bar{n}_m}\ket{\bar{\rho},\bar{s}_{\bar{\rho}}}_R,
\end{equation}
but now with the eigenstates $\bar{J}^{\bar{a}_1}_{-\bar{n}_1}\cdots \bar{J}^{\bar{a}_m}_{-\bar{n}_m}\ket{\bar{\rho},\bar{s}_{\bar{\rho}}}_R$, which are constructed in a manner analogous to Eq.~\eqref{eq:jbasis}. 
(The primary states of the left-moving theory will now be denoted $\left|\rho,s_\rho\right\rangle_L$.)
We denote the conformal weights of the primary states in the $R$ sector in the representation $\bar{\rho}$ by $\bar{h}_{\bar{\rho}}$, and the level of the state is then $\bar{K}_R = \sum_{i=1}^m \bar{n}_i$. 

In the non-chiral CFT, which we will sometimes also refer to, in short, as the ``doubled $\mathrm{SU}(3)_1 \otimes \overline{\mathrm{SU}(3)_1}$ 
CFT''
\footnote{Not to be confused with the $\mathrm{SU}(3)_1 \otimes \overline{\mathrm{SU}(3)_1}$ doubled Chern-Simons theory, the latter being a (2+1)-dimensional topological gapped theory, while the former is a (1+1)-dimensional gapless CFT.},
the states can be organized into multiplets of the global, diagonal $\mathrm{SU}(3)$ symmetry acting simultaneously on $\mathrm{SU}(3)$ and $\overline{\mathrm{SU}(3)}$, which is now the total $\mathrm{SU}(3)$ found from the decompositions of tensor products of the corresponding $\mathrm{SU}(3)$ multiplets from the left- and right-moving chiral theories. 
It is this $\mathrm{SU}(3)$ symmetry that can be determined from the PEPS, as described in the previous section.
With this in mind, the Hamiltonian for the overall theory becomes 
\begin{equation}
	\label{eq:htotal}
 	H_{\text{total}} = H_{\text{doubled}} + H_{\text{interaction}},
\end{equation}
where 
\begin{equation}
	\label{eq:doubledh}
    H_{\text{doubled}} = H_L + \bar{H}_R = \frac{2\pi}{N} \left(v_L L_0 + \overline{v}_R \bar{L}_0 - (v_L+\overline{v}_R) \frac{c}{24}\right) = v_L P_L + \overline{v}_R \bar{P}_R - \frac{\pi c}{12N}(v_L+\overline{v}_R),
\end{equation}
is the non-chiral CFT, and $H_{\text{interaction}}$ represents a globally $\mathrm{SU}(3)$-symmetric coupling term between the left- and right-moving sectors of the CFT.
This interaction term will eventually generate a gap, but as discussed above we will be mostly interested in the limit  where the system size is much smaller than the inverse gap (proportional to the associated correlation length). 
In this limit, as already mentioned, the underlying gapless features of the CFT will appear.
We also have 
\begin{align}
    \label{eq:doubledp}
    P = \bar{P}_R - P_L & & \text{ and }&  & K = \bar{K}_R - K_L
\end{align}
for the total momentum and level, respectively.

Here we consider the case of strong time-reversal symmetry breaking, reflected in vastly different velocities of the right- and the left- moving branches of the CFT, $v_L \gg \overline{v}_R$. 
In such a case, due to the substantial energy penalty assigned to any but the lowest-lying states in the \emph{high-velocity left-moving} theory, we see that the $\mathrm{SU}(3)$ representations (multiplets) available to us at low energies will come from the tensor products of the representations of \emph{only the primary} states (no descendants) of the left-moving theory $\{|\bm{1}\rangle_L,|\bm{3}\rangle_L,|\overline{\bm{3}}\rangle_L\}$ with the representations of the \emph{full} low-energy multiplet content (see Table \ref{table:chiralmultiplets}), \emph{i.e., including the descendant states}, of the three \emph{low-velocity right-moving} primary state sectors (with lowest states $\{|\bm{1}\rangle_R,|\bm{3}\rangle_R,|\overline{\bm{3}}\rangle_R\}$). 
Some of the lowest levels of the multiplet content of the non-chiral $L$/$R$ theory are recorded in Tables \ref{table:doubledmultiplets0}-\ref{table:doubledmultiplets2}, where we refer to the left-moving theory as the ``fast" theory and the right-moving theory as the ``slow" theory. 
Note that the levels $K$ denoted in these tables, which are also those that will be exhibited in the spectra of Section \ref{sec:results}, are equal to the levels $\bar{K}_R$ of the $R$ side, since the $L$ contributions are only from the primary states, which have level $K_L = 0$. 
[See Eqs.~\eqref{eq:l0eigen}, \eqref{eq:RightDescendants}, and \eqref{eq:doubledp}.]

\begin{table}
	\centering
	\begin{tabular}{c|c|c|c|c}
		& \multicolumn{2}{c|}{Charge $q=0$}  & \multirow{2}{*}{Level $K$} & \multirow{2}{*}{Multiplet content} \\
		\text{Flux} & \text{Fast ($L$)} & \text{Slow ($R$)} &  &  \\
		\hline
		\multirow{6}{*}{$\phi = 0$} & \multirow{4}{*}{$\bm{1}$} & \multirow{4}{*}{$\bm{1}$} & 0 & $\bm{1}$ \\
		&  &  & 1 & $\bm{8}$ \\
		&  &  & 2 & $\bm{1}+2(\bm{8})$ \\
		&  &  & 3 & $2(\bm{1})+3(\bm{8})+\bm{10}+\overline{\bm{10}}$ \\
		& \multirow{2}{*}{$\kappa = 0$} & \multirow{2}{*}{$\bar{\kappa} = 0$} & 4 & $3(\bm{1})+6(\bm{8})+\bm{10}+\overline{\bm{10}}+\bm{27}$ \\
		&  &  & 5 & $4(\bm{1})+10(\bm{8})+3(\bm{10})+3(\overline{\bm{10}})+2(\bm{27})$ \\
		\hline
		\multirow{6}{*}{$\phi = 1$} & \multirow{4}{*}{$\overline{\bm{3}}$} & \multirow{4}{*}{$\bm{3}$} & 0 & $\bm{1}+\bm{8}$ \\
		&  &  & 1 & $\bm{1}+2(\bm{8})+\overline{\bm{10}}$ \\
		&  &  & 2 & $2(\bm{1})+4(\bm{8})+\bm{10}+\overline{\bm{10}}+\bm{27}$ \\
		&  &  & 3 & $3(\bm{1})+8(\bm{8})+2(\bm{10})+3(\overline{\bm{10}})+2(\bm{27})$ \\
		& \multirow{2}{*}{$\kappa = -1$} & \multirow{2}{*}{$\bar{\kappa} = +1$} & 4 &
		$6(\bm{1})+14(\bm{8})+4(\bm{10})+5(\overline{\bm{10}})+5(\bm{27})+\overline{\bm{35}}
		$ \\
		&  &  & 5 &
		$9(\bm{1})+24(\bm{8})+8(\bm{10})+10(\overline{\bm{10}})+9(\bm{27})+\bm{35}+2
		(\overline{\bm{35}})$ \\
		\hline
		\multirow{6}{*}{$\phi = 2$} & \multirow{4}{*}{$\bm{3}$} & \multirow{4}{*}{$\overline{\bm{3}}$} & 0 & $\bm{1}+\bm{8}$ \\
		&  &  & 1 & $\bm{1}+2(\bm{8})+\bm{10}$ \\
		&  &  & 2 & $2(\bm{1})+4(\bm{8})+\bm{10}+\overline{\bm{10}}+\bm{27}$ \\
		&  &  & 3 & $3(\bm{1})+8(\bm{8})+3(\bm{10})+2(\overline{\bm{10}})+2(\bm{27})$ \\
		& \multirow{2}{*}{$\kappa = +1$} & \multirow{2}{*}{$\bar{\kappa} = -1$} & 4 &
		$6(\bm{1})+14(\bm{8})+5(\bm{10})+4(\overline{\bm{10}})+5(\bm{27})+\bm{35}$
		\\
		&  &  & 5 &
		$9(\bm{1})+24(\bm{8})+10(\bm{10})+8(\overline{\bm{10}})+9(\bm{27})+2(\bm{35})+\overline{\bm{35}}$ 
	\end{tabular}
	\caption{The multiplet content of the sectors of the doubled theory with charge 0. Note that the first ($\phi = 0$) section of the table, in which the fast-moving theory is in the trivial $\bm{1}$ primary sector, has the same multiplet content as that of the $\bm{1}$ primary state sector of the chiral $\mathrm{SU}(3)_1$ theory in Table \ref{table:chiralmultiplets}.}	
	\label{table:doubledmultiplets0}
\end{table}
\begin{table}
    \centering
	\begin{tabular}{c|c|c|c|c}
		& \multicolumn{2}{c|}{Charge $q = 1$}  & \multirow{2}{*}{Level $K$} & \multirow{2}{*}{Multiplet content} \\
		\text{Flux} & \text{Fast ($L$)} & \text{Slow ($R$)} &  &  \\
		\hline
		\multirow{6}{*}{$\phi = 0$} & \multirow{4}{*}{$\overline{\bm{3}}$} & \multirow{4}{*}{$\overline{\bm{3}}$} & 0 & $\bm{3}+\overline{\bm{6}}$ \\
		&  &  & 1 & $2(\bm{3})+\overline{\bm{6}}+\bm{15}$ \\
		&  &  & 2 & $3(\bm{3})+3(\overline{\bm{6}})+2(\bm{15})+\bm{24}$ \\
		&  &  & 3 & $6(\bm{3})+5(\overline{\bm{6}})+5(\bm{15})+2(\bm{24})$ \\
		& \multirow{2}{*}{$\kappa = -1$} & \multirow{2}{*}{$\bar{\kappa} = -1$} & 4 &
		$10(\bm{3})+10(\overline{\bm{6}})+9(\bm{15})+\bm{15}'+4(\bm{24})+\bm{42}$
		\\
		&  &  & 5 &
		$17(\bm{3})+16(\overline{\bm{6}})+17(\bm{15})+2(\bm{15}')+\bm{21}+8(\bm{24})+2(\bm{42})$ \\
		\hline
		\multirow{6}{*}{$\phi = 1$} & \multirow{4}{*}{$\bm{3}$} & \multirow{4}{*}{$\bm{1}$} & 0 & $\bm{3}$ \\
		&  &  & 1 & $\bm{3}+\overline{\bm{6}}+\bm{15}$ \\
		&  &  & 2 & $3(\bm{3})+2(\overline{\bm{6}})+2(\bm{15})$ \\
		&  &  & 3 & $5(\bm{3})+4(\overline{\bm{6}})+4(\bm{15})+\bm{15}'+\bm{24}$ \\
		& \multirow{2}{*}{$\kappa = +1$} & \multirow{2}{*}{$\bar{\kappa} = 0$} & 4 &
		$9(\bm{3})+7(\overline{\bm{6}})+8(\bm{15})+\bm{15}'+2(\bm{24})+\bm{42}$ \\
		&  &  & 5 &
		$14(\bm{3})+13(\overline{\bm{6}})+15(\bm{15})+3(\bm{15}')+5(\bm{24})+2(\bm{42})$ \\
		\hline
		\multirow{6}{*}{$\phi = 2$} & \multirow{4}{*}{$\bm{1}$} & \multirow{4}{*}{$\bm{3}$} & 0 & $\bm{3}$ \\
		&  &  & 1 & $\bm{3}+\overline{\bm{6}}$ \\
		&  &  & 2 & $2(\bm{3})+\overline{\bm{6}}+\bm{15}$ \\
		&  &  & 3 & $3(\bm{3})+3(\overline{\bm{6}})+2(\bm{15})$ \\
		& \multirow{2}{*}{$\kappa = 0$} & \multirow{2}{*}{$\bar{\kappa} = +1$} & 4 & $6(\bm{3})+4(\overline{\bm{6}})+4(\bm{15})+\bm{24}$ \\
		&  &  & 5 & $9(\bm{3})+8(\overline{\bm{6}})+7(\bm{15})+\bm{15}'+2(\bm{24})$ 
	\end{tabular}
    \caption{The multiplet content of the sectors of the doubled theory with charge 1. Note that the final ($\phi = 2$) section of the table, in which the fast-moving theory is in the trivial $\bm{1}$ primary sector, has the same multiplet content as that of the $\bm{3}$ primary state sector of the chiral $\mathrm{SU}(3)_1$ theory in Table \ref{table:chiralmultiplets}. }	
	\label{table:doubledmultiplets1}
\end{table}
\begin{table}
    \centering
	\begin{tabular}{c|c|c|c|c}
		& \multicolumn{2}{c|}{Charge $q = 2$}  & \multirow{2}{*}{Level $K$} & \multirow{2}{*}{Multiplet content} \\
		\text{Flux} & \text{Fast ($L$)} & \text{Slow ($R$)} &  &  \\
		\hline
		\multirow{6}{*}{$\phi = 0$} & \multirow{4}{*}{$\bm{3}$} & \multirow{4}{*}{$\bm{3}$} & 0 & $\overline{\bm{3}}+\bm{6}$ \\
		&  &  & 1 & $2(\overline{\bm{3}})+\bm{6}+\overline{\bm{15}}$ \\
		&  &  & 2 & $3(\overline{\bm{3}})+3(\bm{6})+2(\overline{\bm{15}})+\overline{\bm{24}}$ \\
		&  &  & 3 & $6(\overline{\bm{3}})+5(\bm{6})+5(\overline{\bm{15}})+2(\overline{\bm{24}})$ \\
		& \multirow{2}{*}{$\kappa = +1$} & \multirow{2}{*}{$\bar{\kappa} = +1$} & 4 &
		$10(\overline{\bm{3}})+10(\bm{6})+9(\overline{\bm{15}})+\overline{\bm{15}}'+4(\overline{\bm{24}})+\overline{\bm{42}}$
		\\
		&  &  & 5 &
		$17(\overline{\bm{3}})+16(\bm{6})+17(\overline{\bm{15}})+2(\overline{\bm{15}}')+\overline{\bm{21}}+8(\overline{\bm{24}})+2(\overline{\bm{42}})$ \\
        \hline
		\multirow{6}{*}{$\phi = 1$} & \multirow{4}{*}{$\bm{1}$} & \multirow{4}{*}{$\overline{\bm{3}}$} & 0 & $\overline{\bm{3}}$ \\
		&  &  & 1 & $\overline{\bm{3}}+\bm{6}$ \\
		&  &  & 2 & $2(\overline{\bm{3}})+\bm{6}+\overline{\bm{15}}$ \\
		&  &  & 3 & $3(\overline{\bm{3}})+3(\bm{6})+2(\overline{\bm{15}})$ \\
		& \multirow{2}{*}{$\kappa = 0$} & \multirow{2}{*}{$\bar{\kappa} = -1$} & 4 & $6(\overline{\bm{3}})+4(\bm{6})+4(\overline{\bm{15}})+\overline{\bm{24}}$ \\
		&  &  & 5 & $9(\overline{\bm{3}})+8(\bm{6})+7(\overline{\bm{15}})+\overline{\bm{15}}'+2(\overline{\bm{24}})$ \\
		\hline
		\multirow{6}{*}{$\phi = 2$} & \multirow{4}{*}{$\overline{\bm{3}}$} & \multirow{4}{*}{$\bm{1}$} & 0 & $\overline{\bm{3}}$ \\
		&  &  & 1 & $\overline{\bm{3}}+\bm{6}+\overline{\bm{15}}$ \\
		&  &  & 2 & $3(\overline{\bm{3}})+2(\bm{6})+2(\overline{\bm{15}})$ \\
		&  &  & 3 & $5(\overline{\bm{3}})+4(\bm{6})+4(\overline{\bm{15}})+\overline{\bm{15}}'+\overline{\bm{24}}$ \\
		& \multirow{2}{*}{$\kappa = -1$} & \multirow{2}{*}{$\bar{\kappa} = 0$} & 4 &
		$9(\overline{\bm{3}})+7(\bm{6})+8(\overline{\bm{15}})+\overline{\bm{15}}'+2(\overline{\bm{24}})+\overline{\bm{42}}$ \\
		&  &  & 5 &
		$14(\overline{\bm{3}})+13(\bm{6})+15(\overline{\bm{15}})+3(\overline{\bm{15}}')+5(\overline{\bm{24}})+2(\overline{\bm{42}})$ 
	\end{tabular}
    \caption{The multiplet content of the sectors of the doubled theory with charge 2. Note that the middle ($\phi = 1$) section of the table in which the fast-moving theory is in the trivial $\bm{1}$ primary sector has the same multiplet content as that of the $\overline{\bm{3}}$ primary state sector of the chiral $\mathrm{SU}(3)_1$ theory, which contains conjugate representations to those found in the $\bm{3}$ primary state sector delineated in Table \ref{table:chiralmultiplets}. }	
	\label{table:doubledmultiplets2}
\end{table}

\begin{figure}[t]
    \includegraphics[scale=2.0,trim=0.4cm 0 0 0]{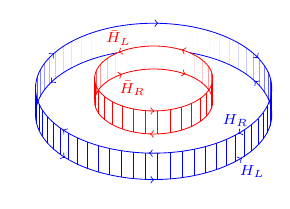}
    \caption{The boundaries of two concentric cut cylinders are depicted along the cut. The outer cylinder colored blue represents the chiral, $\mathrm{SU}(3)_1$ part of the doubled Chern-Simons theory, while the inner, red cylinder represents the anti-chiral $\overline{\mathrm{SU}(3)_1}$ part. The blue cylinder boundary hosts a left-moving edge (on the bottom), with boundary Hamiltonian $H_L$, and a right-moving edge (on the top), with boundary Hamiltonian $H_R$. Meanwhile, the red cylinder boundary hosts a right-moving edge (on the bottom), with boundary Hamiltonian $\bar{H}_R$, and a left-moving edge (on the top), with boundary Hamiltonian $\bar{H}_L$. Short vertical bars between the edges in each cylinder represent the coupling between them (i.e., between $H_L$ and $H_R$, or between $\bar{H}_R$ and $\bar{H}_L$), present in the original system, that gaps the ground state.}
    \label{fig:boundarycylinders}
\end{figure}

\subsection{The conformal boundary state description of the entanglement
spectrum of the (2+1)-dimensional topological theory}
\label{sec:cftbdy}

The previous subsection outlines how we can write down a Hamiltonian $H_{\text{total}}$, Eq.~\eqref{eq:htotal}, that is the sum of a left-moving CFT Hamiltonian $H_L$, a right-moving CFT Hamiltonian $\bar{H}_R$, and their interactions $H_{\text{interaction}}$. 
In this section, we wish to describe the relationship of this $H_{\text{total}}$ to the numerically computed entanglement Hamiltonian $H_{1D}$. 
For a \emph{chiral} topological state described by \emph{chiral} Chern-Simons theory (as opposed to \emph{doubled} Chern-Simons theory), leading to a \emph{chiral} CFT Hamiltonian (such as either $H_L$ or $\bar{H}_R$) considered on its own (and consequently with no $H_{\text{interaction}}$ either), this relationship amounts to the statement of the Li-Haldane correspondence, as touched upon in Section \ref{sec:esmes}. 
Below, we will first briefly review in a bit more detail how this works from the conformal boundary state point of view, an approach which we then use to generalize the correspondence to the case of a non-chiral topological state described by doubled Chern-Simons theory, which will lead to the appearance of a (non-chiral) Hamiltonian $H_{\text{total}}$. 

This point of view was first exhibited in Ref.~\cite{Qi2012}. 
Our brief recapitulation of the chiral CFT case here will follow the discussion in Ref.~\cite{ArildsenLudwig2022}. 
Consider the case of a chiral topological state on a cylinder such that, when the cylindrical geometry is bipartitioned, the theory of one of the two physical edges (specifically, the left-moving edge), produced by a physical cut at the location of that bipartition, is governed by the CFT Hamiltonian $H_L$ above. 
We can take the so also generated decoupled, counter-propagating, right-moving edge to have Hamiltonian $H_R$.
(Note that this is distinct from $\bar{H}_R$, as we are currently only working with opposite edges of a \emph{chiral} topological state---we will return to $\bar{H}_R$ below.) 
Consider a quantum quench, which begins at time $t = 0$ with the full Hamiltonian of what will become the partition, before the cut, i.e., a Hamiltonian including terms for $H_L$ and $H_R$, as well as any coupling between them present in the original system, which gaps the ground state of the system composed of the two counter-propagating edges produced by the physical cut;
compare the blue parts of Fig.~\ref{fig:boundarycylinders}, where the short bars represent such a coupling.
At times $t > 0$ the quench proceeds to a dynamics with a decoupled Hamiltonian $H_L + H_R$. 
In effect, this quench enacts the physical cutting of the cylinder. 
The initial condition of the quench at $t = 0$ will then be the topological ground state $|G\rangle$ of the entire state on the cylinder. 
The state $|G\rangle$ is a state in the Hilbert space of a non-chiral CFT on which $H_L$ and $H_R$ act, and can be understood to be related to a conformally invariant fixed point boundary state $|G_*\rangle$ of this CFT by a renormalization group (RG) flow into the deep infrared.
$|G_*\rangle$ itself is not normalizable, but the ``deformation" of $|G_*\rangle$ to the actual ground state $|G\rangle$ by irrelevant boundary operators
\footnote{moving the state within the basin in attraction
of the boundary fixed point away from that fixed point}
will have finite norm.
The set of irrelevant boundary operators consists of the boundary energy momentum tensor, and others~\cite{Cardy2016}.
The effect of the additional irrelevant  operators, besides the boundary energy momentum tensor, is manifest in splittings of the levels of the chiral entanglement spectrum~\cite{ArildsenLudwig2022}. 
For now, we will first just take the boundary energy momentum tensor of the CFT as our (currently sole) irrelevant boundary  operator
\footnote{This is the approach of Refs.~\cite{Calabrese2006,Calabrese2007} and others, which we will follow here first for simplicity.}. 
The boundary energy momentum tensor has integrals $H_L$ [on the left-moving edge, Eq.~\eqref{eq:cfth}] and $H_R$ [on the right-moving edge, Eq.~\eqref{eq:cfthr}]. 
Therefore we can represent 
\begin{equation}
	\label{eq:qig}
	|G\rangle \propto e^{-\tau_0 (H_L + H_R)}|G_*\rangle,
\end{equation}
for some ``extrapolation length'' $\tau_0$. 
From this, we can then proceed to obtain the reduced density matrix in a particular sector, reduced by tracing out all the right-moving degrees of freedom which are located (in a suitable convention) on the left-hand side of the cut (the upper blue edge in Fig.~\ref{fig:boundarycylinders}): 
that is~\cite{Qi2012},
\begin{equation}
    \rho_{L,a} = \Tr_R \left(|G_a\rangle\langle G_a|\right) \propto \mathcal{P}_a e^{-4\tau_0 H_L} \mathcal{P}_a,
\end{equation}
where $|G\rangle_a$, the ground state in the topological flux sector $a$, corresponds in the language of Section \ref{sec:mes} to a particular minimally entangled state. 
($\mathcal{P}_a$ is the projector onto the sector $a$.)
We then see that the spectrum $\text{spec}(-\ln \rho_{L,a})$ is the entanglement spectrum in that sector, which can now be seen to exhibit a correspondence with the characteristic degeneracies, and here the $\mathrm{SU}(3)$ multiplet content, of the left-moving chiral CFT governed by the Hamiltonian $H_L$. 

We return to the case of $H_{\text{total}}$, of Eq.~\eqref{eq:htotal}, 
which is needed for the (non-chiral) PEPS discussed here. 
As opposed to the chiral wavefunction considered in the previous paragraph, when we consider instead this \emph{non-chiral}  PEPS wavefunction on the surface of a bipartitioned cylinder, the Hamiltonians describing the theory on the two sides of the physical cut will now be $H_{\text{total}}$ and $\bar{H}_{\text{total}}$, respectively. 
Compare Fig.~\ref{fig:boundarycylinders}, where the blue parts represent the cut of the \emph{chiral} $\mathrm{SU}(3)_1$ part of the doubled $\mathrm{SU}(3)_1\otimes\overline{\mathrm{SU}(3)_1}$ Chern-Simons theory (reviewed above), while the red parts represent the cut of the \emph{anti-chiral} $\overline{\mathrm{SU}(3)_1}$ part: 
From Eqs.~\eqref{eq:htotal} and \eqref{eq:doubledh}, and their
\emph{anti-chiral} counter-propagating counterparts, we have that
\begin{align}
    \label{eq:doubledhtotal}
    H_{\text{total}} &= H_L + \bar{H}_R + H_{\text{interaction}} \\
    \bar{H}_{\text{total}} &= \bar{H}_L + H_R + \bar{H}_{\text{interaction}}.
    \label{eq:doubledhbartotal}
\end{align}
Note that, referring again to Fig.~\ref{fig:boundarycylinders}, 
on the left-hand side of the cut reside (in a suitable convention) the upper blue and red counter-propagating edge states (denoted by $H_R$ and $\overline{H}_L$, respectively), 
whereas on the right-hand side of the cut reside the lower blue and red counter-propagating edge states (denoted by $H_L$ and $\overline{H}_R$, respectively). 
Upon tracing over the degrees of freedom on the left-hand side of the cut we are left with the lower pair of counter-propagating blue and red edge states (denoted by $H_L$ and $\overline{H}_R$), residing on the right-hand side of the cut. 
Since the microscopic PEPS wave function will contain interactions between degrees of freedom arising from the chiral and the anti-chiral parts of the doubled Chern-Simons topological field theory 
(reflected, e.g., in the fact that this wave function only possesses global diagonal $\mathrm{SU}(3)$, and not $\mathrm{SU}(3)\times\overline{\mathrm{SU}(3)}$ symmetry), 
there will be interaction terms present between the (blue and red) counter-propagating edge states on either side of the cut, as displayed in Eqs.~\eqref{eq:doubledhtotal} and \eqref{eq:doubledhbartotal} above.
For these two equations to encompass the whole low-energy entanglement spectrum, including splittings which arise from within each of the chiral and anti-chiral parts of the doubled Chern-Simons topological field theory (of the kind discussed in the previous paragraph), rather than due to the interaction between them, we need to include the integrals of the additional irrelevant boundary operators mentioned above as well. 
The specific operators are not crucial here
\footnote{They are spelled out in detail for a number of chiral
cases in Ref.~\cite{ArildsenLudwig2022}.}, 
so we will simply say that we can formally include their integrals into the definition of what we call
$H_L$, $\bar{H}_R$, $\bar{H}_L$, and $H_R$ of Eqs.~\eqref{eq:doubledhtotal} and \eqref{eq:doubledhbartotal}.
In the absence of $H_{\text{interaction}}$, we could simply carry out the analysis of the previous paragraph concerning the chiral case for decoupled $H_L$ and $\bar{H}_R$ in parallel.
The resulting entanglement spectrum would possess the characteristic degeneracies, and (in our case) the total $\mathrm{SU}(3)$ multiplet content, of $H_{\text{doubled}} = H_L + \bar{H}_R$.
These multiplets consist, as discussed in the previous section, of the tensor products of the $\mathrm{SU}(3)$ multiplets from the (in this case decoupled) left- and right-moving branches of the entanglement spectrum.
For the decoupled $H_{\text{doubled}}$, the multiplets of the total $\mathrm{SU}(3)$ corresponding to a given tensor product will all be degenerate in energy.
Introducing a non-zero (and, recall, $\mathrm{SU}(3)$-symmetric) $H_{\text{interaction}}$, however, can split this type of degeneracy, in cases where both of the multiplets factoring into the tensor product have a dimension greater than one.
That is to say, the $\mathrm{SU}(3) \otimes \overline{\mathrm{SU}(3)}$ symmetry of $H_{\text{doubled}}$ (from the $\mathrm{SU}(3)$ symmetry of $H_L$ and $\overline{\mathrm{SU}(3)}$ symmetry of $\bar{H}_R$) is broken to the total (diagonal) $\mathrm{SU}(3)$ by adding $H_{\text{interaction}}$. 
Thus, the presence of splittings among the multiplets in the entanglement spectrum that come from the same tensor product of multiplets in the two, left- and right-moving, chiral branches serves as evidence (in addition to the previously mentioned gapped nature of the entanglement spectrum seen in Ref.~\cite{kurecic:su3_sl}) that $H_{\text{interaction}}$ is non-zero.
In the case where the entanglement gap (proportional to the inverse correlation length) induced by the interaction $H_{\text{interaction}}$ is sufficiently small as compared to the inverse system size (length of the circular cut), however, the entanglement spectrum will, as already mentioned, be close to what it would be in the ideal non-interacting doubled case, up to the likewise weak but clearly visible splittings of the otherwise degenerate groups of multiplets from the same tensor product which are, as discussed above, a consequence of these interactions.
This is the case that we find here---the conformal boundary state approach is able to guide our understanding of the underlying physics of the entanglement spectrum as expected in this limit, despite the gapped nature of the actual, no longer chiral boundary theory along the cut.

\section{Results}
\label{sec:results}

\begin{figure}[H]
	\centering
	\subfigure[]{\label{fig:doubledes00} \includegraphics[scale=0.32]{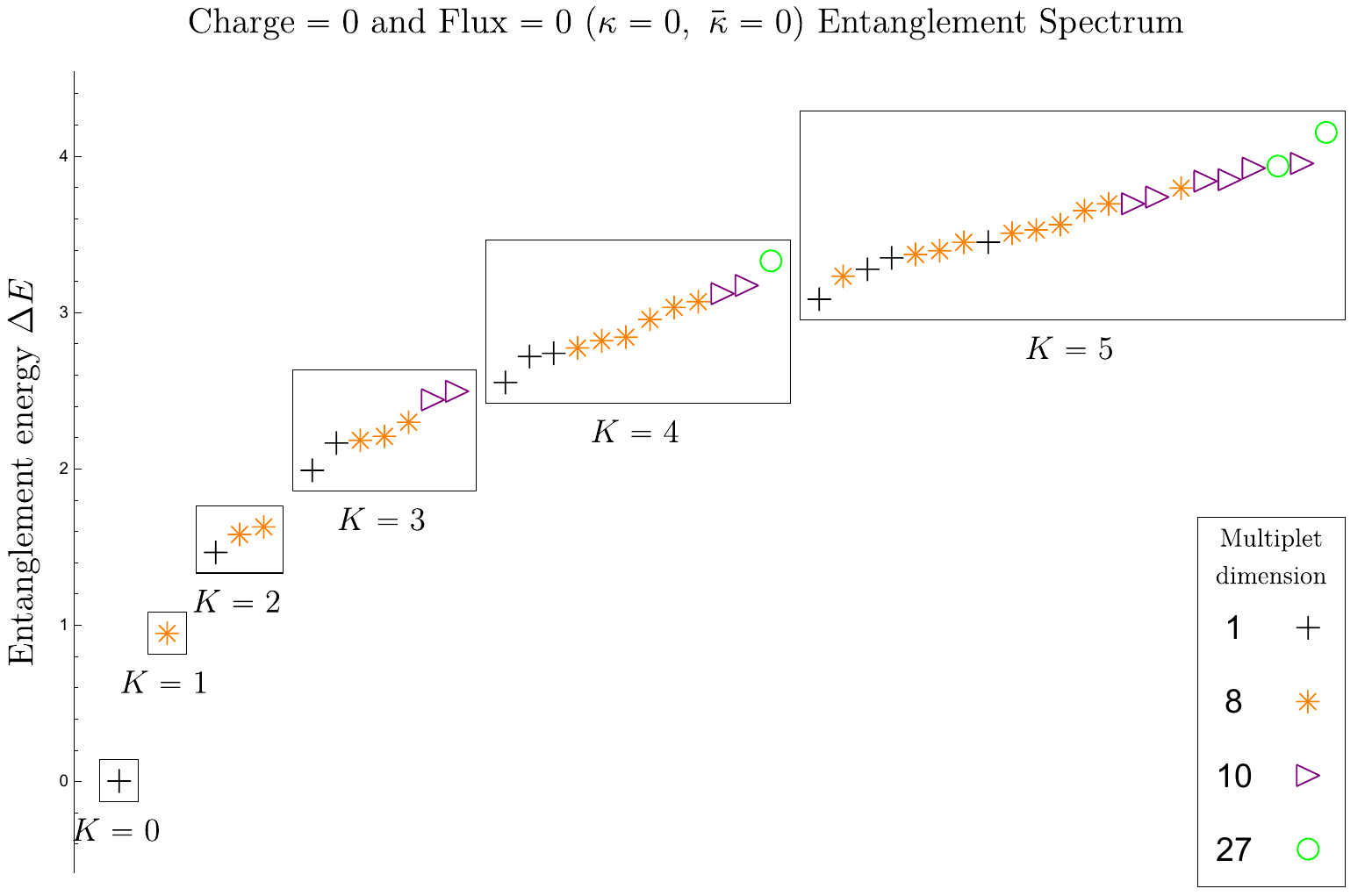}}
	\subfigure[]{\label{fig:doubledes01} \includegraphics[scale=0.32]{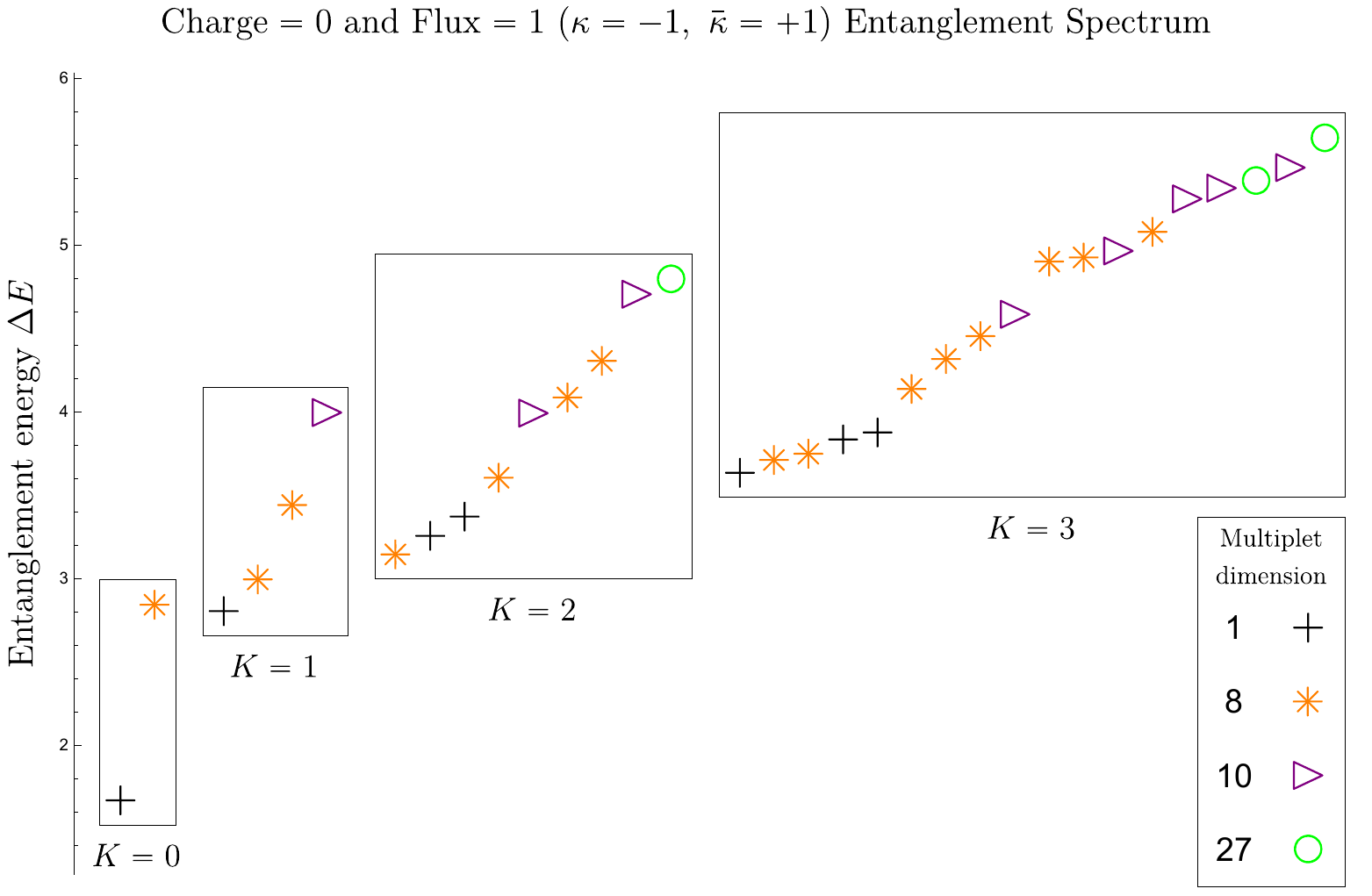}}
	\subfigure[]{\label{fig:doubledes02} \includegraphics[scale=0.32]{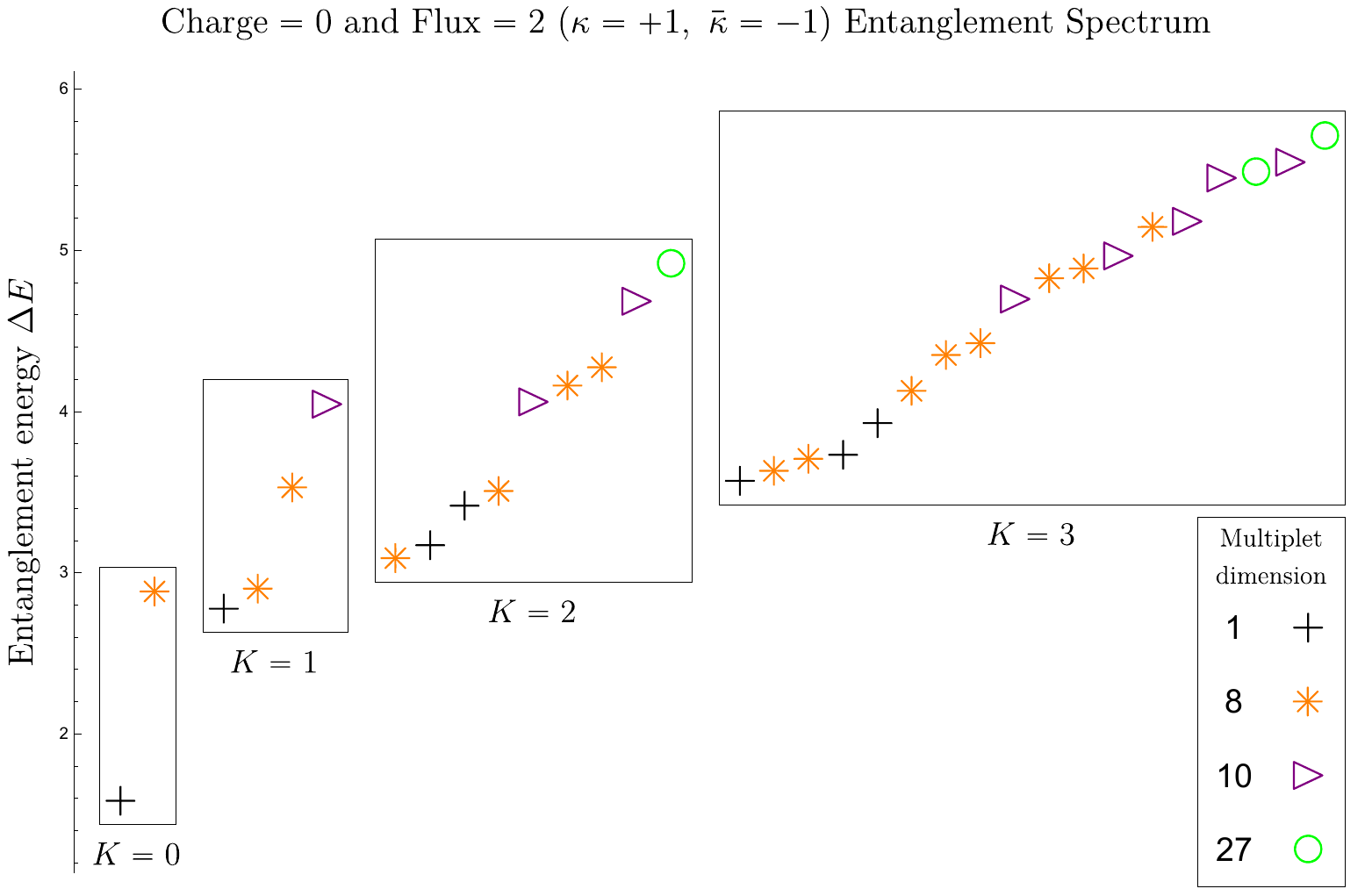}}
	\caption{The entanglement spectra of the PEPS obtained at charge $q = 0$ and with flux $\phi = 0$, 1, and 2, in (a), (b), and (c) respectively, corresponding to $(\kappa, \bar{\kappa})=$~$(0,0)$, $(-1,+1)$, and $(+1, -1)$, on a cylinder with circumference $N=6$.
	The multiplet dimensions and multiplicities correspond precisely to those of Table \ref{table:doubledmultiplets0}.
	The horizontal separation of the data points within each box at level $K$ has been artificially added in order to more clearly show overlapping data.}
	\label{fig:doubledes0}
\end{figure}

\begin{figure}[H]
	\centering
	\subfigure[]{\label{fig:doubledes10} \includegraphics[scale=0.33]{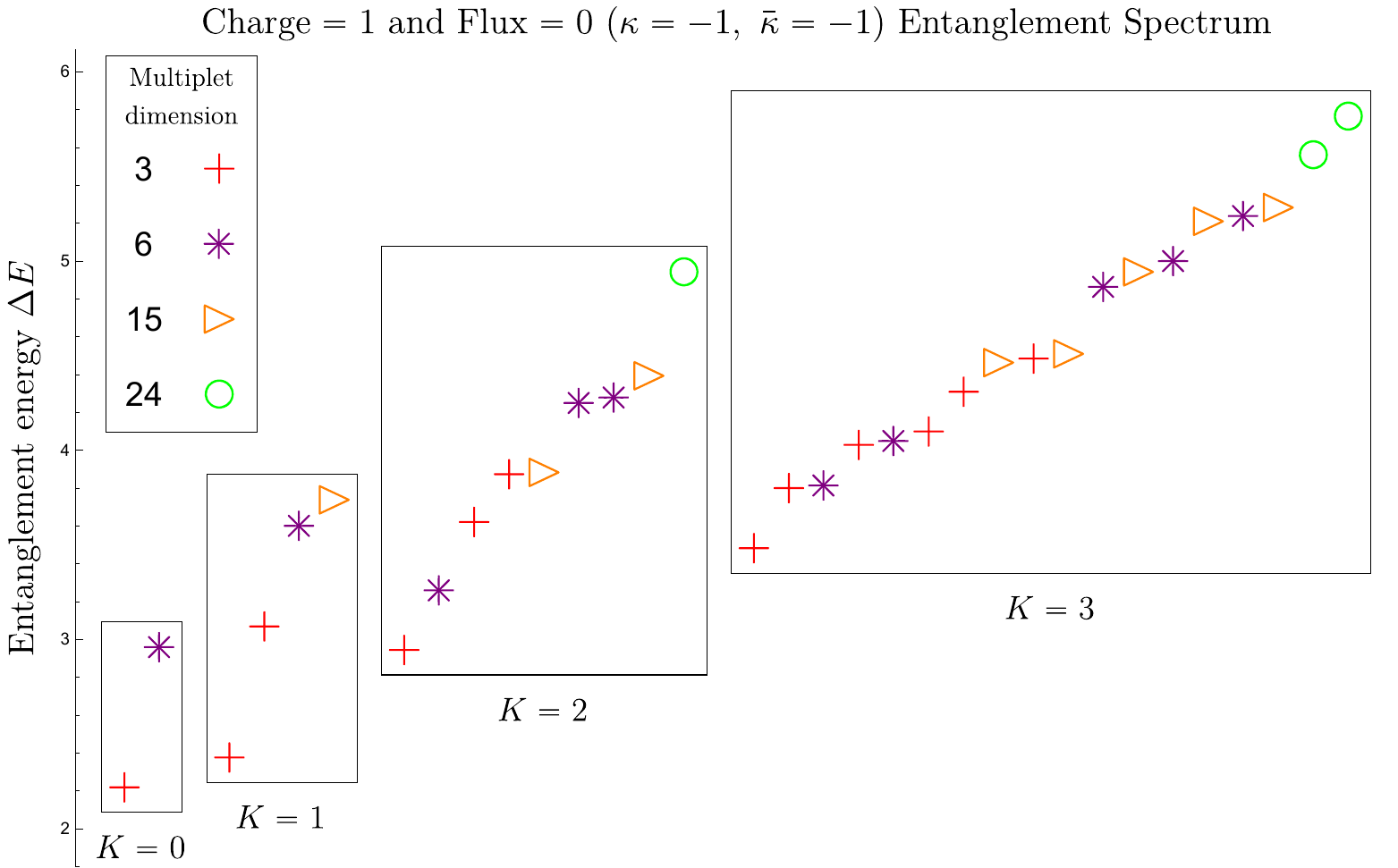}}
	\subfigure[]{\label{fig:doubledes11} \includegraphics[scale=0.33]{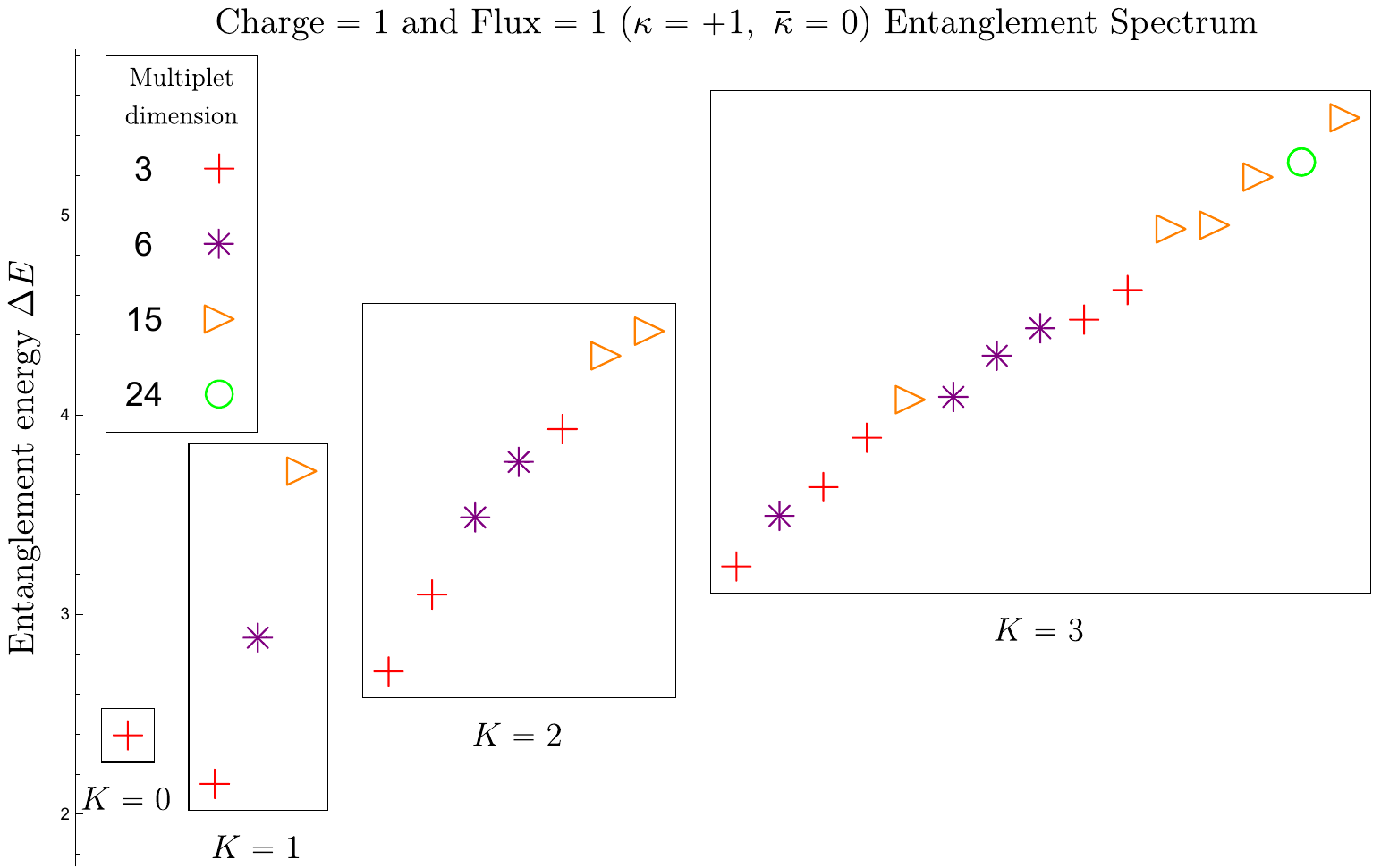}}
	\subfigure[]{\label{fig:doubledes12} \includegraphics[scale=0.33]{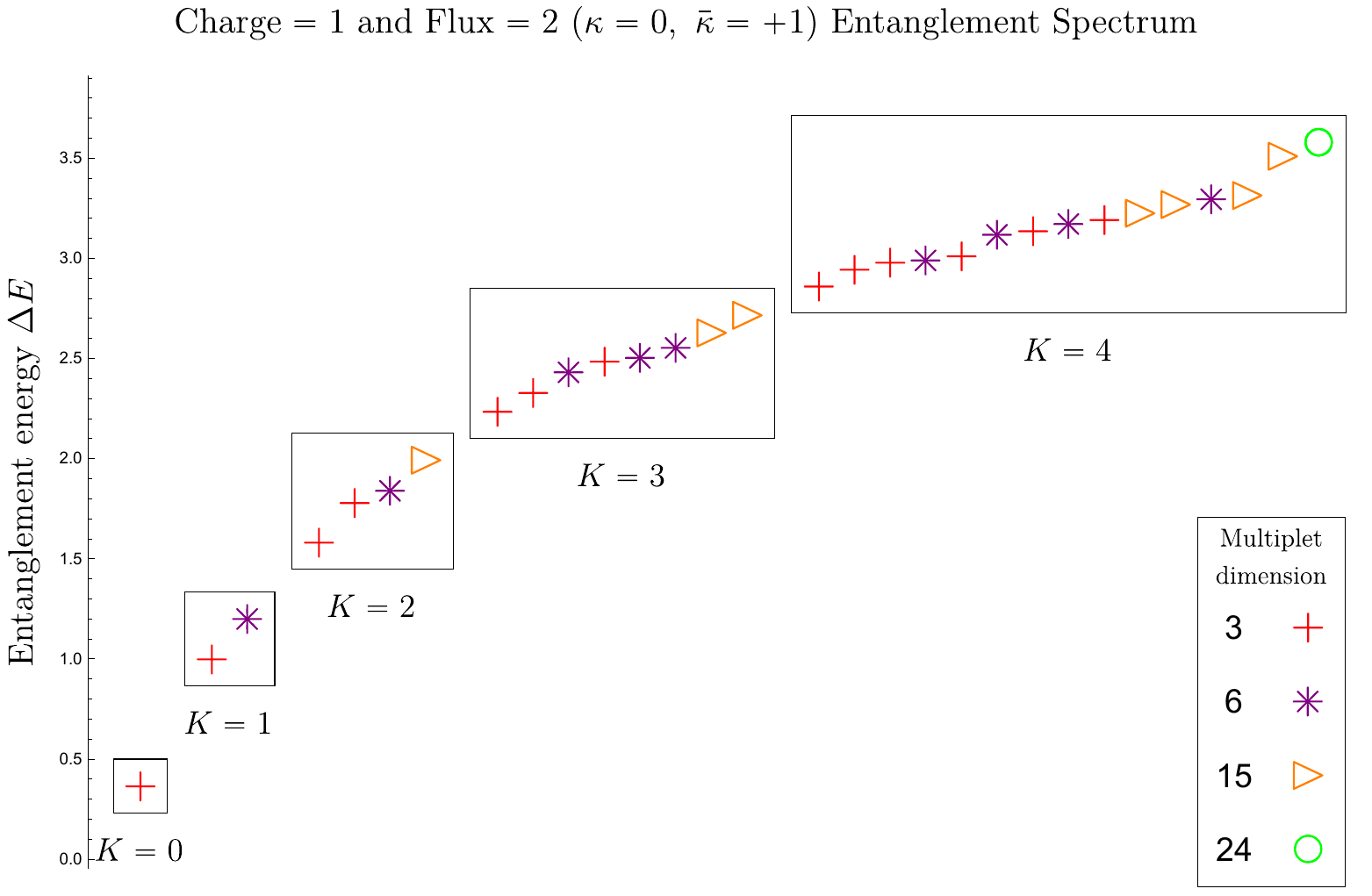}}
	\caption{The entanglement spectra obtained at charge $q = 1$ and with flux $\phi = 0$, 1, and 2, in (a), (b), and (c) respectively, corresponding to $(\kappa, \bar{\kappa})=$~$(-1,-1)$, $(+1,0)$, and $(0, +1)$, on a cylinder with circumference $N=6$.
	The multiplet dimensions and multiplicities correspond precisely to those of Table \ref{table:doubledmultiplets1}.
	The horizontal separation of the data points within each box at level $K$ has been artificially added in order to more clearly show overlapping data.}
	\label{fig:doubledes1}
\end{figure}

\begin{figure}[H]
	\centering
	\subfigure[]{\label{fig:doubledes20} \includegraphics[scale=0.33]{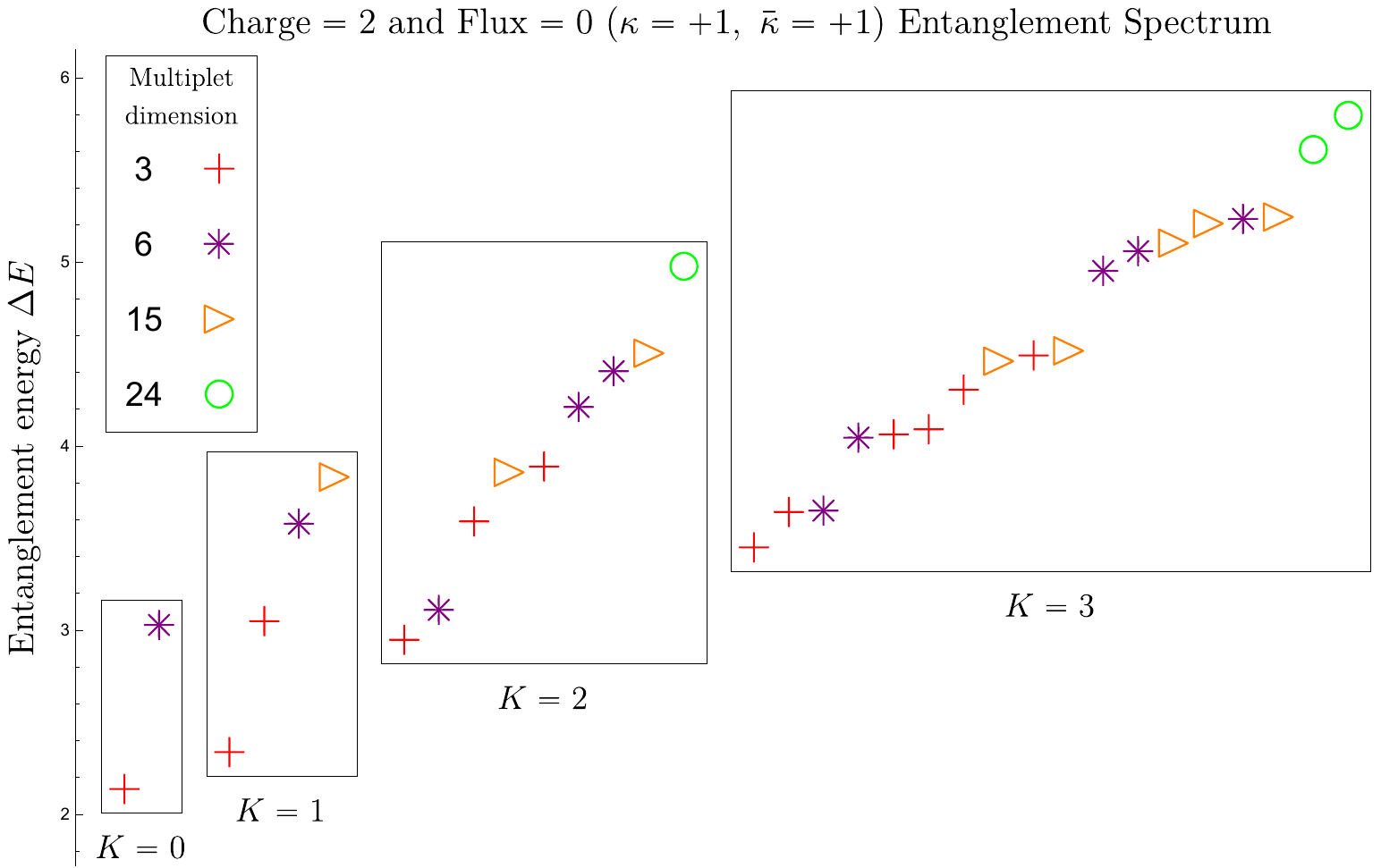}}
	\subfigure[]{\label{fig:doubledes21} \includegraphics[scale=0.33]{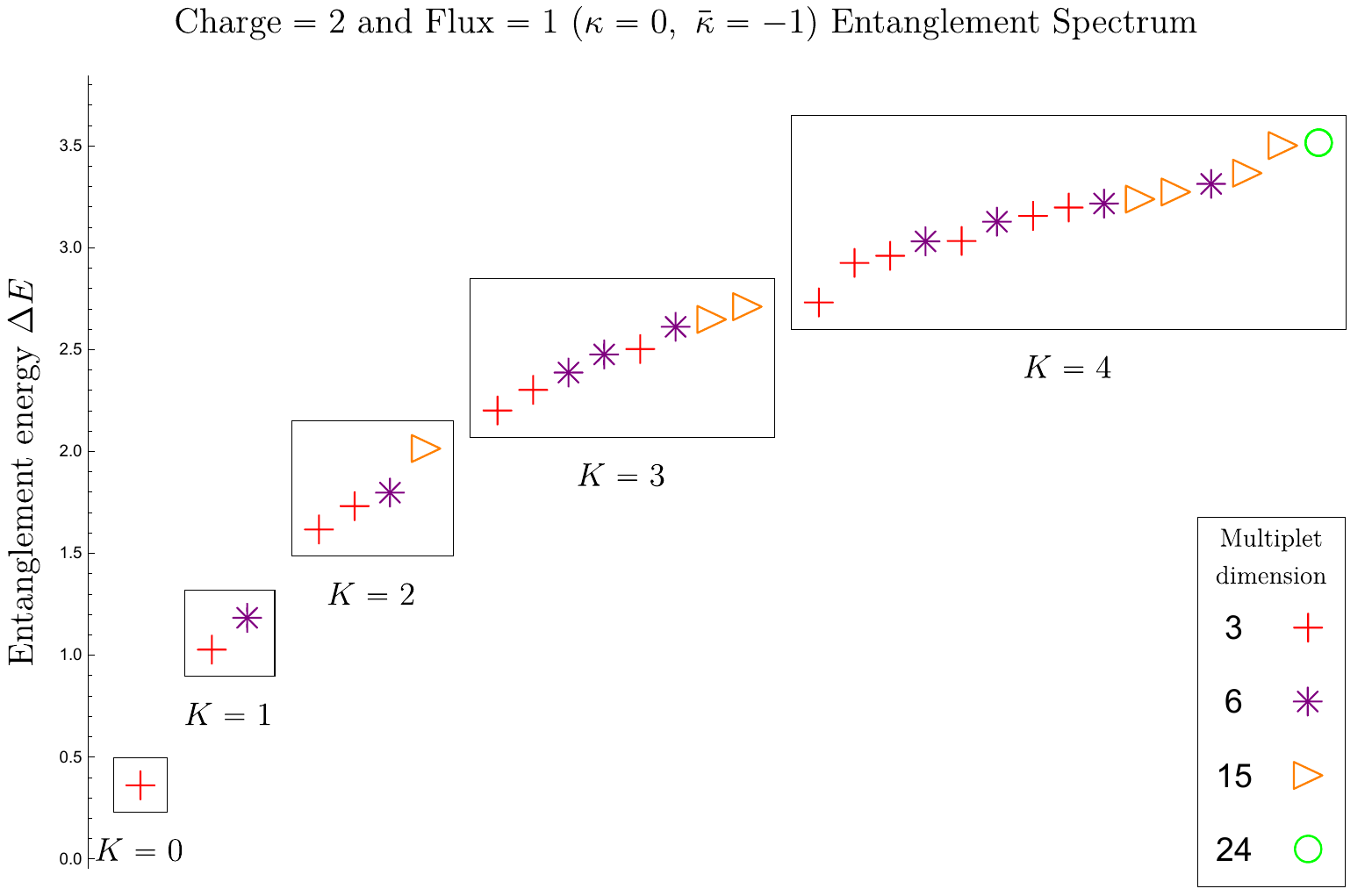}}
	\subfigure[]{\label{fig:doubledes22} \includegraphics[scale=0.33]{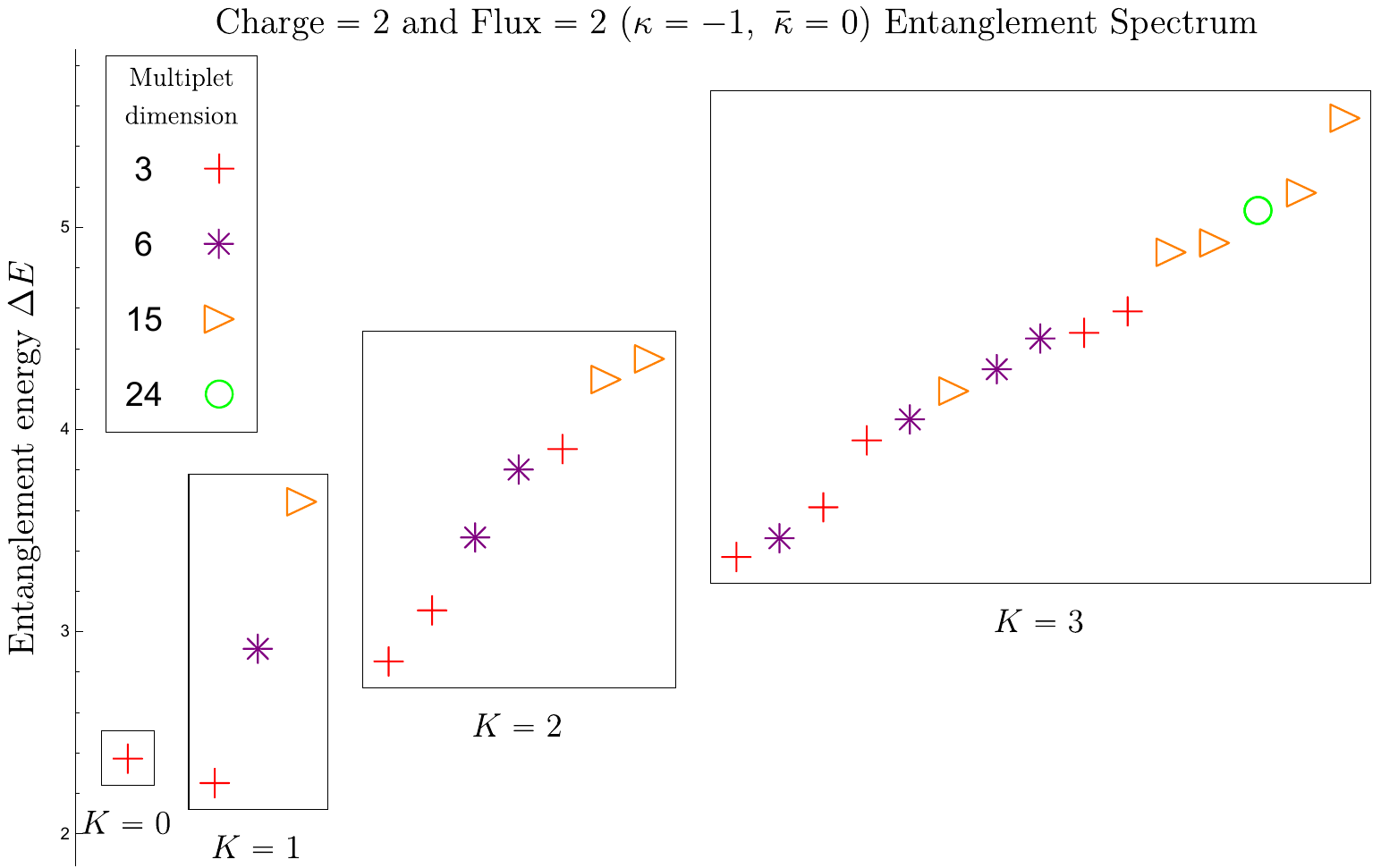}}
	\caption{The entanglement spectra obtained at charge $q = 2$ and with flux $\phi = 0$, 1, and 2, in (a), (b), and (c) respectively, corresponding to $(\kappa, \bar{\kappa})=$~$(+1,+1)$, $(0,-1)$, and $(-1, 0)$, on a cylinder with circumference $N=6$.
	The multiplet dimensions and multiplicities correspond precisely to those of Table \ref{table:doubledmultiplets2}.
	The horizontal separation of the data points within each box at level $K$ has been artificially added in order to more clearly show overlapping data.}
	\label{fig:doubledes2}
\end{figure}

The dimensionalities of the $\mathrm{SU}(3)$ multiplets occurring in each charge and flux topological sector of the entanglement spectrum calculated from the PEPS of Section \ref{sec:peps}, at circumference $N = 6$, are shown in Figs.~\ref{fig:doubledes0}-\ref{fig:doubledes2}.
These results are in exact agreement with those we would expect from the indicated charge projection and flux threading of the PEPS based on the mapping between anyon types in the Drinfeld double $D(\mathbb{Z}_3)$ and doubled $\mathrm{SU}(3)_1 \otimes \overline{\mathrm{SU}(3)_1}$ Chern-Simons theory, described in Table \ref{table:anyoncomparison}. 
In the notation of Table \ref{table:anyoncomparison}, we take $\kappa$ and $\bar{\kappa}$ to correspond to the ``fast" and ``slow" (left- and right-moving) chiral branches, respectively, of the CFT underlying (as explained in the previous section) the entanglement spectrum, with the anyon types 0, 1, and $-1$ corresponding to the primary state sectors in $\mathrm{SU}(3)$ representations $\bm{1}$, $\bm{3}$, and $\overline{\bm{3}}$. 
We found in Eq.~\eqref{eq:charge} that the charge is given by $q=\kappa+\bar{\kappa}~({\rm mod}~3)$, and we use this to sort the nine sectors into three groups based on their charge. 
The multiplet content of the sectors with charge 0 is found in Table \ref{table:doubledmultiplets0}, while that of sectors with charge 1 and 2 can be found in Tables \ref{table:doubledmultiplets1} and \ref{table:doubledmultiplets2}. 
Within each table of the same charge, the flux [$\phi = \kappa-\bar{\kappa}~({\rm mod}~3)$ from Eq.~\eqref{eq:flux}] then uniquely specifies the pair of chiral branch primary states (``fast" minus ``slow") that determines which tensor products of the $\mathrm{SU}(3)$ representation content of branches we expect in particular. 

Looking at Figs.~\ref{fig:doubledes0}-\ref{fig:doubledes2}, we can now see that the dimensionality and multiplicity of the multiplets in each sector indeed correspond with the multiplet content for the anyonic sectors of the doubled theory, as described in Tables \ref{table:doubledmultiplets0}-\ref{table:doubledmultiplets2}. 
The data of Figs.~\ref{fig:doubledes0}-\ref{fig:doubledes2} consists of entanglement \emph{excitation} energies $\Delta E$ relative to the baseline entanglement energy $E_0$ of the ground state, as in Eq.~\eqref{eq:def-delta-ei}. 
For the entanglement spectrum we analyze, $E_0$ is the entanglement energy of the primary state (the state at level $K = 0$) of the singlet $|\bm{1}\rangle_L \otimes |\bm{1}\rangle_R$ of the charge 0, flux 0 sector.
Thus, for that state, as can be observed in Fig.~\ref{fig:doubledes00}, the plotted entanglement (excitation) energy is $\Delta E_{0,00} = 0$. 
Note that here, and for the rest of this paper, we use the notation $\Delta E_{K,q\phi}$, defined as follows
\footnote{When there are several degenerate multiplets at level $K$ in the charge $q$, flux $\phi$ sector, the labeling of the corresponding eigenvalues of the entanglement Hamiltonian would require an additional
multiplicity label (these degeneracies will be split in entanglement energy, as seen in the Figures below); however, in the remainder of this paper we will use the explicit
notation $\Delta E_{K,q\phi}$ only for non-degenerate eigenvalues.}:
\begin{align}
\begin{split}
    \Delta E_{K,q\phi} = \textrm{numerical entanglement excitation energy} \\ 
    \textrm{at level $K$ in the charge $q$, flux $\phi$ sector} \quad
\end{split}
\end{align}

The sectors where the ``fast" side is in the trivial $|\bm{1}\rangle_L$ (singlet) sector will simply have the counting of the chiral $\overline{\mathrm{SU}(3)_1}$ CFT sector corresponding to the ``slow" $R-$side, as is apparent from comparing the counting of Figs.~\ref{fig:doubledes00}, \ref{fig:doubledes12}, and \ref{fig:doubledes21} to Table \ref{table:chiralmultiplets}.
Indeed, as $L_0|\bm{1}\rangle_L = h_{\bm{1}}|\bm{1}\rangle_L = 0$, and we have effectively subtracted off the terms constant in $\bar{L}_0$ in the entanglement excitation energy $\Delta E$, we see that the only piece of the Hamiltonian of the non-chiral (``doubled'') CFT of Eq.~\eqref{eq:doubledh} that will contribute to $\Delta E$ in these particular sectors will be $\frac{2\pi}{N}\overline{v}_R \bar{L}_0$, neglecting the effect of $H_{\text{interaction}}$.
The numerical entanglement energy of the state transforming in the irreducible representation $\bm{3}$ at level $K = 0$ in the charge 1, flux 2 sector [where $(\kappa, \bar{\kappa})=$ $(0,+1)$] is $\Delta E_{0,12} \approx 0.360$, while that transforming in the representation $\overline{\bm{3}}$ at $K = 0$ in the charge 2, flux 1 sector [where $(\kappa, \bar{\kappa})=$ $(0,-1)$] is $\Delta E_{0,21} \approx 0.358$.
Meanwhile, the conformal weight of the $|\bm{3}\rangle_R$ and $|\overline{\bm{3}}\rangle_R$ sectors is $\bar{h}_{\bm{3}} = \bar{h}_{\overline{\bm{3}}} = 1/3$, so the results for $\Delta E_{0,12}$ and $\Delta E_{0,21}$ are quite close to the corresponding eigenvalues of $\bar{L}_0$ we expect for these multiplets.
Further, the $\bm{8}$ multiplet at $K = 1$ in the charge 0, flux 0 sector [where $(\kappa, \bar{\kappa})=$ $(0,0)$], corresponding to the eight current descendants $\bar{J}_{-1}^{\bar{a}}|\bm{1}\rangle_L \otimes |\bm{1}\rangle_R$, has eigenvalue 1 under $\bar{L}_0$, and we have $\Delta E_{1,00}\approx 0.945$.
Thus, empirically, it would seem that at these lowest energy levels, we can approximate $\frac{2\pi}{N}\overline{v}_R \approx 1$. 

We can also consider the charge 1, flux 1 and charge 2, flux 2 sectors [where $(\kappa, \bar{\kappa})=$ $(+1,0)$ and $(\kappa, \bar{\kappa})=$ $(-1,0)$, respectively], which have $\bm{3}$ and $\overline{\bm{3}}$ multiplets, respectively, at $K = 0$.
Here, it is the ``slow'' side that is, at least for the lowest-energy state, in the $\bm{1}$ representation. 
Thus the ``slow" primary states of these sectors will have $\bar{h}_{\bm{1}} = 0$, and so the contribution of the non-chiral $L$/$R$ ``doubled'' Hamiltonian of Eq.~\eqref{eq:doubledh} to the entanglement excitation energy $\Delta E$ of the lowest-energy states in these sectors will instead reduce to $\frac{2\pi}{N}v_L L_0$, again neglecting the effect of $H_{\text{interaction}}$, and subject to the assumption that $v_L$ is uniform across the $\ket{\bm 1}_L$, $\ket{\bm 3}_L$, and $\ket{\overline{\bm 3}}_L$ sectors
\footnote{Only the primary states of the ``fast''-moving side factor into the low-energy part of the entanglement spectrum that we see, so unlike on the ``slow'' side for $\overline{v}_R$ in the $\protect\ket{\bm{1}}_R$ sector, it is not possible to deduce $v_L$ for the $\protect\ket{\bm{1}}_L$ sector from states within the same sector on the fast-moving side. 
Thus we must assume that $v_L$ is the same, or close to it, in all three sectors, such that setting $\Delta E_{0,00} = 0$ to obtain the entanglement excitation energy $\Delta E$ removes an identical constant term [see Eq.~\protect\eqref{eq:doubledh}] from all sectors.}.
The primaries in both charge 1, flux 1 and charge 2, flux 2 sectors of the $L$/$R$ doubled theory have $L_0$ eigenvalue $h_{\bm{3}} = h_{\overline{\bm{3}}} = 1/3$. The numerical entanglement energies we find for these states are $\Delta E_{0,11} \approx 2.39$ and $\Delta E_{0,22} \approx 2.37$. 

Note that these numerical values for the entanglement energy are dependent on sector-by-sector normalizations of the entanglement spectrum, which we have taken to be equal---this is discussed in Section \ref{sec:pepses}, which notes that this is a consequence of the locality of the entanglement Hamiltonian of a PEPS as outlined in Ref.~\cite{schuch:topo-top}.
With this understanding, we can make direct comparisons between the numerical entanglement energies of the different sectors. In the earlier case of the $(h,\bar{h}) = (0,1/3)$ primary states, we had $\Delta E_{0,12} \approx \Delta E_{0,21} \sim \frac{2\pi}{3N}\overline{v}_R$, while in the case of the $(h,\bar{h}) = (1/3,0)$ primary states we have $\Delta E_{0,11} \approx \Delta E_{0,22} \sim \frac{2\pi}{3N}v_L$
\footnote{This is again subject to the assumption that $v_L$ is uniform across all primary state sectors. Then $v_L$ can be found from our knowledge of the conformal weight of the primaries, which is the approach that is taken here.}.
Taking the ratio of the numerical entanglement energies $\Delta E_{0,11} \approx 2.39$ and $\Delta E_{0,22} \approx 2.37$ of the latter to the entanglement energies $\Delta E_{0,12} \approx 0.360$ and $\Delta E_{0,21} \approx 0.358$ of the former therefore yields a rough estimate of $v_L/\overline{v}_R \sim 6.6$. 
Within the first 200 eigenvalues of the entanglement spectrum that we consider, no entanglement excitation energy $\Delta E$ greater than 6 occurs. 
Since the units are approximately those appropriate to the ``slow"-moving theory, with $\frac{2\pi}{N}\overline{v}_R \sim 1$, this shows that in the range of numerical data for the entanglement energies we consider, we would not expect to see evidence of \emph{any} descendant states from the ``fast"-moving theory, as their lowest contribution to the entanglement energy (from states with with $\bar{L}_0$ eigenvalue 1) is approximately $\frac{2\pi}{N}v_L \sim 6.6 \frac{2\pi}{N}\overline{v}_R \sim 6.6$.
This is in accord with the obtained result that all of the entanglement spectrum data is consistent with the expected countings for the $L$/$R$ doubled theory including only the primary states on the ``fast"-moving side; hence our focus on the $v_L \gg \overline{v}_R$ case for the description of that theory in Section \ref{sec:countings}.

We note that an independent, rough and less accurate estimate of the velocity ratio can be obtained from the entanglement spectrum data in the thermodynamic limit $N=\infty$ (obtained by using an iMPS excitation ansatz), plotted in Fig.~4(b) of Ref.~\cite{kurecic:su3_sl}.
These data exhibit two branches of a ``dispersion" relation which have (as already stressed in that reference) different slopes for the  right- and the left-moving branch.
Estimating the ratio of those slopes from these plots
\footnote{especially the green dispersion curve in the middle of the plot giving a value $v_L/\overline{v}_R \sim 6.1$, or the blue curves giving a range of values $v_L/\overline{v}_R \sim 5.4 -7.3$; these two kinds of curves have a longest discernible steep-slope segment, while the red curves are less suitable as they are shifted upwards and thus have a shorter steep-slope segment.}
yields values of the velocity ratio roughly in the range $v_L/\overline{v}_R \sim 5.4 - 7.3$, values which are in rough agreement with the value $v_L/\overline{v}_R \sim 6.6$ obtained from the independent considerations detailed in the previous paragraph. 
The basic rough agreement of the value of the velocity ratio obtained by using independent methods serves as a confirmation of the consistency of our approach and analysis.

One noticeable characteristic of the entanglement spectrum data we show here is the great degree of similarity between the spectra in conjugate sectors. 
Conjugate sectors are those where both the ``slow" and ``fast" primary states have $\mathrm{SU}(3)$ representations conjugate to one another, resulting in an overall multiplet content of conjugate representations. 
These conjugate pairs of sectors include the following: the charge 0, flux 1 and flux 2 sectors [where $(\kappa,\bar{\kappa})=$~$(-1,+1), (+1,-1)$]; 
the charge 1 and charge 2, flux 0 sectors [where $(\kappa,\bar{\kappa})=$~$(-1,-1), (+1,+1)$]; 
the charge 1, flux 1 and charge 2, flux 2 sectors [where $(\kappa,\bar{\kappa})=$~$(+1,0), (-1,0)$];
and the charge 1, flux 2 and charge 2, flux 1 sectors [where $(\kappa,\bar{\kappa})=$~$(0,+1), (0,-1)$].
This can be verified from Tables \ref{table:doubledmultiplets0}-\ref{table:doubledmultiplets2}.
Since the countings in the entanglement spectrum only show the dimension of the multiplets, we observe the same multiplet countings in the various levels $K$ in the spectra of each of these conjugate pairs of sectors in Figs.~\ref{fig:doubledes0}-\ref{fig:doubledes2}.
This alone is not at all unexpected, and a clear consequence of the spectra exhibiting the countings of the  non-chiral (``doubled'') CFT reflecting the underlying doubled Chern-Simons topological field theory in the entanglement spectrum.
Yet the commonality extends beyond this in a non-trivial way: the relative orderings and energy levels of the multiplets within each level $K$ are substantially similar between each pair of conjugate sectors' spectra. 

We end this section by a corollary of our analysis for the nature of the low-lying entanglement spectrum at finite size in the limit of very large velocity ratio $v_L/\overline{v}_R$ of non-chiral topological quantum states (including PEPS).
It is a fact that reliable data for numerical entanglement spectra are typically only available for entanglement excitation energies $\Delta E_i$ below some threshold $\Delta E_{thresh}$.
If data for the spectrum in the thermodynamic limit $N \gg 1$ are not available (the method used in this limit in Ref.~\cite{kurecic:su3_sl} requires special technology),
one may ask whether such a finite-size entanglement spectrum alone allows us to decide whether the underlying (2+1)-dimensional topological quantum state is chiral or not. 
Let us assume that, as in the case discussed in the present paper, the system is actually non-chiral, but ``close-to-chiral'' with a velocity ratio $v_L/\overline{v}_R$ which is so large that,
in the language of the previous paragraph, the entanglement excitation energies 
$\Delta E_{0,11} \approx \Delta E_{0,22} \sim \frac{2\pi}{3N}v_L$ 
[located in the $(\kappa, \bar{\kappa})= (1,0)$ and $(-1,0)$ sectors]
lie above the threshold $\Delta E_{thresh}$ and we have no usable data on them.
Since these are the only entanglement excitation energies that have a non-singlet primary in the high-velocity left-moving ``L" sector but a singlet primary in the low-velocity right-moving ``R" sector of the CFT, any of the other sectors which have a non-singlet primary in the high-velocity sector will have even higher entanglement excitation energies and will thus also lie beyond the threshold. 
In other words, all sectors with $(\kappa, \bar{\kappa})$ where $\kappa \not = 0$ will be above threshold
[these correspond to $(q, \phi)=(0,1), (0,2), (1,0), (1, 1), (2, 0), (2,2)$].
Thus, in this situation, the only states in the finite size entanglement spectrum that will be below threshold, and thus be visible, are the states in the sectors $(\kappa,\bar{\kappa})=$ $(0,\bar{\kappa})$: these are precisely the states (and the only states) exhibiting a purely chiral finite size entanglement spectrum, based on considerations of Li-Haldane state counting. 
Thus, from the point of view of Li-Haldane state counting alone, such a finite size entanglement spectrum of the non-chiral, but ``close-to-chiral'' topological theory would be indistinguishable from that of a corresponding purely chiral topological theory.
But there is the possibility that a detailed analysis of
\emph{splittings} of the multiplets could, even in this case, exhibit unique features that would distinguish a close-to-chiral finite size entanglement spectrum, as discussed, from a purely chiral one.
An analysis of such splittings of the entanglement spectrum is not part of this present paper.
However, this question and a corresponding analysis will be pursued in future work.

Summarizing the present section, the results presented therein illustrate the successful correspondence of the entanglement spectrum of the PEPS we consider to the framework of Section \ref{sec:cftes}. And we see that in addition to providing a full understanding of the Li-Haldane counting of the low-lying entanglement spectrum at finite size in the present non-chiral case exhibiting counter-propagating branches, this framework allows us to extract quantitative data of interest from the spectrum.

\newpage

\section{Discussion and Outlook}
In this paper we have shown that the entanglement spectrum illuminates characteristics of the topological order of the non‐chiral PEPS model of Ref.~\cite{kurecic:su3_sl}, which we considered in this work.
While the low-lying levels of the entanglement spectrum appeared to have Li-Haldane counting of the multiplicities of $\mathrm{SU}(3)$ representations at given momentum characteristic of a chiral spin liquid, i.e., of the chiral $\mathrm{SU}(3)_1$ CFT, in the $q=0$ and $\phi=0$ sector investigated by Ref.~\cite{kurecic:su3_sl} (and shown in Fig.~\ref{fig:doubledes00} of the present paper), 
and also in the $q=1, \phi=2$ and $q=2$, $\phi=1$ charge/flux sectors (as shown in Fig.~\ref{fig:doubledes12} and Fig.~\ref{fig:doubledes21} of the present paper),
we have demonstrated that the full entanglement spectrum of the PEPS is characterized in all nine anyonic sectors by the multiplicities of $\mathrm{SU}(3)$ representations characteristic instead of the tensor product of two chiral, left- and right-moving $\mathrm{SU}(3)_1$ CFTs, one of which has a substantially higher velocity than the other.
The resulting Li-Haldane counting of multiplicities in the remaining six sectors turns out to be  very different from that of a corresponding chiral theory. 
We have also shown that the ratio of the two velocities, a key quantity of this system, can be extracted from the finite-size entanglement spectrum, without having to work in the thermodynamic limit. 
We obtain our results by generalizing the conformal boundary state description of the entanglement spectrum, introduced in Ref.~\cite{Qi2012} and extended in Ref.~\cite{ArildsenLudwig2022}, to the case of a non-chiral topological state.
While the resulting entanglement spectrum is gapped, for system sizes much smaller than the inverse gap it is characterized by consideration of the (1+1)-dimensional CFT that reflects the characteristics of the underlying (2+1)-dimensional topological state:
this is the regime of the numerical finite-size data from Ref.~\cite{kurecic:su3_sl} that we analyze.
The effects that we demonstrate are induced by the difference in velocities provide new insight into the structure of finite-size entanglement spectra of PEPS models in this ``close-to-chiral" regime.
This deepened understanding of what occurs in a ``close-to-chiral" PEPS model may be able to shed light on the problem of chiral PEPS
(underscored in the case of non-interacting fermionic PEPS, with the corresponding no-go theorem of Refs.~\cite{Dubail2015,Wahl2013}). 

Indeed, such a close-to-chiral (2+1)-dimensional topological state represents an instructive model which is, in a sense, intermediate between a chiral and a completely non-chiral topological state (with equal velocities for the two branches). 
This also suggests that a possibly interesting direction to explore might be the limit of such close-to-chiral models where the velocity of the high-velocity branch (in our case the left-moving branch) diverges, and the other velocity remains finite.
In this limit, the models will effectively exhibit entanglement spectra with one chirality. 
One way of varying this velocity ratio could be by tuning the relative weights of the different terms in the objects \eqref{eq:def-tau} and \eqref{eq:def-P} which define the PEPS wavefunction; for instance, in Ref.~\cite{kurecic:su3_sl}, it was observed that changing the prefactor of the rightmost term in \eqref{eq:def-P} [equivalently, changing the weight of $\bar{\bm 3}$ in \eqref{eq:def-tau}] has such an effect (cf.\ Fig.~3 vs.\ Fig.~4b of Ref.~\cite{kurecic:su3_sl}). A systematic investigation of which choice of parameters results in a wavefunction which is closest to chiral is an interesting open problem left for future work.

One further avenue of investigation is the additional information contained in these finite-size entanglement spectra, beyond the Li-Haldane countings of $\mathrm{SU}(3)$ representations that this work considers: the {\it level splittings} in the entanglement spectrum.
This refers to the fact that those states in the entanglement spectrum which are degenerate in {\it momentum} (this degeneracy being the focus of ``Li-Haldane counting''), are in general not degenerate in {\it entanglement energy}, i.e., they exhibit ``level splittings".
As the present paper demonstrates, it can often be difficult to infer chirality directly from the accessible subset of Li-Haldane countings alone. But accounting for the {\it splittings} in the entanglement spectrum provides more insight, as the behavior of splittings turns out to differ between the non-chiral PEPS state considered here and a truly chiral Abelian $\mathrm{SU}(3)$ PEPS state \cite{Arildsen2023}. It should also be noted that this analysis clarifies that while the non-chiral PEPS discussed in the present paper is a representative of an important class of possible scenarios for ``close-to-chiral" PEPS, the existence of this non-chiral PEPS does not by any means preclude the possibility of PEPS representing truly chiral states.

Finally, this work could also be naturally generalized to $(2+1)$-dimensional topological phases described by doubled $\mathrm{SU}(N)$ Chern-Simons theories at level $k$ with $N>3$ and/or $k>1$, including possible corresponding PEPS models.

\acknowledgements

This work was supported in part by the National Science Foundation under Grant No.\ DMR-1309667 (AWWL), 
and by the European Research Council (ERC) under the European Union's Horizon 2020 research and 
innovation programme through Grant No.~863476 (ERC-CoG SEQUAM). MJA acknowledges financial support from the PNRR MUR project PE0000023-NQST. The computational results presented have been achieved in part using the Vienna Scientific Cluster (VSC).

\newpage

\appendix

\section{Derivation of possible mappings between the $D(\mathbb{Z}_3)$ and $\mathrm{SU}(3)_1 \otimes \overline{\mathrm{SU}(3)_1}$ theories}
\label{app:matrixmap}

To find the possible mappings between the anyons of the $D(\mathbb{Z}_3)$ and doubled $\mathrm{SU}(3)_1 \otimes \overline{\mathrm{SU}(3)_1}$ Chern-Simons theories as described in Section~\ref{sec:anyons}, we make use of the $S$ and $T$ matrices found in Eqs.~\eqref{eq:sdz3}-\eqref{eq:tdsu3}. 
We then obtain the equations
\begin{align}
    \label{eq:sexponent}
    -q_1\phi_2 - q_2 \phi_1 &= \kappa_1 \kappa_2 -\bar{\kappa}_1\bar{\kappa}_2 \\
    \label{eq:texponent1}
    q_1 \phi_1 &= \kappa_1^2 - \bar{\kappa}_1^2 \\
    \label{eq:texponent2}
    q_2 \phi_2 &= \kappa_2^2 - \bar{\kappa}_2^2 . 
\end{align}
Note that these equations relate integers modulo 3.
If we consider the case where $q_1 \phi_1 = 1$, we therefore see that we must have $q_1 = \phi_1$. By Eq.~\eqref{eq:texponent1}, then, $\kappa_1^2 - \bar{\kappa}_1^2 = 1$, which means that $\kappa_1 \neq 0$, while $\bar{\kappa}_1 = 0$.
Then Eq.~\eqref{eq:sexponent} implies that we have 
\begin{equation}
    -q_1(\phi_2 + q_2) = \kappa_1 \kappa_2,
\end{equation}
or, since we know that $\kappa_1 \neq 0$,
\begin{equation}
   \kappa_2 =  -\frac{q_1}{\kappa_1}(\phi_2 + q_2).
\end{equation}
$\kappa_2$ must be dependent only on $\phi_2$ and $q_2$, so $\frac{q_1}{\kappa_1}$ is a constant, at least in this particular case where $q_1 \phi_1 = 1$.
In this case we also have $q_1 \neq 0$, so $\frac{q_1}{\kappa_1} = \pm 1$. Thus $\kappa_2 = \mp(\phi_2 + q_2)$, a relation which will apply in general. 
To find a similar relation for $\bar{\kappa}_2$, we instead consider the case where $q_1 \phi_1 = -1$, in which we must have $q_1 = -\phi_1$. 
Similar logic as above then shows that 
\begin{equation}
   \bar{\kappa}_2 =  -\frac{q_1}{\bar{\kappa}_1}(q_2-\phi_2),
\end{equation}
and so $\frac{q_1}{\bar{\kappa}_1} = \pm 1$, which in turn implies that $\bar{\kappa}_2 =  \mp(q_2-\phi_2)$, a relation that again applies in general.
We have therefore found that 
\begin{align}
    \alpha \kappa &= -(q+\phi) \\
    \beta \bar{\kappa} &=  -(q-\phi),
\end{align}
where $\alpha^2 = \beta^2 = 1$. 
Thus
\begin{align}
    q &= \alpha \kappa + \beta \bar{\kappa} \\
    \phi &= \alpha \kappa - \beta \bar{\kappa}.
\end{align}
This provides the general form of the map between the anyons of the two theories. The choice of $\alpha$ and $\beta$ amounts to the choice of sign of $\kappa$ (and $\bar{\kappa}$) for the two non-trivial sectors of the chiral $\mathrm{SU}(3)_1$ (and the anti-chiral $\overline{\mathrm{SU}(3)_1}$) Chern-Simons theory.
We will take $\alpha = \beta = 1$, and thus we obtain Eqs.~\eqref{eq:charge} and \eqref{eq:flux}.

\bibliographystyle{apsrev4-2}
\bibliography{su3_doubling_bibliography}

\begin{thebibliography}{53}%
\makeatletter
\providecommand \@ifxundefined [1]{%
 \@ifx{#1\undefined}
}%
\providecommand \@ifnum [1]{%
 \ifnum #1\expandafter \@firstoftwo
 \else \expandafter \@secondoftwo
 \fi
}%
\providecommand \@ifx [1]{%
 \ifx #1\expandafter \@firstoftwo
 \else \expandafter \@secondoftwo
 \fi
}%
\providecommand \natexlab [1]{#1}%
\providecommand \enquote  [1]{``#1''}%
\providecommand \bibnamefont  [1]{#1}%
\providecommand \bibfnamefont [1]{#1}%
\providecommand \citenamefont [1]{#1}%
\providecommand \href@noop [0]{\@secondoftwo}%
\providecommand \href [0]{\begingroup \@sanitize@url \@href}%
\providecommand \@href[1]{\@@startlink{#1}\@@href}%
\providecommand \@@href[1]{\endgroup#1\@@endlink}%
\providecommand \@sanitize@url [0]{\catcode `\\12\catcode `\$12\catcode
  `\&12\catcode `\#12\catcode `\^12\catcode `\_12\catcode `\%12\relax}%
\providecommand \@@startlink[1]{}%
\providecommand \@@endlink[0]{}%
\providecommand \url  [0]{\begingroup\@sanitize@url \@url }%
\providecommand \@url [1]{\endgroup\@href {#1}{\urlprefix }}%
\providecommand \urlprefix  [0]{URL }%
\providecommand \Eprint [0]{\href }%
\providecommand \doibase [0]{https://doi.org/}%
\providecommand \selectlanguage [0]{\@gobble}%
\providecommand \bibinfo  [0]{\@secondoftwo}%
\providecommand \bibfield  [0]{\@secondoftwo}%
\providecommand \translation [1]{[#1]}%
\providecommand \BibitemOpen [0]{}%
\providecommand \bibitemStop [0]{}%
\providecommand \bibitemNoStop [0]{.\EOS\space}%
\providecommand \EOS [0]{\spacefactor3000\relax}%
\providecommand \BibitemShut  [1]{\csname bibitem#1\endcsname}%
\let\auto@bib@innerbib\@empty
\bibitem [{Note1()}]{Note1}%
  \BibitemOpen
  \bibinfo {note} {See, e.g., Ref.~\cite {KitaevAnnalsPhysics2006}}\BibitemShut
  {NoStop}%
\bibitem [{\citenamefont {Li}\ and\ \citenamefont {Haldane}(2008)}]{Li2008}%
  \BibitemOpen
  \bibfield  {author} {\bibinfo {author} {\bibfnamefont {H.}~\bibnamefont
  {Li}}\ and\ \bibinfo {author} {\bibfnamefont {F.~D.~M.}\ \bibnamefont
  {Haldane}},\ }\href {https://doi.org/10.1103/PhysRevLett.101.010504}
  {\bibfield  {journal} {\bibinfo  {journal} {Phys. Rev. Lett.}\ }\textbf
  {\bibinfo {volume} {101}},\ \bibinfo {pages} {010504} (\bibinfo {year}
  {2008})},\ \Eprint {https://arxiv.org/abs/arXiv:0805.0332} {arXiv:0805.0332}
  \BibitemShut {NoStop}%
\bibitem [{\citenamefont {Kitaev}\ and\ \citenamefont
  {Preskill}(2006)}]{Kitaev2006}%
  \BibitemOpen
  \bibfield  {author} {\bibinfo {author} {\bibfnamefont {A.}~\bibnamefont
  {Kitaev}}\ and\ \bibinfo {author} {\bibfnamefont {J.}~\bibnamefont
  {Preskill}},\ }\href {https://doi.org/10.1103/PhysRevLett.96.110404}
  {\bibfield  {journal} {\bibinfo  {journal} {Phys. Rev. Lett.}\ }\textbf
  {\bibinfo {volume} {96}},\ \bibinfo {pages} {110404} (\bibinfo {year}
  {2006})},\ \Eprint {https://arxiv.org/abs/arXiv:hep-th/0510092}
  {arXiv:hep-th/0510092} \BibitemShut {NoStop}%
\bibitem [{\citenamefont {Verstraete}\ \emph {et~al.}(2006)\citenamefont
  {Verstraete}, \citenamefont {Wolf}, \citenamefont {Perez-Garcia},\ and\
  \citenamefont {Cirac}}]{Verstraete2006}%
  \BibitemOpen
  \bibfield  {author} {\bibinfo {author} {\bibfnamefont {F.}~\bibnamefont
  {Verstraete}}, \bibinfo {author} {\bibfnamefont {M.~M.}\ \bibnamefont
  {Wolf}}, \bibinfo {author} {\bibfnamefont {D.}~\bibnamefont {Perez-Garcia}},\
  and\ \bibinfo {author} {\bibfnamefont {J.~I.}\ \bibnamefont {Cirac}},\ }\href
  {https://doi.org/10.1103/PhysRevLett.96.220601} {\bibfield  {journal}
  {\bibinfo  {journal} {Phys. Rev. Lett.}\ }\textbf {\bibinfo {volume} {96}},\
  \bibinfo {pages} {220601} (\bibinfo {year} {2006})},\ \Eprint
  {https://arxiv.org/abs/arXiv:quant-ph/0601075} {arXiv:quant-ph/0601075}
  \BibitemShut {NoStop}%
\bibitem [{\citenamefont {{Schuch}}\ \emph {et~al.}(2010)\citenamefont
  {{Schuch}}, \citenamefont {{Cirac}},\ and\ \citenamefont
  {{P{\'e}rez-Garc{\'{\i}}a}}}]{schuch:peps-sym}%
  \BibitemOpen
  \bibfield  {author} {\bibinfo {author} {\bibfnamefont {N.}~\bibnamefont
  {{Schuch}}}, \bibinfo {author} {\bibfnamefont {I.}~\bibnamefont {{Cirac}}},\
  and\ \bibinfo {author} {\bibfnamefont {D.}~\bibnamefont
  {{P{\'e}rez-Garc{\'{\i}}a}}},\ }\href
  {https://doi.org/10.1016/j.aop.2010.05.008} {\bibfield  {journal} {\bibinfo
  {journal} {Ann. Phys. (NY)}\ }\textbf {\bibinfo {volume} {325}},\ \bibinfo
  {pages} {2153} (\bibinfo {year} {2010})},\ \Eprint
  {https://arxiv.org/abs/arXiv:1001.3807} {arXiv:1001.3807} \BibitemShut
  {NoStop}%
\bibitem [{\citenamefont {Dubail}\ and\ \citenamefont
  {Read}(2015)}]{Dubail2015}%
  \BibitemOpen
  \bibfield  {author} {\bibinfo {author} {\bibfnamefont {J.}~\bibnamefont
  {Dubail}}\ and\ \bibinfo {author} {\bibfnamefont {N.}~\bibnamefont {Read}},\
  }\href {https://doi.org/10.1103/PhysRevB.92.205307} {\bibfield  {journal}
  {\bibinfo  {journal} {Phys. Rev. B}\ }\textbf {\bibinfo {volume} {92}},\
  \bibinfo {pages} {205307} (\bibinfo {year} {2015})},\ \Eprint
  {https://arxiv.org/abs/arXiv:1307.7726} {arXiv:1307.7726} \BibitemShut
  {NoStop}%
\bibitem [{\citenamefont {Wahl}\ \emph {et~al.}(2013)\citenamefont {Wahl},
  \citenamefont {Tu}, \citenamefont {Schuch},\ and\ \citenamefont
  {Cirac}}]{Wahl2013}%
  \BibitemOpen
  \bibfield  {author} {\bibinfo {author} {\bibfnamefont {T.~B.}\ \bibnamefont
  {Wahl}}, \bibinfo {author} {\bibfnamefont {H.-H.}\ \bibnamefont {Tu}},
  \bibinfo {author} {\bibfnamefont {N.}~\bibnamefont {Schuch}},\ and\ \bibinfo
  {author} {\bibfnamefont {J.~I.}\ \bibnamefont {Cirac}},\ }\href
  {https://doi.org/10.1103/PhysRevLett.111.236805} {\bibfield  {journal}
  {\bibinfo  {journal} {Phys. Rev. Lett.}\ }\textbf {\bibinfo {volume} {111}},\
  \bibinfo {pages} {236805} (\bibinfo {year} {2013})},\ \Eprint
  {https://arxiv.org/abs/arXiv:1308.0316} {arXiv:1308.0316} \BibitemShut
  {NoStop}%
\bibitem [{\citenamefont {Gorshkov}\ \emph {et~al.}(2010)\citenamefont
  {Gorshkov}, \citenamefont {Hermele}, \citenamefont {Gurarie}, \citenamefont
  {Xu}, \citenamefont {Julienne}, \citenamefont {Ye}, \citenamefont {Zoller},
  \citenamefont {Demler}, \citenamefont {Lukin},\ and\ \citenamefont
  {Rey}}]{Gorshkov2010}%
  \BibitemOpen
  \bibfield  {author} {\bibinfo {author} {\bibfnamefont {A.~V.}\ \bibnamefont
  {Gorshkov}}, \bibinfo {author} {\bibfnamefont {M.}~\bibnamefont {Hermele}},
  \bibinfo {author} {\bibfnamefont {V.}~\bibnamefont {Gurarie}}, \bibinfo
  {author} {\bibfnamefont {C.}~\bibnamefont {Xu}}, \bibinfo {author}
  {\bibfnamefont {P.~S.}\ \bibnamefont {Julienne}}, \bibinfo {author}
  {\bibfnamefont {J.}~\bibnamefont {Ye}}, \bibinfo {author} {\bibfnamefont
  {P.}~\bibnamefont {Zoller}}, \bibinfo {author} {\bibfnamefont
  {E.}~\bibnamefont {Demler}}, \bibinfo {author} {\bibfnamefont {M.~D.}\
  \bibnamefont {Lukin}},\ and\ \bibinfo {author} {\bibfnamefont {A.~M.}\
  \bibnamefont {Rey}},\ }\href {https://doi.org/10.1038/nphys1535} {\bibfield
  {journal} {\bibinfo  {journal} {Nat. Phys.}\ }\textbf {\bibinfo {volume}
  {6}},\ \bibinfo {pages} {289} (\bibinfo {year} {2010})},\ \Eprint
  {https://arxiv.org/abs/arXiv:0905.2610} {arXiv:0905.2610} \BibitemShut
  {NoStop}%
\bibitem [{\citenamefont {Nataf}\ \emph {et~al.}(2016)\citenamefont {Nataf},
  \citenamefont {Lajk\'o}, \citenamefont {Wietek}, \citenamefont {Penc},
  \citenamefont {Mila},\ and\ \citenamefont {L\"auchli}}]{Nataf2016}%
  \BibitemOpen
  \bibfield  {author} {\bibinfo {author} {\bibfnamefont {P.}~\bibnamefont
  {Nataf}}, \bibinfo {author} {\bibfnamefont {M.}~\bibnamefont {Lajk\'o}},
  \bibinfo {author} {\bibfnamefont {A.}~\bibnamefont {Wietek}}, \bibinfo
  {author} {\bibfnamefont {K.}~\bibnamefont {Penc}}, \bibinfo {author}
  {\bibfnamefont {F.}~\bibnamefont {Mila}},\ and\ \bibinfo {author}
  {\bibfnamefont {A.~M.}\ \bibnamefont {L\"auchli}},\ }\href
  {https://doi.org/10.1103/PhysRevLett.117.167202} {\bibfield  {journal}
  {\bibinfo  {journal} {Phys. Rev. Lett.}\ }\textbf {\bibinfo {volume} {117}},\
  \bibinfo {pages} {167202} (\bibinfo {year} {2016})},\ \Eprint
  {https://arxiv.org/abs/arXiv:1601.00958} {arXiv:1601.00958} \BibitemShut
  {NoStop}%
\bibitem [{\citenamefont {{Cazalilla}}\ and\ \citenamefont
  {{Rey}}(2014)}]{Cazalilla2014}%
  \BibitemOpen
  \bibfield  {author} {\bibinfo {author} {\bibfnamefont {M.~A.}\ \bibnamefont
  {{Cazalilla}}}\ and\ \bibinfo {author} {\bibfnamefont {A.~M.}\ \bibnamefont
  {{Rey}}},\ }\href {https://doi.org/10.1088/0034-4885/77/12/124401} {\bibfield
   {journal} {\bibinfo  {journal} {Rep. Prog. Phys.}\ }\textbf {\bibinfo
  {volume} {77}},\ \bibinfo {eid} {124401} (\bibinfo {year} {2014})},\ \Eprint
  {https://arxiv.org/abs/1403.2792} {arXiv:1403.2792} \BibitemShut {NoStop}%
\bibitem [{\citenamefont {Pagano}\ \emph {et~al.}(2014)\citenamefont {Pagano},
  \citenamefont {Mancini}, \citenamefont {Cappellini}, \citenamefont
  {Lombardi}, \citenamefont {Sch{\"a}fer}, \citenamefont {Hu}, \citenamefont
  {Liu}, \citenamefont {Catani}, \citenamefont {Sias}, \citenamefont
  {Inguscio},\ and\ \citenamefont {Fallani}}]{Pagano2014}%
  \BibitemOpen
  \bibfield  {author} {\bibinfo {author} {\bibfnamefont {G.}~\bibnamefont
  {Pagano}}, \bibinfo {author} {\bibfnamefont {M.}~\bibnamefont {Mancini}},
  \bibinfo {author} {\bibfnamefont {G.}~\bibnamefont {Cappellini}}, \bibinfo
  {author} {\bibfnamefont {P.}~\bibnamefont {Lombardi}}, \bibinfo {author}
  {\bibfnamefont {F.}~\bibnamefont {Sch{\"a}fer}}, \bibinfo {author}
  {\bibfnamefont {H.}~\bibnamefont {Hu}}, \bibinfo {author} {\bibfnamefont
  {X.-J.}\ \bibnamefont {Liu}}, \bibinfo {author} {\bibfnamefont
  {J.}~\bibnamefont {Catani}}, \bibinfo {author} {\bibfnamefont
  {C.}~\bibnamefont {Sias}}, \bibinfo {author} {\bibfnamefont {M.}~\bibnamefont
  {Inguscio}},\ and\ \bibinfo {author} {\bibfnamefont {L.}~\bibnamefont
  {Fallani}},\ }\href {https://doi.org/10.1038/nphys2878} {\bibfield  {journal}
  {\bibinfo  {journal} {Nat. Phys.}\ }\textbf {\bibinfo {volume} {10}},\
  \bibinfo {pages} {198} (\bibinfo {year} {2014})},\ \Eprint
  {https://arxiv.org/abs/arXiv:1408.0928} {arXiv:1408.0928} \BibitemShut
  {NoStop}%
\bibitem [{\citenamefont {Taie}\ \emph {et~al.}(2012)\citenamefont {Taie},
  \citenamefont {Yamazaki}, \citenamefont {Sugawa},\ and\ \citenamefont
  {Takahashi}}]{Taie2012}%
  \BibitemOpen
  \bibfield  {author} {\bibinfo {author} {\bibfnamefont {S.}~\bibnamefont
  {Taie}}, \bibinfo {author} {\bibfnamefont {R.}~\bibnamefont {Yamazaki}},
  \bibinfo {author} {\bibfnamefont {S.}~\bibnamefont {Sugawa}},\ and\ \bibinfo
  {author} {\bibfnamefont {Y.}~\bibnamefont {Takahashi}},\ }\href
  {https://doi.org/10.1038/nphys2430} {\bibfield  {journal} {\bibinfo
  {journal} {Nat. Phys.}\ }\textbf {\bibinfo {volume} {8}},\ \bibinfo {pages}
  {825} (\bibinfo {year} {2012})},\ \Eprint
  {https://arxiv.org/abs/arXiv:1208.4883} {arXiv:1208.4883} \BibitemShut
  {NoStop}%
\bibitem [{\citenamefont {Hofrichter}\ \emph {et~al.}(2016)\citenamefont
  {Hofrichter}, \citenamefont {Riegger}, \citenamefont {Scazza}, \citenamefont
  {H\"ofer}, \citenamefont {Fernandes}, \citenamefont {Bloch},\ and\
  \citenamefont {F\"olling}}]{Hofrichter2016}%
  \BibitemOpen
  \bibfield  {author} {\bibinfo {author} {\bibfnamefont {C.}~\bibnamefont
  {Hofrichter}}, \bibinfo {author} {\bibfnamefont {L.}~\bibnamefont {Riegger}},
  \bibinfo {author} {\bibfnamefont {F.}~\bibnamefont {Scazza}}, \bibinfo
  {author} {\bibfnamefont {M.}~\bibnamefont {H\"ofer}}, \bibinfo {author}
  {\bibfnamefont {D.~R.}\ \bibnamefont {Fernandes}}, \bibinfo {author}
  {\bibfnamefont {I.}~\bibnamefont {Bloch}},\ and\ \bibinfo {author}
  {\bibfnamefont {S.}~\bibnamefont {F\"olling}},\ }\href
  {https://doi.org/10.1103/PhysRevX.6.021030} {\bibfield  {journal} {\bibinfo
  {journal} {Phys. Rev. X}\ }\textbf {\bibinfo {volume} {6}},\ \bibinfo {pages}
  {021030} (\bibinfo {year} {2016})},\ \Eprint
  {https://arxiv.org/abs/arXiv:1511.07287} {arXiv:1511.07287} \BibitemShut
  {NoStop}%
\bibitem [{\citenamefont {Taie}\ \emph {et~al.}(2022)\citenamefont {Taie},
  \citenamefont {Ibarra-Garc{\'\i}a-Padilla}, \citenamefont {Nishizawa},
  \citenamefont {Takasu}, \citenamefont {Kuno}, \citenamefont {Wei},
  \citenamefont {Scalettar}, \citenamefont {Hazzard},\ and\ \citenamefont
  {Takahashi}}]{Taie2022}%
  \BibitemOpen
  \bibfield  {author} {\bibinfo {author} {\bibfnamefont {S.}~\bibnamefont
  {Taie}}, \bibinfo {author} {\bibfnamefont {E.}~\bibnamefont
  {Ibarra-Garc{\'\i}a-Padilla}}, \bibinfo {author} {\bibfnamefont
  {N.}~\bibnamefont {Nishizawa}}, \bibinfo {author} {\bibfnamefont
  {Y.}~\bibnamefont {Takasu}}, \bibinfo {author} {\bibfnamefont
  {Y.}~\bibnamefont {Kuno}}, \bibinfo {author} {\bibfnamefont {H.-T.}\
  \bibnamefont {Wei}}, \bibinfo {author} {\bibfnamefont {R.~T.}\ \bibnamefont
  {Scalettar}}, \bibinfo {author} {\bibfnamefont {K.~R.~A.}\ \bibnamefont
  {Hazzard}},\ and\ \bibinfo {author} {\bibfnamefont {Y.}~\bibnamefont
  {Takahashi}},\ }\href {https://doi.org/10.1038/s41567-022-01725-6} {\bibfield
   {journal} {\bibinfo  {journal} {Nat. Phys.}\ }\textbf {\bibinfo {volume}
  {18}},\ \bibinfo {pages} {1356} (\bibinfo {year} {2022})},\ \Eprint
  {https://arxiv.org/abs/arXiv:2010.07730} {arXiv:2010.07730} \BibitemShut
  {NoStop}%
\bibitem [{\citenamefont {Zhang}\ \emph {et~al.}(2014)\citenamefont {Zhang},
  \citenamefont {Bishof}, \citenamefont {Bromley}, \citenamefont {Kraus},
  \citenamefont {Safronova}, \citenamefont {Zoller}, \citenamefont {Rey},\ and\
  \citenamefont {Ye}}]{Zhang2014}%
  \BibitemOpen
  \bibfield  {author} {\bibinfo {author} {\bibfnamefont {X.}~\bibnamefont
  {Zhang}}, \bibinfo {author} {\bibfnamefont {M.}~\bibnamefont {Bishof}},
  \bibinfo {author} {\bibfnamefont {S.~L.}\ \bibnamefont {Bromley}}, \bibinfo
  {author} {\bibfnamefont {C.~V.}\ \bibnamefont {Kraus}}, \bibinfo {author}
  {\bibfnamefont {M.~S.}\ \bibnamefont {Safronova}}, \bibinfo {author}
  {\bibfnamefont {P.}~\bibnamefont {Zoller}}, \bibinfo {author} {\bibfnamefont
  {A.~M.}\ \bibnamefont {Rey}},\ and\ \bibinfo {author} {\bibfnamefont
  {J.}~\bibnamefont {Ye}},\ }\href {https://doi.org/10.1126/science.1254978}
  {\bibfield  {journal} {\bibinfo  {journal} {Science}\ }\textbf {\bibinfo
  {volume} {345}},\ \bibinfo {pages} {1467} (\bibinfo {year} {2014})},\ \Eprint
  {https://arxiv.org/abs/arXiv:1403.2964} {arXiv:1403.2964} \BibitemShut
  {NoStop}%
\bibitem [{\citenamefont {Scazza}\ \emph {et~al.}(2014)\citenamefont {Scazza},
  \citenamefont {Hofrichter}, \citenamefont {H{\"o}fer}, \citenamefont
  {De~Groot}, \citenamefont {Bloch},\ and\ \citenamefont
  {F{\"o}lling}}]{Scazza2014}%
  \BibitemOpen
  \bibfield  {author} {\bibinfo {author} {\bibfnamefont {F.}~\bibnamefont
  {Scazza}}, \bibinfo {author} {\bibfnamefont {C.}~\bibnamefont {Hofrichter}},
  \bibinfo {author} {\bibfnamefont {M.}~\bibnamefont {H{\"o}fer}}, \bibinfo
  {author} {\bibfnamefont {P.~C.}\ \bibnamefont {De~Groot}}, \bibinfo {author}
  {\bibfnamefont {I.}~\bibnamefont {Bloch}},\ and\ \bibinfo {author}
  {\bibfnamefont {S.}~\bibnamefont {F{\"o}lling}},\ }\href
  {https://doi.org/10.1038/nphys3061} {\bibfield  {journal} {\bibinfo
  {journal} {Nat. Phys.}\ }\textbf {\bibinfo {volume} {10}},\ \bibinfo {pages}
  {779} (\bibinfo {year} {2014})},\ \Eprint
  {https://arxiv.org/abs/arXiv:1403.4761} {arXiv:1403.4761} \BibitemShut
  {NoStop}%
\bibitem [{\citenamefont {Cappellini}\ \emph {et~al.}(2014)\citenamefont
  {Cappellini}, \citenamefont {Mancini}, \citenamefont {Pagano}, \citenamefont
  {Lombardi}, \citenamefont {Livi}, \citenamefont {Siciliani~de Cumis},
  \citenamefont {Cancio}, \citenamefont {Pizzocaro}, \citenamefont {Calonico},
  \citenamefont {Levi}, \citenamefont {Sias}, \citenamefont {Catani},
  \citenamefont {Inguscio},\ and\ \citenamefont {Fallani}}]{Cappellini2014}%
  \BibitemOpen
  \bibfield  {author} {\bibinfo {author} {\bibfnamefont {G.}~\bibnamefont
  {Cappellini}}, \bibinfo {author} {\bibfnamefont {M.}~\bibnamefont {Mancini}},
  \bibinfo {author} {\bibfnamefont {G.}~\bibnamefont {Pagano}}, \bibinfo
  {author} {\bibfnamefont {P.}~\bibnamefont {Lombardi}}, \bibinfo {author}
  {\bibfnamefont {L.}~\bibnamefont {Livi}}, \bibinfo {author} {\bibfnamefont
  {M.}~\bibnamefont {Siciliani~de Cumis}}, \bibinfo {author} {\bibfnamefont
  {P.}~\bibnamefont {Cancio}}, \bibinfo {author} {\bibfnamefont
  {M.}~\bibnamefont {Pizzocaro}}, \bibinfo {author} {\bibfnamefont
  {D.}~\bibnamefont {Calonico}}, \bibinfo {author} {\bibfnamefont
  {F.}~\bibnamefont {Levi}}, \bibinfo {author} {\bibfnamefont {C.}~\bibnamefont
  {Sias}}, \bibinfo {author} {\bibfnamefont {J.}~\bibnamefont {Catani}},
  \bibinfo {author} {\bibfnamefont {M.}~\bibnamefont {Inguscio}},\ and\
  \bibinfo {author} {\bibfnamefont {L.}~\bibnamefont {Fallani}},\ }\href
  {https://doi.org/10.1103/PhysRevLett.113.120402} {\bibfield  {journal}
  {\bibinfo  {journal} {Phys. Rev. Lett.}\ }\textbf {\bibinfo {volume} {113}},\
  \bibinfo {pages} {120402} (\bibinfo {year} {2014})},\ \Eprint
  {https://arxiv.org/abs/arXiv:1406.6642} {arXiv:1406.6642} \BibitemShut
  {NoStop}%
\bibitem [{\citenamefont {Ozawa}\ \emph {et~al.}(2018)\citenamefont {Ozawa},
  \citenamefont {Taie}, \citenamefont {Takasu},\ and\ \citenamefont
  {Takahashi}}]{Ozawa2018}%
  \BibitemOpen
  \bibfield  {author} {\bibinfo {author} {\bibfnamefont {H.}~\bibnamefont
  {Ozawa}}, \bibinfo {author} {\bibfnamefont {S.}~\bibnamefont {Taie}},
  \bibinfo {author} {\bibfnamefont {Y.}~\bibnamefont {Takasu}},\ and\ \bibinfo
  {author} {\bibfnamefont {Y.}~\bibnamefont {Takahashi}},\ }\href
  {https://doi.org/10.1103/PhysRevLett.121.225303} {\bibfield  {journal}
  {\bibinfo  {journal} {Phys. Rev. Lett.}\ }\textbf {\bibinfo {volume} {121}},\
  \bibinfo {pages} {225303} (\bibinfo {year} {2018})},\ \Eprint
  {https://arxiv.org/abs/arXiv:1801.05962} {arXiv:1801.05962} \BibitemShut
  {NoStop}%
\bibitem [{\citenamefont {Zhang}\ \emph {et~al.}(2021)\citenamefont {Zhang},
  \citenamefont {Sheng},\ and\ \citenamefont {Vishwanath}}]{Zhang2021}%
  \BibitemOpen
  \bibfield  {author} {\bibinfo {author} {\bibfnamefont {Y.-H.}\ \bibnamefont
  {Zhang}}, \bibinfo {author} {\bibfnamefont {D.~N.}\ \bibnamefont {Sheng}},\
  and\ \bibinfo {author} {\bibfnamefont {A.}~\bibnamefont {Vishwanath}},\
  }\href {https://doi.org/10.1103/PhysRevLett.127.247701} {\bibfield  {journal}
  {\bibinfo  {journal} {Phys. Rev. Lett.}\ }\textbf {\bibinfo {volume} {127}},\
  \bibinfo {pages} {247701} (\bibinfo {year} {2021})},\ \Eprint
  {https://arxiv.org/abs/arXiv:2103.09825} {arXiv:2103.09825} \BibitemShut
  {NoStop}%
\bibitem [{\citenamefont {Kure\ifmmode \check{c}\else
  \v{c}\fi{}i\ifmmode~\acute{c}\else \'{c}\fi{}}\ \emph
  {et~al.}(2019)\citenamefont {Kure\ifmmode \check{c}\else
  \v{c}\fi{}i\ifmmode~\acute{c}\else \'{c}\fi{}}, \citenamefont
  {Vanderstraeten},\ and\ \citenamefont {Schuch}}]{kurecic:su3_sl}%
  \BibitemOpen
  \bibfield  {author} {\bibinfo {author} {\bibfnamefont {I.}~\bibnamefont
  {Kure\ifmmode \check{c}\else \v{c}\fi{}i\ifmmode~\acute{c}\else \'{c}\fi{}}},
  \bibinfo {author} {\bibfnamefont {L.}~\bibnamefont {Vanderstraeten}},\ and\
  \bibinfo {author} {\bibfnamefont {N.}~\bibnamefont {Schuch}},\ }\href
  {https://doi.org/10.1103/PhysRevB.99.045116} {\bibfield  {journal} {\bibinfo
  {journal} {Phys. Rev. B}\ }\textbf {\bibinfo {volume} {99}},\ \bibinfo
  {pages} {045116} (\bibinfo {year} {2019})},\ \Eprint
  {https://arxiv.org/abs/arXiv:1805.11628} {arXiv:1805.11628} \BibitemShut
  {NoStop}%
\bibitem [{Note2()}]{Note2}%
  \BibitemOpen
  \bibinfo {note} {In the case under consideration, the entanglement spectrum,
  just as a physical boundary of our topological state, will eventually be
  gapped in the limit of large system size (cylinder circumference). However,
  we will investigate the entanglement spectrum at system sizes much smaller
  than the inverse entanglement gap (proportional to the correlation length).
  In this limit the entanglement spectrum (as well as the spectrum of a
  physical boundary) reflects the spectrum of the underlying gapless CFT, and
  thus the entanglement spectrum serves as a direct diagnostic of the
  underlying Topological Field Theory.}\BibitemShut {Stop}%
\bibitem [{\citenamefont {Qi}\ \emph {et~al.}(2012)\citenamefont {Qi},
  \citenamefont {Katsura},\ and\ \citenamefont {Ludwig}}]{Qi2012}%
  \BibitemOpen
  \bibfield  {author} {\bibinfo {author} {\bibfnamefont {X.-L.}\ \bibnamefont
  {Qi}}, \bibinfo {author} {\bibfnamefont {H.}~\bibnamefont {Katsura}},\ and\
  \bibinfo {author} {\bibfnamefont {A.~W.~W.}\ \bibnamefont {Ludwig}},\ }\href
  {https://doi.org/10.1103/PhysRevLett.108.196402} {\bibfield  {journal}
  {\bibinfo  {journal} {Phys. Rev. Lett.}\ }\textbf {\bibinfo {volume} {108}},\
  \bibinfo {pages} {196402} (\bibinfo {year} {2012})},\ \Eprint
  {https://arxiv.org/abs/arXiv:1103.5437} {arXiv:1103.5437} \BibitemShut
  {NoStop}%
\bibitem [{\citenamefont {Arildsen}\ and\ \citenamefont
  {Ludwig}(2022)}]{ArildsenLudwig2022}%
  \BibitemOpen
  \bibfield  {author} {\bibinfo {author} {\bibfnamefont {M.~J.}\ \bibnamefont
  {Arildsen}}\ and\ \bibinfo {author} {\bibfnamefont {A.~W.~W.}\ \bibnamefont
  {Ludwig}},\ }\href {https://doi.org/10.1103/PhysRevB.106.035138} {\bibfield
  {journal} {\bibinfo  {journal} {Phys. Rev. B}\ }\textbf {\bibinfo {volume}
  {106}},\ \bibinfo {pages} {035138} (\bibinfo {year} {2022})},\ \Eprint
  {https://arxiv.org/abs/arXiv:2107.02545} {arXiv:2107.02545} \BibitemShut
  {NoStop}%
\bibitem [{\citenamefont {Kitaev}(2003)}]{Kitaev2003}%
  \BibitemOpen
  \bibfield  {author} {\bibinfo {author} {\bibfnamefont {A.}~\bibnamefont
  {Kitaev}},\ }\href
  {https://doi.org/https://doi.org/10.1016/S0003-4916(02)00018-0} {\bibfield
  {journal} {\bibinfo  {journal} {Ann. Phys. (NY)}\ }\textbf {\bibinfo {volume}
  {303}},\ \bibinfo {pages} {2} (\bibinfo {year} {2003})},\ \Eprint
  {https://arxiv.org/abs/arXiv:quant-ph/9707021} {arXiv:quant-ph/9707021}
  \BibitemShut {NoStop}%
\bibitem [{\citenamefont {Bakalov}\ and\ \citenamefont
  {Kirillov}(2001)}]{Bakalov2001}%
  \BibitemOpen
  \bibfield  {author} {\bibinfo {author} {\bibfnamefont {B.}~\bibnamefont
  {Bakalov}}\ and\ \bibinfo {author} {\bibfnamefont {A.}~\bibnamefont
  {Kirillov}},\ }\href@noop {} {\emph {\bibinfo {title} {Lectures on tensor
  categories and modular functors}}},\ Vol.~\bibinfo {volume} {21}\ (\bibinfo
  {publisher} {American Mathematical Society},\ \bibinfo {address} {Providence,
  RI},\ \bibinfo {year} {2001})\BibitemShut {NoStop}%
\bibitem [{Note3()}]{Note3}%
  \BibitemOpen
  \bibinfo {note} {Its topological order can be related, e.g., to the
  topological order of the stacking of a fractional quantum Hall state at
  filling fraction $\nu =1/3$ and its time-reversed partner.}\BibitemShut
  {Stop}%
\bibitem [{\citenamefont {Nayak}\ \emph {et~al.}(2008)\citenamefont {Nayak},
  \citenamefont {Simon}, \citenamefont {Stern}, \citenamefont {Freedman},\ and\
  \citenamefont {Das~Sarma}}]{Nayak2008}%
  \BibitemOpen
  \bibfield  {author} {\bibinfo {author} {\bibfnamefont {C.}~\bibnamefont
  {Nayak}}, \bibinfo {author} {\bibfnamefont {S.~H.}\ \bibnamefont {Simon}},
  \bibinfo {author} {\bibfnamefont {A.}~\bibnamefont {Stern}}, \bibinfo
  {author} {\bibfnamefont {M.}~\bibnamefont {Freedman}},\ and\ \bibinfo
  {author} {\bibfnamefont {S.}~\bibnamefont {Das~Sarma}},\ }\href
  {https://doi.org/10.1103/RevModPhys.80.1083} {\bibfield  {journal} {\bibinfo
  {journal} {Rev. Mod. Phys.}\ }\textbf {\bibinfo {volume} {80}},\ \bibinfo
  {pages} {1083} (\bibinfo {year} {2008})},\ \Eprint
  {https://arxiv.org/abs/arXiv:0707.1889} {arXiv:0707.1889} \BibitemShut
  {NoStop}%
\bibitem [{\citenamefont {Zhang}\ \emph {et~al.}(2012)\citenamefont {Zhang},
  \citenamefont {Grover}, \citenamefont {Turner}, \citenamefont {Oshikawa},\
  and\ \citenamefont {Vishwanath}}]{Zhang2012}%
  \BibitemOpen
  \bibfield  {author} {\bibinfo {author} {\bibfnamefont {Y.}~\bibnamefont
  {Zhang}}, \bibinfo {author} {\bibfnamefont {T.}~\bibnamefont {Grover}},
  \bibinfo {author} {\bibfnamefont {A.}~\bibnamefont {Turner}}, \bibinfo
  {author} {\bibfnamefont {M.}~\bibnamefont {Oshikawa}},\ and\ \bibinfo
  {author} {\bibfnamefont {A.}~\bibnamefont {Vishwanath}},\ }\href
  {https://doi.org/10.1103/PhysRevB.85.235151} {\bibfield  {journal} {\bibinfo
  {journal} {Phys. Rev. B}\ }\textbf {\bibinfo {volume} {85}},\ \bibinfo
  {pages} {235151} (\bibinfo {year} {2012})},\ \Eprint
  {https://arxiv.org/abs/arXiv:1111.2342} {arXiv:1111.2342} \BibitemShut
  {NoStop}%
\bibitem [{\citenamefont {Poilblanc}\ \emph {et~al.}(2015)\citenamefont
  {Poilblanc}, \citenamefont {Cirac},\ and\ \citenamefont
  {Schuch}}]{poilblanc:kl-peps-1}%
  \BibitemOpen
  \bibfield  {author} {\bibinfo {author} {\bibfnamefont {D.}~\bibnamefont
  {Poilblanc}}, \bibinfo {author} {\bibfnamefont {J.~I.}\ \bibnamefont
  {Cirac}},\ and\ \bibinfo {author} {\bibfnamefont {N.}~\bibnamefont
  {Schuch}},\ }\href {https://doi.org/10.1103/PhysRevB.91.224431} {\bibfield
  {journal} {\bibinfo  {journal} {Phys. Rev. B}\ }\textbf {\bibinfo {volume}
  {91}},\ \bibinfo {pages} {224431} (\bibinfo {year} {2015})},\ \Eprint
  {https://arxiv.org/abs/arXiv:1504.05236} {arXiv:1504.05236} \BibitemShut
  {NoStop}%
\bibitem [{\citenamefont {Poilblanc}\ \emph {et~al.}(2016)\citenamefont
  {Poilblanc}, \citenamefont {Schuch},\ and\ \citenamefont
  {Affleck}}]{Poilblanc2016}%
  \BibitemOpen
  \bibfield  {author} {\bibinfo {author} {\bibfnamefont {D.}~\bibnamefont
  {Poilblanc}}, \bibinfo {author} {\bibfnamefont {N.}~\bibnamefont {Schuch}},\
  and\ \bibinfo {author} {\bibfnamefont {I.}~\bibnamefont {Affleck}},\ }\href
  {https://doi.org/10.1103/PhysRevB.93.174414} {\bibfield  {journal} {\bibinfo
  {journal} {Phys. Rev. B}\ }\textbf {\bibinfo {volume} {93}},\ \bibinfo
  {pages} {174414} (\bibinfo {year} {2016})},\ \Eprint
  {https://arxiv.org/abs/arXiv:1602.05969} {arXiv:1602.05969} \BibitemShut
  {NoStop}%
\bibitem [{Note4()}]{Note4}%
  \BibitemOpen
  \bibinfo {note} {Note that the mapping does not preserve lattice angles.
  However, this is only relevant if we want to consider a torus, or if we want
  to consider momenta with a component along the cylinder axis. For our
  scenario, where we consider the system on a cylinder and only care about
  momenta around the torus, the shearing of the unit cell has no
  consequences.}\BibitemShut {Stop}%
\bibitem [{\citenamefont {Zaletel}\ and\ \citenamefont
  {Vishwanath}(2015)}]{zaletel:su2-semion}%
  \BibitemOpen
  \bibfield  {author} {\bibinfo {author} {\bibfnamefont {M.~P.}\ \bibnamefont
  {Zaletel}}\ and\ \bibinfo {author} {\bibfnamefont {A.}~\bibnamefont
  {Vishwanath}},\ }\href {https://doi.org/10.1103/PhysRevLett.114.077201}
  {\bibfield  {journal} {\bibinfo  {journal} {Phys.\ Rev.\ Lett.}\ }\textbf
  {\bibinfo {volume} {114}},\ \bibinfo {pages} {077201} (\bibinfo {year}
  {2015})},\ \Eprint {https://arxiv.org/abs/arXiv:1410.2894} {arXiv:1410.2894}
  \BibitemShut {NoStop}%
\bibitem [{\citenamefont {{{\c{S}}ahino{\u{g}}lu}}\ \emph
  {et~al.}(2021)\citenamefont {{{\c{S}}ahino{\u{g}}lu}}, \citenamefont
  {{Williamson}}, \citenamefont {{Bultinck}}, \citenamefont {{Mari{\"e}n}},
  \citenamefont {{Haegeman}}, \citenamefont {{Schuch}},\ and\ \citenamefont
  {{Verstraete}}}]{sahinoglu:mpo-injectivity}%
  \BibitemOpen
  \bibfield  {author} {\bibinfo {author} {\bibfnamefont {M.~B.}\ \bibnamefont
  {{{\c{S}}ahino{\u{g}}lu}}}, \bibinfo {author} {\bibfnamefont
  {D.}~\bibnamefont {{Williamson}}}, \bibinfo {author} {\bibfnamefont
  {N.}~\bibnamefont {{Bultinck}}}, \bibinfo {author} {\bibfnamefont
  {M.}~\bibnamefont {{Mari{\"e}n}}}, \bibinfo {author} {\bibfnamefont
  {J.}~\bibnamefont {{Haegeman}}}, \bibinfo {author} {\bibfnamefont
  {N.}~\bibnamefont {{Schuch}}},\ and\ \bibinfo {author} {\bibfnamefont
  {F.}~\bibnamefont {{Verstraete}}},\ }\href
  {https://doi.org/10.1007/s00023-020-00992-4} {\bibfield  {journal} {\bibinfo
  {journal} {Ann. Henri Poincar\'e}\ }\textbf {\bibinfo {volume} {22}},\
  \bibinfo {pages} {563} (\bibinfo {year} {2021})},\ \Eprint
  {https://arxiv.org/abs/arXiv:1409.2150} {arXiv:1409.2150} \BibitemShut
  {NoStop}%
\bibitem [{\citenamefont {{Bultinck}}\ \emph {et~al.}(2017)\citenamefont
  {{Bultinck}}, \citenamefont {{Mari{\"e}n}}, \citenamefont {{Williamson}},
  \citenamefont {{{\c S}ahino{\u g}lu}}, \citenamefont {{Haegeman}},\ and\
  \citenamefont {{Verstraete}}}]{bultinck:mpo-anyons}%
  \BibitemOpen
  \bibfield  {author} {\bibinfo {author} {\bibfnamefont {N.}~\bibnamefont
  {{Bultinck}}}, \bibinfo {author} {\bibfnamefont {M.}~\bibnamefont
  {{Mari{\"e}n}}}, \bibinfo {author} {\bibfnamefont {D.~J.}\ \bibnamefont
  {{Williamson}}}, \bibinfo {author} {\bibfnamefont {M.~B.}\ \bibnamefont {{{\c
  S}ahino{\u g}lu}}}, \bibinfo {author} {\bibfnamefont {J.}~\bibnamefont
  {{Haegeman}}},\ and\ \bibinfo {author} {\bibfnamefont {F.}~\bibnamefont
  {{Verstraete}}},\ }\href {https://doi.org/10.1016/j.aop.2017.01.004}
  {\bibfield  {journal} {\bibinfo  {journal} {Ann. Phys. (NY)}\ }\textbf
  {\bibinfo {volume} {378}},\ \bibinfo {pages} {183} (\bibinfo {year}
  {2017})},\ \Eprint {https://arxiv.org/abs/arXiv:1511.08090}
  {arXiv:1511.08090} \BibitemShut {NoStop}%
\bibitem [{\citenamefont {Schuch}\ \emph {et~al.}(2013)\citenamefont {Schuch},
  \citenamefont {Poilblanc}, \citenamefont {Cirac},\ and\ \citenamefont
  {P\'erez-Garc\'{\i}a}}]{schuch:topo-top}%
  \BibitemOpen
  \bibfield  {author} {\bibinfo {author} {\bibfnamefont {N.}~\bibnamefont
  {Schuch}}, \bibinfo {author} {\bibfnamefont {D.}~\bibnamefont {Poilblanc}},
  \bibinfo {author} {\bibfnamefont {J.~I.}\ \bibnamefont {Cirac}},\ and\
  \bibinfo {author} {\bibfnamefont {D.}~\bibnamefont {P\'erez-Garc\'{\i}a}},\
  }\href {https://doi.org/10.1103/PhysRevLett.111.090501} {\bibfield  {journal}
  {\bibinfo  {journal} {Phys. Rev. Lett.}\ }\textbf {\bibinfo {volume} {111}},\
  \bibinfo {pages} {090501} (\bibinfo {year} {2013})},\ \Eprint
  {https://arxiv.org/abs/arXiv:1210.5601} {arXiv:1210.5601} \BibitemShut
  {NoStop}%
\bibitem [{Note5()}]{Note5}%
  \BibitemOpen
  \bibinfo {note} {As long as we either choose boundary conditions with a fixed
  $\protect \mathrm {SU}(3)$ irrep, or consider the limit of an infinitely long
  cylinder and use that the transfer operator has a unique fixed point in each
  topological sector, which is given in the topological phase~\cite
  {schuch:topo-top}}\BibitemShut {NoStop}%
\bibitem [{\citenamefont {Cirac}\ \emph {et~al.}(2021)\citenamefont {Cirac},
  \citenamefont {P\'erez-Garc\'{\i}a}, \citenamefont {Schuch},\ and\
  \citenamefont {Verstraete}}]{cirac:tn-review-2021}%
  \BibitemOpen
  \bibfield  {author} {\bibinfo {author} {\bibfnamefont {J.~I.}\ \bibnamefont
  {Cirac}}, \bibinfo {author} {\bibfnamefont {D.}~\bibnamefont
  {P\'erez-Garc\'{\i}a}}, \bibinfo {author} {\bibfnamefont {N.}~\bibnamefont
  {Schuch}},\ and\ \bibinfo {author} {\bibfnamefont {F.}~\bibnamefont
  {Verstraete}},\ }\href {https://doi.org/10.1103/RevModPhys.93.045003}
  {\bibfield  {journal} {\bibinfo  {journal} {Rev. Mod. Phys.}\ }\textbf
  {\bibinfo {volume} {93}},\ \bibinfo {pages} {045003} (\bibinfo {year}
  {2021})},\ \Eprint {https://arxiv.org/abs/arXiv:2011.12127}
  {arXiv:2011.12127} \BibitemShut {NoStop}%
\bibitem [{Note6()}]{Note6}%
  \BibitemOpen
  \bibinfo {note} {When, as was mentioned before, $N$ is much smaller than the
  inverse entanglement gap}\BibitemShut {NoStop}%
\bibitem [{\citenamefont {Kass}\ \emph {et~al.}(1990)\citenamefont {Kass},
  \citenamefont {Moody}, \citenamefont {Patera},\ and\ \citenamefont
  {Slansky}}]{Kass1990}%
  \BibitemOpen
  \bibfield  {author} {\bibinfo {author} {\bibfnamefont {S.}~\bibnamefont
  {Kass}}, \bibinfo {author} {\bibfnamefont {R.}~\bibnamefont {Moody}},
  \bibinfo {author} {\bibfnamefont {J.}~\bibnamefont {Patera}},\ and\ \bibinfo
  {author} {\bibfnamefont {R.}~\bibnamefont {Slansky}},\ }\href@noop {} {\emph
  {\bibinfo {title} {Affine Lie Algebras, Weight Multiplicities, and Branching
  Rules}}},\ Vol.~\bibinfo {volume} {2}\ (\bibinfo  {publisher} {University of
  California Press},\ \bibinfo {year} {1990})\BibitemShut {NoStop}%
\bibitem [{\citenamefont {Knizhnik}\ and\ \citenamefont
  {Zamolodchikov}(1984)}]{Knizhnik1984}%
  \BibitemOpen
  \bibfield  {author} {\bibinfo {author} {\bibfnamefont {V.}~\bibnamefont
  {Knizhnik}}\ and\ \bibinfo {author} {\bibfnamefont {A.}~\bibnamefont
  {Zamolodchikov}},\ }\href
  {https://doi.org/https://doi.org/10.1016/0550-3213(84)90374-2} {\bibfield
  {journal} {\bibinfo  {journal} {Nucl. Phys. B}\ }\textbf {\bibinfo {volume}
  {247}},\ \bibinfo {pages} {83 } (\bibinfo {year} {1984})}\BibitemShut
  {NoStop}%
\bibitem [{Note7()}]{Note7}%
  \BibitemOpen
  \bibinfo {note} {Not to be confused with the $\protect \mathrm {SU}(3)_1
  \otimes \protect \overline {\protect \mathrm {SU}(3)_1}$ doubled Chern-Simons
  theory, the latter being a (2+1)-dimensional topological gapped theory, while
  the former is a (1+1)-dimensional gapless CFT.}\BibitemShut {Stop}%
\bibitem [{Note8()}]{Note8}%
  \BibitemOpen
  \bibinfo {note} {Moving the state within the basin in attraction of the
  boundary fixed point away from that fixed point}\BibitemShut {NoStop}%
\bibitem [{\citenamefont {Cardy}(2016)}]{Cardy2016}%
  \BibitemOpen
  \bibfield  {author} {\bibinfo {author} {\bibfnamefont {J.}~\bibnamefont
  {Cardy}},\ }\href
  {https://doi.org/https://doi.org/10.1088/1742-5468/2016/02/023103} {\bibfield
   {journal} {\bibinfo  {journal} {J. Stat. Mech.: Theory Exp.}\ }\textbf
  {\bibinfo {volume} {2016}}\bibfield  {number} {\bibinfo  {number} { (2)},\
  \bibinfo {pages} {023103}},\ }\Eprint
  {https://arxiv.org/abs/arXiv:1507.07266} {arXiv:1507.07266} \BibitemShut
  {NoStop}%
\bibitem [{Note9()}]{Note9}%
  \BibitemOpen
  \bibinfo {note} {This is the approach of Refs.~\cite
  {Calabrese2006,Calabrese2007} and others, which we will follow here first for
  simplicity.}\BibitemShut {Stop}%
\bibitem [{Note10()}]{Note10}%
  \BibitemOpen
  \bibinfo {note} {They are spelled out in detail for a number of chiral cases
  in Ref.~\cite {ArildsenLudwig2022}.}\BibitemShut {Stop}%
\bibitem [{Note11()}]{Note11}%
  \BibitemOpen
  \bibinfo {note} {When there are several degenerate multiplets at level $K$ in
  the charge $q$, flux $\phi $ sector, the labeling of the corresponding
  eigenvalues of the entanglement Hamiltonian would require an additional
  multiplicity label (these degeneracies will be split in entanglement energy,
  as seen in the Figures below); however, in the remainder of this paper we
  will use the explicit notation $\Delta E_{K,q\phi }$ only for non-degenerate
  eigenvalues.}\BibitemShut {Stop}%
\bibitem [{Note12()}]{Note12}%
  \BibitemOpen
  \bibinfo {note} {Only the primary states of the ``fast''-moving side factor
  into the low-energy part of the entanglement spectrum that we see, so unlike
  on the ``slow'' side for $\protect \overline {v}_R$ in the $\protect \ket
  {\protect \bm {1}}_R$ sector, it is not possible to deduce $v_L$ for the
  $\protect \ket {\protect \bm {1}}_L$ sector from states within the same
  sector on the fast-moving side. Thus we must assume that $v_L$ is the same,
  or close to it, in all three sectors, such that setting $\Delta E_{0,00} = 0$
  to obtain the entanglement excitation energy $\Delta E$ removes an identical
  constant term [see Eq.~\protect \eqref {eq:doubledh}] from all
  sectors.}\BibitemShut {Stop}%
\bibitem [{Note13()}]{Note13}%
  \BibitemOpen
  \bibinfo {note} {This is again subject to the assumption that $v_L$ is
  uniform across all primary state sectors. Then $v_L$ can be found from our
  knowledge of the conformal weight of the primaries, which is the approach
  that is taken here.}\BibitemShut {Stop}%
\bibitem [{Note14()}]{Note14}%
  \BibitemOpen
  \bibinfo {note} {Especially the green dispersion curve in the middle of the
  plot giving a value $v_L/\protect \overline {v}_R \sim 6.1$, or the blue
  curves giving a range of values $v_L/\protect \overline {v}_R \sim 5.4 -7.3$;
  these two kinds of curves have a longest discernible steep-slope segment,
  while the red curves are less suitable as they are shifted upwards and thus
  have a shorter steep-slope segment.}\BibitemShut {Stop}%
\bibitem [{\citenamefont {Arildsen}\ \emph {et~al.}(2023)\citenamefont
  {Arildsen}, \citenamefont {Chen}, \citenamefont {Schuch},\ and\ \citenamefont
  {Ludwig}}]{Arildsen2023}%
  \BibitemOpen
  \bibfield  {author} {\bibinfo {author} {\bibfnamefont {M.~J.}\ \bibnamefont
  {Arildsen}}, \bibinfo {author} {\bibfnamefont {J.-Y.}\ \bibnamefont {Chen}},
  \bibinfo {author} {\bibfnamefont {N.}~\bibnamefont {Schuch}},\ and\ \bibinfo
  {author} {\bibfnamefont {A.~W.~W.}\ \bibnamefont {Ludwig}},\ }\href@noop {}
  {\bibinfo {title} {{Entanglement Spectrum as a diagnostic of chirality of
  Topological Spin Liquids: Analysis of an $\mathrm{SU}(3)$ PEPS}}} (\bibinfo
  {year} {2023}),\ \Eprint {https://arxiv.org/abs/2305.13240} {arXiv:2305.13240
  [cond-mat.str-el]} \BibitemShut {NoStop}%
\bibitem [{\citenamefont {Kitaev}(2006)}]{KitaevAnnalsPhysics2006}%
  \BibitemOpen
  \bibfield  {author} {\bibinfo {author} {\bibfnamefont {A.}~\bibnamefont
  {Kitaev}},\ }\href
  {https://doi.org/https://doi.org/10.1016/j.aop.2005.10.005} {\bibfield
  {journal} {\bibinfo  {journal} {Ann. Phys. (NY)}\ }\textbf {\bibinfo {volume}
  {321}},\ \bibinfo {pages} {2} (\bibinfo {year} {2006})},\ \bibinfo {note}
  {{January Special Issue}},\ \Eprint
  {https://arxiv.org/abs/arXiv:cond-mat/0506438} {arXiv:cond-mat/0506438}
  \BibitemShut {NoStop}%
\bibitem [{\citenamefont {Calabrese}\ and\ \citenamefont
  {Cardy}(2006)}]{Calabrese2006}%
  \BibitemOpen
  \bibfield  {author} {\bibinfo {author} {\bibfnamefont {P.}~\bibnamefont
  {Calabrese}}\ and\ \bibinfo {author} {\bibfnamefont {J.}~\bibnamefont
  {Cardy}},\ }\href {https://doi.org/10.1103/PhysRevLett.96.136801} {\bibfield
  {journal} {\bibinfo  {journal} {Phys. Rev. Lett.}\ }\textbf {\bibinfo
  {volume} {96}},\ \bibinfo {pages} {136801} (\bibinfo {year} {2006})},\
  \Eprint {https://arxiv.org/abs/arXiv:cond-mat/0601225}
  {arXiv:cond-mat/0601225} \BibitemShut {NoStop}%
\bibitem [{\citenamefont {Calabrese}\ and\ \citenamefont
  {Cardy}(2007)}]{Calabrese2007}%
  \BibitemOpen
  \bibfield  {author} {\bibinfo {author} {\bibfnamefont {P.}~\bibnamefont
  {Calabrese}}\ and\ \bibinfo {author} {\bibfnamefont {J.}~\bibnamefont
  {Cardy}},\ }\href {https://doi.org/10.1088/1742-5468/2007/06/P06008}
  {\bibfield  {journal} {\bibinfo  {journal} {J. Stat. Mech.: Theory Exp.}\
  }\textbf {\bibinfo {volume} {2007}}\bibfield  {number} {\bibinfo  {number} {
  (06)},\ \bibinfo {pages} {P06008}},\ }\Eprint
  {https://arxiv.org/abs/arXiv:0704.1880} {arXiv:0704.1880} \BibitemShut
  {NoStop}%
\end{thebibliography}%

\end{document}